\documentclass[final,3p,twocolumn,authoryear,times]{elsarticle}

\usepackage[utf8]{inputenc}
\usepackage{graphicx}
\usepackage{lscape}
\usepackage{natbib}
\usepackage[normalem]{ulem}

\newcommand{\msun}{M$_{\odot}$}
\newcommand{\rsun}{R$_{\odot}$}
\newcommand{\amlt}{\alpha_{\rm MLT}}
\newcommand{\teff}{$T_{\rm eff}$}
\newcommand{\tauf}{\tau_{\rm fit}}
\newcommand{\lbol}{$L_{\rm bol}$}
\newcommand{\logg}{$\log g$}

\newcommand{\ha}{H$\alpha$}
\newcommand{\lha}{$L_{\rm H\alpha}$}

\journal{New Astronomy Reviews}

\begin{document}

\begin{frontmatter}
\title{Empirical Tests of Pre--Main-Sequence Stellar Evolution Models
with Eclipsing Binaries}

\author[vdb,fsk]{Keivan G.\ Stassun}
\ead{keivan.stassun@vanderbilt.edu}

\author[upp]{Gregory A.\ Feiden}
\ead{gregory.a.feiden@gmail.com}

\author[cfa]{Guillermo Torres}
\ead{gtorres@cfa.harvard.edu}

\address[vdb]{Department of Physics \& Astronomy, Vanderbilt University,
             1807 Station B, Nashville, TN 37235, USA}
\address[fsk]{Physics Department, Fisk University, 1000 17th Avenue 
              North, Nashville, TN 37208, USA}
\address[upp]{Department of Physics \& Astronomy, Uppsala University,
              Box 516, SE-751 20 Uppsala, Sweden}
\address[cfa]{Harvard-Smithsonian Center for Astrophysics, 60 Garden 
              Street, Cambridge, MA 02138, USA}

\begin{abstract}
We examine the performance of standard pre--main-sequence (PMS) stellar evolution 
models against the accurately measured properties
of a benchmark sample of 26 PMS stars in 13 eclipsing binary (EB) systems
having masses 0.04--4.0 \msun\ and nominal ages $\approx$1--20 Myr.
We provide a definitive compilation of all 
fundamental properties for the EBs, with a careful and consistent reassessment
of observational uncertainties. We also provide a definitive compilation of
the various PMS model sets, including physical ingredients
and limits of applicability.
No set of model isochrones is able to successfully 
reproduce all of the measured properties of all of the EBs.
In the H-R diagram, the masses inferred for the individual stars by the models
are accurate to better than 10\% at $\gtrsim$1 \msun, but below 1 \msun\ they
are discrepant by 50--100\%.
Adjusting the observed radii and temperatures using empirical relations 
for the effects of magnetic activity helps to resolve the discrepancies in 
a few cases, but fails as a general solution.
We find evidence that the failure of the models to match
the data is linked to the triples in the EB sample; at least half of the 
EBs possess tertiary companions. 
Excluding the triples, the models reproduce the stellar masses 
to better than $\sim$10\% in the H-R diagram, down to 0.5 \msun,
below which the current sample is fully contaminated by tertiaries.
We consider several mechanisms 
by which a tertiary might cause changes in the EB properties and
thus corrupt the agreement with stellar model predictions.
We show that the energies of the tertiary orbits are comparable to 
that needed to potentially explain the scatter in the EB properties
through injection of heat, 
perhaps involving tidal interaction.
It seems from the evidence at hand that this mechanism, however it
operates in detail, has more influence on the
surface properties of the stars than on their internal structure, as the
lithium abundances are broadly in good agreement with model predictions.
The EBs that are members of young clusters appear individually coeval to 
within 20\%, but collectively show an apparent age spread of $\sim$50\%,
suggesting true age spreads in young clusters. However, this 
apparent spread in the EB ages may also be the 
result of scatter in the EB properties induced by tertiaries.
%
\end{abstract}


\end{frontmatter}

\section{Introduction\label{sec:intro}}

\subsection{Eclipsing binaries as tests of stellar models}
Eclipsing binary (EB) stars have long served as fundamental benchmarks
in stellar astrophysics. Through spectroscopic and photometric analysis
of an EB, it is possible to empirically measure the fundamental physical
properties of the component stars to a high degree of both precision and
accuracy with almost no theoretical assumptions
\citep[e.g.,][]{Andersen:91}. 
For example, through radial velocities measured from a set of double-lined
spectra, the component masses can be directly determined, with an accuracy
of $\sim$1\% in the best cases
\citep[e.g.,][]{Morales2009a}. 
Indeed, the number of EBs with component masses and radii determined with
an accuracy of better than 3\% is now approximately 100 \citep{Torres:10}. 

Such accurately determined stellar properties enable stringent tests
of stellar models. For main-sequence EBs, these tests generally find that the models perform very well at
masses $\gtrsim 1$ \msun, but at lower masses there are important
discrepancies. For example, low-mass EBs exhibit systematically
lower effective temperatures (\teff) and systematically larger 
radii ($R$) than predicted by standard stellar models 
\citep[see, e.g.,][]{Torres2002, Ribas2003, Lopezm2005, Torres2013}. 
This may be the result of the magnetic activity that is
often observed among low-mass stars. 
While there is not yet a clear
consensus in the literature as to the physical mechanism that drives
the connection between activity, \teff\ suppression, and $R$
inflation, 
several studies have suggested that the effect can be
empirically related to activity
indicators such as \ha\ and X-ray emission 
\citep[e.g.,][]{Lopezm2007, Morales2008, Stassun2012}.
If correct,
these relations might allow empirical corrections to be made
to the inferred masses of active, low-mass stars, and they suggest
that there are missing physical ingredients in standard
models, most notably magnetic fields. 

In the last 15 years, new generations of theoretical stellar evolution 
models including the effects of magnetic fields have been developed
\citep{Antona2000,MM01,FC12b} that are better able to reconcile the 
observed \teff\ and $R$ of some observed low-mass EBs \citep{MM10,
MM11,FC13}. For instance, \citet{FC12b} 
are able to successfully model the previously vexing EB
EF\,Aqr by incorporating magnetic fields with strengths of 
3--5 kG in their stellar models in a fully physically consistent
fashion. Similarly, \citet{FC13} find that YY\,Gem requires models 
with field strengths of several hundred Gauss driven by convective energy
to reconcile the observed inflated $R$ and suppressed \teff.

At the same time, application of these new models questions the degree 
to which the observed properties for most EBs in fact systematically
deviate from expectation. Potentially underestimated systematic 
errors in EB measurements and uncertainties in age and/or metallicity 
could be lurking in the data. \citet{FC12} find that more than
90\% of the benchmark EBs they considered have properties fully consistent 
with the Dartmouth stellar models when the previously assumed EB ages 
and metallicities are carefully re-examined. For instance, 
the low-mass EB CU~Cnc, considered a quintessential example of
activity-inflated $R$, is found to in fact be well modeled by
standard models if its age is several Gyr as opposed to the
commonly assumed $\sim$400 Myr; essentially, CU~Cnc may not
be a member of the Castor moving group as has been commonly assumed
\citep{FC13}.

Thus, while the effects of magnetic activity appear to be important
for some extremely active low-mass EBs \citep{Stassun2012,FC12b,FC13},
it is proving valuable to carefully
reexamine the ability of standard stellar models to reproduce the
majority of EBs when confronted with realistic observational
uncertainties and more accurate priors on other relevant stellar
properties, such as the stellar ages. 

\subsection{Pre--main-sequence eclipsing binaries}

The complexity of the stellar physics intrinsic to low-mass stars
is exacerbated for low-mass stars in the pre--main-sequence (PMS)
phase of evolution. PMS stars are expected to be fully convective
in the early stages of PMS evolution, making the difficult physics
of convection central to the modeling problem. Very young stars
are generally rapidly rotating, at least as compared to their
main sequence counterparts which have begun to spin down, and thus
almost always exhibit some indicator of strong magnetic activity
\citep[e.g., periodic variability attributed to spots, strong
H$\alpha$ and X-ray emission, etc.;][]{Stassun1999,Stassun:04,Stassun:06},
making any effects of strong surface magnetic fields ubiquitous.
All PMS stars have presumably undergone a phase of vigorous 
accretion, if they are not still actively accreting, making any
effects of previous accretion history salient 
\citep[e.g.,][]{Baraffe2009,Simon2009,Baraffe2010}.
Finally, PMS stars by definition have not yet established
full
main-sequence hydrostatic equilibrium and are still 
contracting. 
This contraction of PMS stars furthermore
suggests that stars in close binary systems may have previously
experienced strong(er) tidal interactions when the stars were even
larger than observed at the present moment. Thus, from a modeling
standpoint alone, PMS stars are complex, dynamic, and may be prone
to effects from their recent evolutionary history 
(i.e., accretion, tides) that may or may
not still be directly observable.

\begin{figure*}[!ht]
    \centering
    \includegraphics[width=0.45\linewidth]{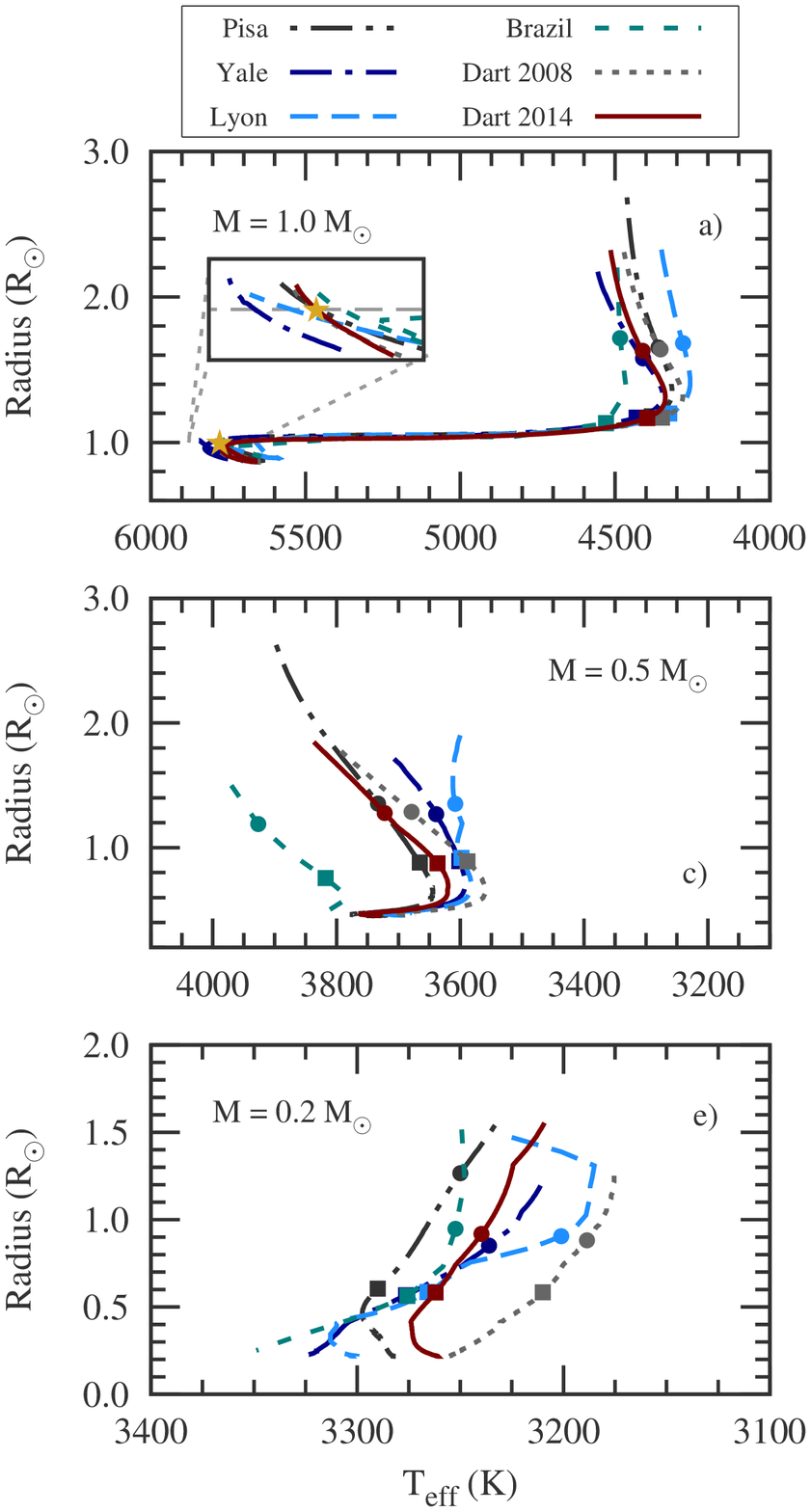} \qquad
    \includegraphics[width=0.45\linewidth]{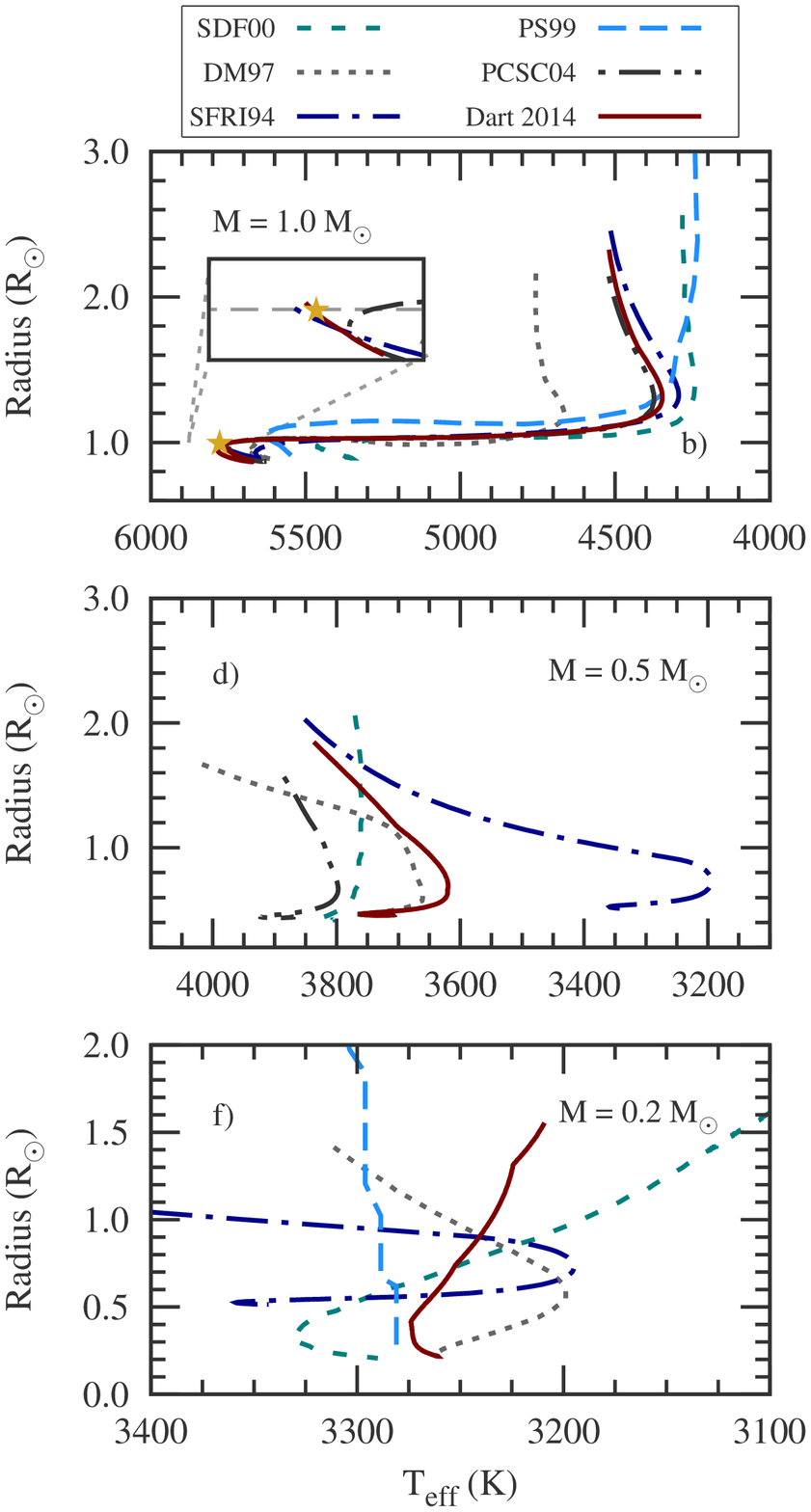}
    \caption{\label{fig:trackoverview} 
     Model evolutionary tracks for stars of 1.0 \msun\ (a and b), 0.5 \msun\ (c and d)
     and 0.2 \msun\ (e and f). Left panels (a, c, e) display tracks for model sets 
     adopted in the review, while right panels (b, d, f) show tracks from discarded model 
     sets. The Dartmouth 2014 models are shown in both the left and right panels for reference.
     Each track begins at an age of 1 Myr and terminates on the main sequence.
     Yellow star symbols in the 1.0 
     \msun\ panels represent the Sun, and small insets highlight the quality of the 
     solar calibration. Two ages, 3 Myr and 10 Myr, are marked along the adopted tracks with a 
     solid circle and solid square, respectively.
    }
\end{figure*}

Observationally, PMS stars are challenging as well. During the first
few Myr of evolution, many if not most PMS stars exhibit variability
and complexities arising from active accretion 
\citep[e.g.,][]{Herbst1994} that are difficult  
if not impossible to separate from magnetic activity. These stars
are also frequently observed within larger star-forming regions
that have strong nebular emission and that subject the individual
stars to a highly dynamical environment including gravitational
interactions with the other stars \citep[e.g.,][]{Bate2009}. 
Furthermore, the determination of basic stellar properties for PMS
stars is often confounded by uncertain circumstellar and/or 
environmental extinction, uncertain distances, and even uncertainties
in the \teff\ scale and bolometric corrections. As a consequence,
there remain ongoing debates in the literature as to the degree of
coevality in young star-forming regions because of large uncertainties
in the observed stellar luminosities \citep[e.g.,][]{Kraus2009,DaRio2010,Jeffries2011}.

Empirical testing and benchmarking of PMS stellar evolution models
is important for multiple reasons. These models are the basis for
the inferred masses and ages of the vast majority of young stars.
Consequently, our understanding of basic stellar astrophysical
ingredients, such as initial mass functions, depends on these models.
Likewise, our understanding of the timescales for the dissipation of
circumstellar disks and for the formation of planets around stars relies entirely
on the age scale implied by the models.
But transforming the directly observed properties of young stars,
such as temperatures and luminosities, into masses and ages using
the current suite of theoretical PMS stellar models results in 
a factor of $\sim$2--3 variation 
in mass and in age due to large differences in 
theoretical evolutionary 
PMS tracks \citep[Figure \ref{fig:trackoverview}, and see, e.g.,][]{Simon2000,Hillenbrand2004,Soderblom2014}. 
For example, given a star with $R = 1.0$~\rsun\ and \teff\ = 3600~K, 
PMS stellar evolution models listed in Tables~\ref{tab:models_in} and 
\ref{tab:models_ex} predict that star to have a mass of 0.33~\msun\ up to 
0.60~\msun\ with an age anywhere from 3 Myr to 10 Myr.
Consequently, the predictions of PMS stellar models remain
very much in need of benchmarking against accurate empirical 
measurements of PMS stellar properties. 

Previous reviews of empirical constraints on PMS models include \citet{Young:01},
\citet{Hillenbrand2004}, \citet{Mathieu2007}, 
\citet{Tognelli2012}, \citet{Gennaro2012}, and \citet{Bell:12}. 
Some of those reviews considered
all PMS stars with dynamical mass determinations, including PMS
EBs, PMS visual binaries with astrometric orbits, and single PMS
stars with rotation curve measurements of their circumstellar disks.
In this review, we focus exclusively on the sample of published,
well characterized PMS EBs, as these provide
direct measurements of both the stellar masses and the stellar radii,
as well as additional empirical determinations of quantities such as
mass ratios and temperature ratios that can be used as stringent
constraints on stellar models. 

Fortunately, the past decade has seen advances in the number of
PMS EBs with which to perform such tests.
Whereas the most recent review on PMS EBs \citep{Mathieu2007}
included only six PMS EB systems,
as of this writing there are now 13 PMS EB systems published with suitably 
well determined properties (see Sect.~\ref{sec:sample}). 
While only three of these (EK\,Cep, RS\,Cha, and V1174\,Ori)
have mass and radius measurements
of such high accuracy as the $\sim$100 main-sequence EBs included in
the review by \citet{Torres:10}, these systems collectively 
possess sufficien\-tly well measured stellar properties to permit a
quantitative assessment of the various PMS stellar model suites
currently available. They also permit an assessment
of the degree to  
which non-standard stellar physics may be required to improve the
performance of the models. 
Moreover, EBs bring the added benefit
of permitting one to test the stellar models in the context of
coevality, or in turn to ascertain the degree to which non-coevality
may exist in young binaries. Indeed, the sample of PMS EBs
that we study includes members of at least two 
young clusters 
with very different ages (Orion Nebula Cluster at $\sim$1--2 Myr,
and h\,Persei at $\sim$13 Myr) so that we may investigate the question
of coevality also across multiple EBs within the same cluster.

Our aim in this paper is not principally to perform a comprehensive
review of the literature on this subject 
(see previous reviews cited above).
Rather, we aim to (a) objectively compile
the fundamental properties of a benchmark sample of PMS EBs in one
place as a resource to the community, 
(b) objectively compile the salient physical ingredients for the
various PMS model suites in one place so that the community may
more easily compare and contrast them, (c) systematically compare the
physical predictions of the models against the EB measurements, and
(d) synthesize the results of these comparisons in an attempt to 
identify the most important physical effects that will be needed to make 
progress in understanding PMS stars and the efficacy of PMS stellar models.

The remainder of the paper is structured as follows:
Sect.~\ref{sec:sample} presents the 13 PMS EBs that we use as our
benchmark sample and describes the basis for their selection.
Sect.~\ref{sec:models} presents the various stellar model sets
that we compare against the measurements, summarizes their physical
ingredients and assumptions, compares them to one another, and
describes the methods by which we compare them to the data. This
section also includes our thoughts regarding which model sets 
should continue to be used and which should be retired.
Sect.~\ref{sec:results} presents the basic results of our attempts
to fit the stellar models to the EB measurements, including an
assessment of which model sets perform best in parameter spaces
of particular interest to observers seeking to utilize the stellar
models for determinations of basic properties for young stars,
and includes an examination of activity effects and lithium abundances.
Sect.~\ref{sec:disc} discusses these results and synthesizes
them around a discussion of various observational issues that will 
need to be resolved as well as various astrophysical effects that
future generations of stellar models may need to incorporate,
including magnetic fields, tidal interaction, and accretion history.
In Sect.~\ref{sec:summary} we end with a summary of our
conclusions and recommendations, and briefly identify key directions
and challenges for future progress.

\section{Sample of PMS Eclipsing Binaries\label{sec:sample}}

In this paper, we will utilize a set of benchmark-grade EBs
at PMS ages for comparison against various theoretical PMS 
stellar evolution models. We restrict our consideration to EBs
that satisfy the following criteria: 
\begin{enumerate}
\item At least one peer-reviewed paper has been published
that includes at least a double-lined
spectroscopic orbit solution providing the two EB component masses 
($M$) and a light curve analysis providing the two EB component radii
($R$) and effective temperatures (\teff).
\item The system includes at least one component whose properties place it
definitively above the nominal zero-age main sequence (ZAMS), and where the
discovery paper(s) identify the system as being likely PMS.
\item Component masses below 5 \msun\ with reported uncertainties 
less than 10\%.
\end{enumerate}



\begin{table*}[!htbp]
\caption{Fundamental properties of young EBs\label{tab:data1}}
\small
\begin{tabular}{l c c c c c c @{}c }
\noalign{\smallskip}\hline
\hline\noalign{\smallskip}
 Star & Per (d)   & Comp & Mass (\msun) & Radius (\rsun) & \teff\ (K) & $q\equiv M_B/M_A$ & Tertiary? \\
      & $V$ (mag) &      &                    &                      &                   & $\Delta T_{\rm eff}$ (K) & \\
\noalign{\smallskip}\hline\noalign{\smallskip}
V615 Per                &13.714  & A & $4.075 \pm 0.055$  & $2.29  \pm  0.14$   &  $15000 \pm  500$ & $0.7801 \pm 0.0098$ & N \\
                        & 13.02  & B & $3.179 \pm 0.051$  & $1.903 \pm  0.094$  &  $12700 \pm  700$ &   $2300 \pm 500$  &  \\

\noalign{\medskip}

TY CrA                  & 2.889  & A & $3.16  \pm 0.08$   & $1.80   \pm 0.10$   & $12000 \pm  500$ & $0.5176 \pm 0.0052$ & Y \\
                        &  9.30  & B & $1.64  \pm 0.04$   & $2.08   \pm 0.14$   &  $4900 \pm  400$ &     $7100 \pm 300$  &  \\
\noalign{\medskip}

V618 Per                & 6.367  & A & $2.332 \pm 0.031$  & $1.64   \pm 0.10$   & $11000 \pm 1000$ & $0.6682 \pm 0.0087$  & N \\
                        & 14.62  & B & $1.558 \pm 0.025$  & $1.32   \pm 0.10$   &  $8100 \pm  700$ &   $2900 \pm 500$  & \\

\noalign{\medskip}

EK Cep                  & 4.428  & A & $2.025 \pm 0.023$  & $1.5800 \pm 0.0065$ &  $9000 \pm  200$ & $0.5540 \pm 0.0039$ & N  \\
                        &  7.85  & B & $1.122 \pm 0.012$  & $1.3153 \pm 0.0057$ &  $5600 \pm  200$ &   $3400 \pm 150$   &   \\
    
\noalign{\medskip}

RS Cha                  & 1.670  & A & $1.854 \pm 0.016$  & $2.138  \pm 0.055$  &  $7640 \pm  180$ & $0.9798 \pm 0.0056$ & Y?  \\
                        &  6.04  & B & $1.817 \pm 0.018$  & $2.339  \pm 0.055$  &  $7240 \pm  170$ &    $400 \pm 30$     &   \\

\noalign{\medskip}

ASAS                & 3.873  & A & $1.375 \pm 0.028$  & $1.83   \pm 0.07$   &  $5100 \pm  100$ & $0.9719 \pm 0.0067$ & N  \\
J052821$+$0338.5    & 11.71  & B & $1.329 \pm 0.020$  & $1.73   \pm 0.07$   &  $4750 \pm  180$ &    $350 \pm 150$    &   \\

\noalign{\medskip}
                        
RX J0529.4+0041         & 3.038  & A & $1.27  \pm 0.01$   & $1.44   \pm 0.10$   &  $5200 \pm  150$ & $0.7305 \pm 0.0025$ & Y   \\
                        & 12.35  & B & $0.93  \pm 0.01$   & $1.35   \pm 0.10$   &  $4220 \pm  150$ &    $980 \pm 50$     &    \\
    
\noalign{\medskip}
            
V1174 Ori               & 2.635  & A & $1.006 \pm 0.013$  & $1.338  \pm 0.011$  &  $4470 \pm  120$ & $0.7231 \pm 0.0055$ & Y  \\
                        & 13.95  & B & $0.7271\pm 0.0096$ & $1.063  \pm 0.011$  &  $3615 \pm  100$ &    $855 \pm 50$    &   \\
        
\noalign{\medskip}
                    
MML 53                  & 2.098  & A & $0.994 \pm 0.030$  & $2.201  \pm 0.071$$^a$  &  $4886 \pm  100$ & $0.863  \pm 0.016$  & Y  \\
                        & 10.78  & B & $0.857 \pm 0.026$  &       ...           &  $4309 \pm  100$ &     ...            &   \\

\noalign{\medskip}

CoRoT                   & 3.875  & A & $0.668 \pm 0.012$  & $1.295  \pm 0.040$  &  $4000 \pm  200$ & $0.7417 \pm 0.0074$ & N  \\
223992193               & 16.69  & B & $0.4953\pm 0.0073$ & $1.107  \pm 0.050$  &  $3750 \pm  200$ &     ...            &   \\
    
\noalign{\medskip}
                    
Par 1802                & 4.674  & A & $0.391 \pm 0.032$  & $1.73   \pm 0.015$  &  $3675 \pm  150$ & $0.985  \pm 0.029$ & Y  \\
                        & 15.38  & B & $0.385 \pm 0.032$  & $1.62   \pm 0.015$  &  $3365 \pm  150$ &    $310 \pm 40$    &   \\

\noalign{\medskip}

JW 380                  & 5.299  & A & $0.262 \pm 0.025$  & $1.189  \pm 0.175$  &  $3590 \pm  150$ & $0.577  \pm 0.032$ & Y  \\
                        & 16.92  & B & $0.151 \pm 0.013$  & $0.897  \pm 0.170$  &  $3120 \pm  100$ &    $470 \pm 70$    &   \\

\noalign{\medskip}

2MASS                   & 9.780  & A & $0.0572\pm 0.0033$ & $0.690  \pm 0.011$  &  $2715 \pm  200$ & $0.639  \pm 0.024$ & N   \\
J05352184$-$0546085     & 13.47K & B & $0.0366\pm 0.0022$ & $0.540  \pm 0.009$  &  $2850 \pm  200$ &  $-135 \pm 30$     &    \\
\noalign{\smallskip}\hline
\end{tabular}
\vskip 5pt
{$^a$ Radius sum; the individual radii have not been determined for this system.}
\end{table*}



\begin{figure*}[!t]
    \centering
    \includegraphics[width=0.8\linewidth]{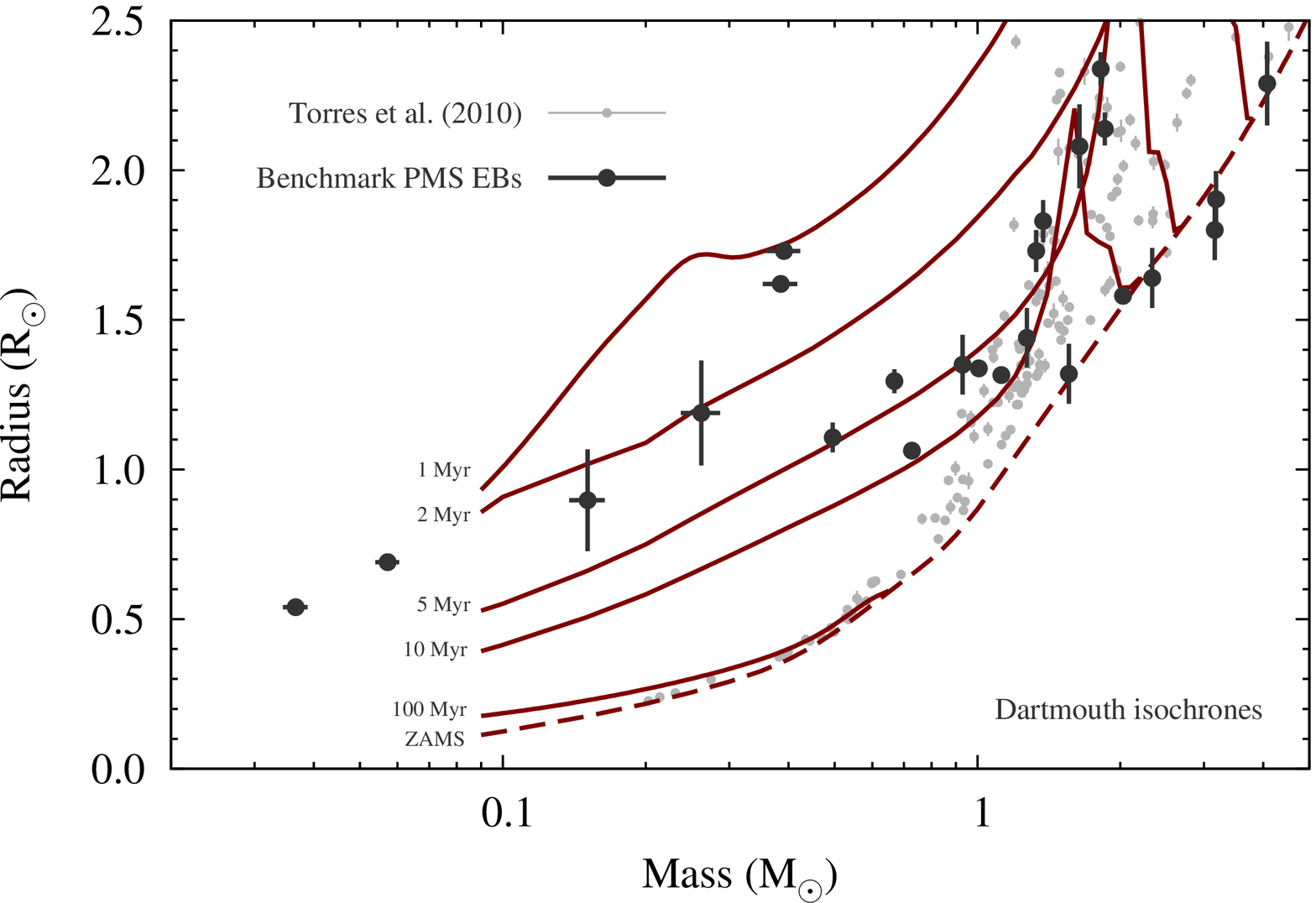}
    \caption{Overview of the benchmark EBs. Isochrones are from the Dartmouth
    2014 series, black points are the benchmark EBs considered
    in this paper, and light grey points are the benchmark EBs from 
    \cite{Torres:10}. The zero-age-main-sequence (ZAMS) is shown as a dashed line.
    Noticeable features in the isochrones include the 
    deuterium burning bump at low-masses around 1~Myr, and the rise and fall of
    stellar radii for stars with masses $\gtrsim$1.2 \msun\ prior to the ZAMS (see Sec.~\ref{sec:models}).
    \label{fig:allebs}}
\end{figure*}

The EBs we selected from the literature 
satisfying the above criteria are summarized in 
Table~\ref{tab:data1} and shown in Figure \ref{fig:allebs}. 
Since the sample is relatively small, we have allowed three exceptions 
to these criteria in cases of special interest. These are the systems
V615\,Per and V618\,Per, which are essentially on the 
ZAMS but still young enough to provide useful tests, and MML\,53, for which the individual
radii have not been measured (although their sum has) but which has a determination of the
Li abundance for both stars that constitutes a useful constraint on models.

We have carefully reviewed the primary references pertaining to each 
of the 13 EB systems in order to confirm that they are otherwise suitable as 
benchmark-grade comparisons to theoretical PMS evolution models,
and notes pertaining to each one based on our review of the 
literature are provided in \ref{sec:ebnotes}. 
This final sample set for
our analysis below comprises 26 individual components in 13 EB 
systems, with masses spanning the range 0.037--4.075 \msun\ and
nominal ages of $\approx$1--20 Myr. 

Table~\ref{tab:data1} also includes the fundamental physical parameters of each
EB, namely $M$, $R$, and \teff, in order of primary star mass ($M_A$).
We do not tabulate bolometric luminosity (\lbol)
or surface gravity (\logg) as these can be 
calculated directly from $M$, $R$, and \teff. We also provide the 
mass ratio ($q \equiv M_B / M_A$) and the
difference in \teff, as these quantities are usually 
directly determined from the EB radial velocity and light curve analysis.
In particular, $\Delta$\teff\ is generally determined more precisely than 
the individual absolute temperatures. Finally, Table~\ref{tab:data1} also 
indicates whether the EBs possess known tertiary companions. 
The distance and nominal age for
each system are collected in Table~\ref{tab:data2} (\ref{sec:ebnotes}). 
These ages are assigned primarily based on
cluster membership (for those EBs that are established kinematic members of
young clusters or associations) 
or else are the nominal age provided in the original papers.

Note that two of the EBs (four stars) in our final sample set
are members of the h~Persei
cluster with a nominal age of $\approx$13 Myr \citep[e.g.,][]{Capilla:02}, 
and three of the EBs (4 stars and 2 brown dwarfs)
are members of the Orion Nebula Cluster (ONC) star-forming region with
nominal age of $\approx$1--2 Myr \citep[e.g.,][]{Hillenbrand1997,Mayne2008,DaRio2010}.

In Table~\ref{tab:data2} we provide for each of these EBs some
ancillary information that we will utilize in our analysis. These data
include \ha\ equivalent widths (EWs) and luminosities (\lha), X-ray fluxes
and luminosities ($F_{\rm X}$ and $L_{\rm X}$), and the
abundances of lithium, $\log N{\rm (Li)}$. \lha\ and $L_{\rm X}$ can be used as tracers of 
chromospheric activity for assessing the impact of magnetic activity on
the stellar properties, whereas Li can be used as an independent measure
of the stellar age since Li is not yet fully depleted in most low-mass 
PMS stars.

\section{Stellar Evolution Models Examined and Methods of Comparison
to Measured Stellar Properties\label{sec:models}}

In this section, we summarize the stellar evolution
models against which we compare the sample of benchmark PMS EBs
from Sect.~\ref{sec:sample}. All of the published and publicly 
available model sets with specific applicability to PMS
evolution for low-mass stars have been considered and are summarized
in Tables~\ref{tab:models_in} and \ref{tab:models_ex}.
Each set of models includes a range of applicability and a set 
of physical ingredients and assumptions as summarized in
Tables~\ref{tab:models_in} and \ref{tab:models_ex}. For context, 
evolutionary tracks for
1~\msun, 0.5~\msun, and 0.2~\msun\ are illustrated in the
\teff--$R$ plane in Figure~\ref{fig:trackoverview}
and a representative set of isochrones are shown in 
Figure~\ref{fig:allebs}\footnote{The isochrones show an increase
and then a decrease in radius as stars approach the ZAMS. 
The initial radius increase is a response
    of the stellar envelope to increased energy output from the $p$--$p$ chain
    and gravitational contraction of the core, producing greater ionization in
    the outer layers and thus reducing the size of the outer convection zone. 
    Strong burning contributions from the CN cycle reverse the increasing 
    radii by impeding core contraction and triggering formation of a convective core. 
    Progress toward the ZAMS is temporarily halted as the CN cycle efficiency is 
    reduced prior to ignition of nitrogen burning and equilibration of the complete 
    CNO cycle.}.

There exists a large number of stellar evolution codes, each typically
designed to suit a particular application 
given that the physical
conditions present inside stars vary over orders of magnitude in pressure
and temperature. Additionally, physical inputs used in stellar evolution
codes (e.g., equation of state, opacities, boundary conditions) are undergoing
constant revision, and consequently so are the stellar evolution models.
Our attempt at a near-comprehensive list of published stellar models with 
applicability to PMS evolution of low-mass stars is given in 
Tables~\ref{tab:models_in} and \ref{tab:models_ex}.


We have endeavored to select a subset of the theoretical calculations that we believe are 
representative of what modelers might consider to be
the current best effort at producing models with realistic physical inputs.
We define the following set of criteria for the inclusion of a model set in our 
review:
\begin{enumerate}
    \item Models are or will be publicly available.
    \item Non-grey surface boundary conditions are used.
    \item Models have been solar-calibrated (i.e., they reproduce the Sun's
    properties at the solar age).
\end{enumerate}
The series of models that meet these criteria are listed in Table \ref{tab:models_in},
while those that do not are listed in Table \ref{tab:models_ex}.
One may certainly debate these particular criteria as a global 
discriminant for adopting a given model set. For instance, it can be
argued that there is no reason, {\it a priori}, to calibrate the 
convective mixing-length parameter 
to the Sun when the focus of the investigation is on cooler, PMS stars. 
Therefore our exclusion of certain model sets is not meant to suggest that 
these models are ``wrong'' or not suitable for use in certain contexts. 
Users of stellar models must carefully
evaluate the applicability of any given model set to the problem at hand,
based on the physical inputs and assumptions that went into the calculations.
The information assembled in Tables \ref{tab:models_in}--\ref{tab:models_ex}
is therefore intended as a guide to users who might otherwise have difficulty
extracting this information readily from the original literature.

\subsection{Physical Ingredients: Criteria\label{sec:criteria}}

The characteristics of stellar evolution models listed above were selected
as discriminants to optimize the validity of the adopted models across 
the mass regime spanned by our benchmark EBs. Surface boundary conditions
defined by non-grey atmospheres are valid across the entire mass regime,
from the sub-stellar regime up to the late-B-type stars. While grey atmosphere
boundary conditions may be applied to higher mass stars ($> 0.9$~\msun)
with relative accuracy, the same cannot be said of the lower mass population.
Below roughly 5000\,K it becomes increasingly important to adopt non-grey
boundary conditions, as convection and radiation both play important roles
in the transport of flux in the molecule-ridden cool star atmospheres. 
Therefore, while non-grey atmospheres would not be a limiting factor for
the applicability of one model set against a given EB, a grey atmosphere
would certainly provide a much less physically appropriate analysis against 
a low-mass star.

A similar argument can be made for the inclusion of modern opacity tables,
although there has generally been widespread adoption of the latest opacity tables,
so no model sets were excluded based on their adopted opacity data. 

Solar calibration of stellar models is a continual source of debate. While
few would argue against calibrating initial abundances of helium and heavy
elements, the mixing length of convection is a highly contentious subject.
For the purposes of this review, we sought to obtain a consistent sample 
of stellar models to permit a fair comparison based on minimal
assumptions and as little tweaking of model parameters as possible. The
simplest means of obtaining this sample is to adopt models that have been
calibrated to the Sun. 

This does raise a specific concern in the context of strong activity or 
magnetic field effects, which might cause departures from the solar convective
calibration. For example, strong surface fields have been suggested to cause
a reduction in surface convective efficiency (effectively, a reduction in the
convective mixing-length parameter) for PMS stars. In our analysis, we
first adopt the models with solar-calibrated convection, and then consider 
the possible effects of magnetic activity separately.

Lastly, it was required that models be publicly available. This criterion
has nothing to do with the physical ingredients and is in no way indicative
of the quality of a given model set. We enforced this criterion to make sure
we were evaluating models that are more accessible to the wider community and
therefore more likely to be adopted. Only in the case of the \citet{Montalban2004}
tracks did we find this to be a restriction.
Most model sets having a large grid are publicly available.

\subsubsection{Adopted Evolutionary Tracks}

In Table~\ref{tab:models_in} we list the
six model sets that satisfied our criteria above
for inclusion: \citet[hereafter Lyon]{BCAH98} models, \citet[hereafter
Dartmouth 2008]{Dotter2008} models, \citet[hereafter Brazil]{Landin2010} 
models, \citet[hereafter Pisa]{Tognelli2011} models, \citet[hereafter 
Yale]{Spada2013} models, and Feiden et al.\ (in prep., hereafter 
Dartmouth 2014) models. 
These are also displayed in Fig.~\ref{fig:trackoverview} (left panels).
Note that the Dartmouth and Yale models are of the same lineage, both having
developed from the Yale Rotating Evolution Code \citep{Guenther1992}. Despite
their common lineage, the physics specific to low-mass stars developed 
independently, as evidenced in Table~\ref{tab:models_in}.
It is {\bf also} important to note that while the Yale and 
Dartmouth 2008 models were not specifically designed for PMS studies, 
the physics included in the models are still valid in the PMS regime. 
Only at sufficiently young ages of around 1 Myr are there noticeable 
deficiencies in the adopted physics due to a lack of deuterium burning. 
The Brazil models are also available in a version that includes rotation 
\citep{Landin2006,Landin2009}, but we have adopted the non-rotating 
models for consistency with the other model sets.

\subsubsection{Discarded Evolutionary Tracks}

Stellar evolution model sets that did not satisfy the criteria for
inclusion are listed in Table~\ref{tab:models_ex}
(and are also displayed in the right panels of 
Fig.~\ref{fig:trackoverview}).
Some widely adopted 
stellar models are contained within this list,
including the \citet{DAntona1997} and \citet{Siess2000}
models. We 
recommend that 
use of these models should be approached with care. 
We briefly discuss why each of these model sets was excluded in 
\ref{sec:tracknotes}.

\subsection{Inter-model Comparison}
\subsubsection{Mass Tracks}

%

\begin{figure*}[!t]
    \includegraphics[width=0.8\linewidth]{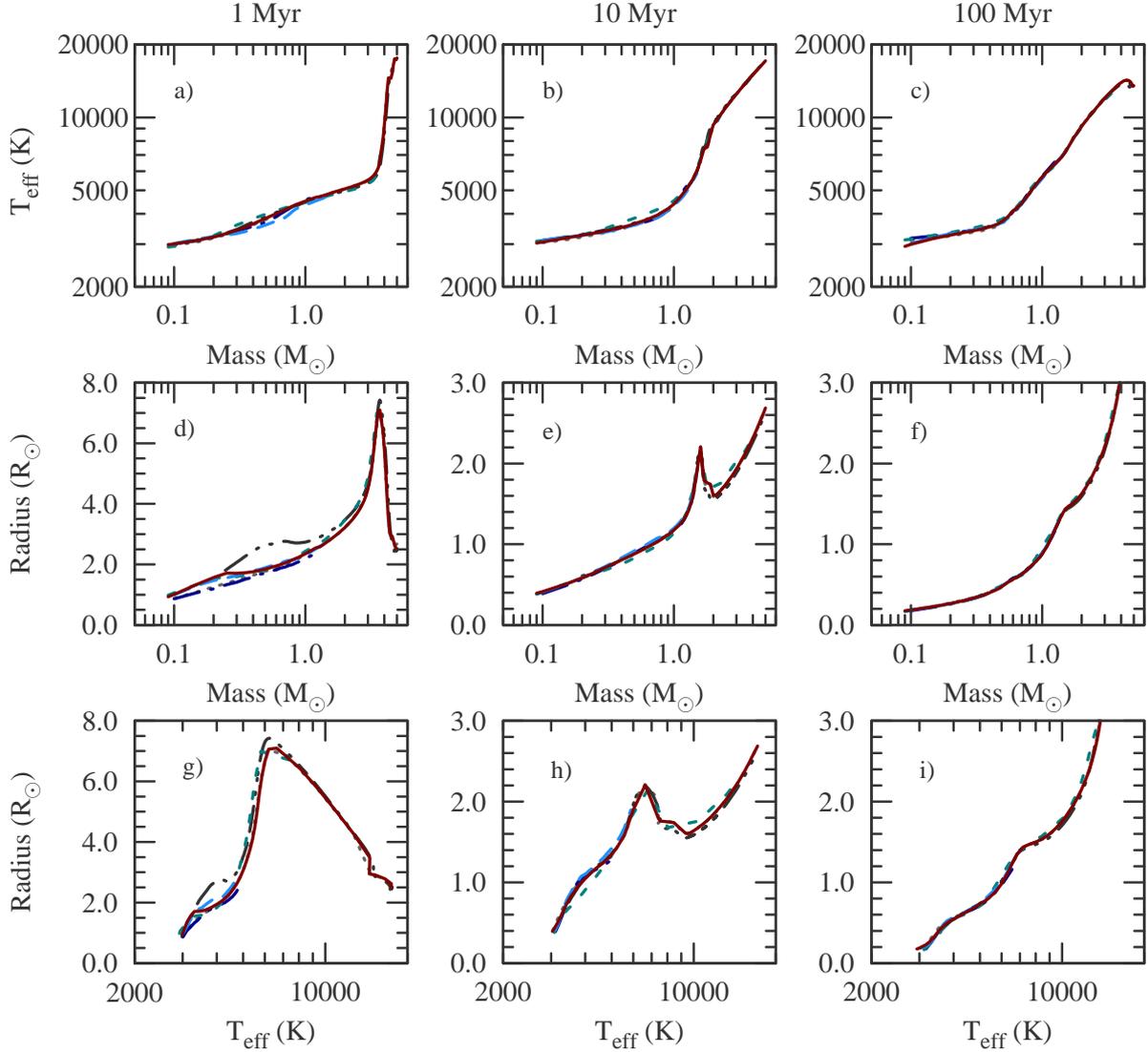}
    \caption{Comparison of accepted model isochrones at 1, 10, and 100~Myr
     in three different theoretical planes. Line styles represent the same 
     stellar models as in Figure \ref{fig:trackoverview}a): Brazil (green,
     short-dashed), Dartmouth (2014; maroon, solid), Dartmouth (2008; grey, 
     dotted), Lyon (light-blue, long-dashed), Pisa (black, dash-double-dotted),
     Yale (dark-blue, dash-dotted). Panels a) -- c) show the mass--\teff\
     plane, d) -- e) show the mass--radius plane, and g) -- i) show the 
     \teff\--radius plane.}
    \label{fig:isooverview}
\end{figure*}


To illustrate how the various criteria affect the predictions of stellar
models, we compare mass tracks and isochrones for all
model sets in Figures~\ref{fig:trackoverview} and
\ref{fig:isooverview}.
As shown in Figure~\ref{fig:trackoverview}(a, c, e), accepted mass 
tracks yield similar predictions at all masses. Noticeable differences for 
1.0~\msun\ tracks occur at the youngest ages along the Hayashi track, typically 
before an age of 10 Myr, which is marked as solid square along each track. After
10 Myr, the model sets exhibit consistent properties on the nearly-horizontal 
Henyey track on the approach to the MS. A similar trend emerges at 0.5~\msun, 
where the tracks produce nearly identical morphologies, but exhibit a spread in 
temperature of $\pm 50$~K in \teff\ along the Hayashi track near 3~Myr. This
spread decreases to about $\pm 25$~K after 10 Myr, just prior to the Henyey track.
The mass track from the Brazil group is a noticeable exception, showing a 
systematically hotter temperature of about 175~K compared to the others.

At 0.2~\msun, the spreads in \teff\ are decreased to about $\pm 25$~K  
along the mass tracks, with exception of the Lyon track at young
ages near 3 Myr and the Dartmouth 2008 track, which is consistently 50~K cooler
than the other tracks. We also see that the Dartmouth 2008, Yale, and Brazil
tracks do not exhibit a hook toward a cooler \teff\ shortly after 10 Myr as
is characteristic of the other models. This can be understood, at least for the 
Dartmouth 2008 models, as a consequence of the fact that they attach the model 
atmosphere boundary conditions where $T =$~\teff. Though this has a minimal
impact on the stellar radius predictions, \teff\ values are typically cooler
by 50\,K compared to models that define the surface boundary conditions
deeper within the star. Given this explanation, it is strange that the 
Brazil models do not exhibit this hook, as they report that boundary conditions
are attached at an optical depth $\tau_{\rm fit} = 10$. We are unable to 
confirm this for the Yale models, as they do not explicitly report
the optical depth where boundary conditions are defined.

Among the discarded tracks, most of the 1.0~\msun\ tracks in Figure~\ref{fig:trackoverview}(b) 
exhibit similar behavior, as was the case for the
included models. This is expected as boundary conditions play a less 
significant role at higher masses than they do for lower masses. One 
noticeable exception is the mass track from the \citet{DAntona1997} series, 
which is nearly 300~K hotter than the other models during 
contraction along the Hay\-ashi track. Given the relative insensitivity of
models to surface boundary conditions, this is a consequence of their 
non-local treatment of convection with FST (Full Spectrum of Turbulence). Also of note in panel (b) 
is the lack of agreement with the Sun by the \citet{Siess2000} models.
This is difficult to understand given that they did perform a solar 
calibration \citep{Siess2000}.

At lower masses, shown in Figures~\ref{fig:trackoverview}(d) and 
\ref{fig:trackoverview}(f), the mass tracks begin to diverge significantly.
Excluded models tend to have systematically hotter \teff s at 0.5~\msun,
with the exception of the \citet{Swenson1994} track, which we do not 
attempt to explain. Hotter \teff s are
a consequence of adopting grey boundary conditions, as has been discussed
previously \citep{CB97}. 
At 0.2~\msun\, mass tracks again display a variety
of morphologies. There are nearly as many different morphologies as there 
are mass tracks, a diversity caused largely by the different treatments of 
surface boundary conditions. 

\subsubsection{Isochrones}

Figure~\ref{fig:isooverview} shows a comparison of solar composition isochrones at three ages
for the adopted model sets. Isochrones from excluded models sets will not be 
discussed. In general there is good agreement between the different isochrone
sets. This is not unexpected given the uniform selection criteria outlined
in Sect.~\ref{sec:criteria}. At hotter \teff s and higher masses, model sets predict
very similar morphologies, although there are some noticeable offsets between
6300~K and 8000~K in the $R$--\teff\ plane that correspond to the ignition of 
the CN cycle at lower temperatures and the nitrogen-burning bump at higher temperatures. 
The latter signals the CNO cycle coming into equilibrium with the ignition of nitrogen
burning. These offsets appears to stem most from radius differences exhibited in panels
(d) -- (e). Significant variations between the model sets applicable in this temperature
regime disappear by the age of 100~Myr as all of the higher mass stars have fully reached
the main-sequence. 

There is near unanimous agreement between the various model sets in the solar
mass regime (\teff\ $\sim$ 4300~K at 1 and 10 Myr and \teff\ $\sim$ 5500~K at 100 Myr). The
Pisa models show systematically larger radii at 1 Myr, but this difference vanishes
by 10 Myr and suggests initial conditions are still important to consider at this age.
Around 10 Myr, the Lyon models have simultaneously larger radii and cooler \teff s by 
about 2\% each, in the vicinity of 1.0~\msun. Though small, the differences are at odds 
with the other model sets, which agree to within 0.5\%. By 100~Myr, all model sets agree 
to within 0.3\%. We have no precise explanation for the larger radii and cooler \teff s 
of the Lyon models at early 
times, but it may be related to a combination of the depth at which the surface 
boundary conditions are fit and the adopted heavy element composition \citep[for a
thorough discussion see][]{Tognelli2011}. The Lyon models adopt the \citet{GN93}
heavy element composition and fit their surface boundary conditions at an optical 
depth $\tauf = 100$. In contrast, the other model sets adopt an overall lower
heavy element abundance, at least for the interior composition \citep{GS98,
Asplund2005}, and fit the surface boundary conditions at lower optical depths 
(see Table~\ref{tab:models_in} for details).

Differences at low masses are more difficult to discern from the panels in Figure
\ref{fig:isooverview} as large relative variations in the predicted fundamental 
properties translate to small absolute variations. However, we observe several
differences among the different groups. First, the Pisa models show a large radius 
offset (up to 35\%) at 1~Myr. This feature seems to be a consequence of the model 
initial conditions, which set the age when deuterium burning occurs; at 1~Myr, 
deuterium burning appears to extend to higher masses in the Pisa models than
in the others. It is difficult to assess which, if any, of the models in this regime 
are correct based on purely theoretical arguments. However, it is important to be 
cautious of models at this age, as they do not ``forget'' their initial conditions until 
sometime between 1 and 10~Myr \citep{Baraffe2002}. In the same manner, the Lyon 
isochrone at 1 Myr appears cooler than the others at a given mass. By 10 Myr the 
temperature difference disppaears, but a slight radius offset 
is still present before the tracks come into agreement with the other sets around 100 Myr. One 
also notices in panel (d) that the Yale and Dartmouth 2008 isochrones are cooler than the others. This is a consequence of the absence of deuterium burning in these model
sets.
 
Also difficult to discern from the isochrones are deviations of the Brazil, Yale, and
Dartmouth 2008 models at the lowest masses at 100 Myr. These isochrones exhibit a 
different concavity between 0.1 and 0.2~\msun, or \teff\ $\approx$ 3500~K. In addition, 
the models show a milder slope in panel (c), corresponding to a steeper slope in
panel (i) below 3200~K. These features are very likely the result of surface boundary 
conditions being attached in regions of the star where convection is sufficiently
non-adiabatic that detailed radiative transfer from non-grey models are required.

\subsection{Methods of Fitting and Comparison against EB Data\label{sec:fitmethod}}

Previous comparisons of stellar models to observed stellar properties
have often been performed in the H-R diagram plane, i.e., fitting model isochrones to
the observed \teff\ and $L$. Here,
the benchmark PMS EBs were fit to stellar evolution isochrones using 
six of the most accurately directly measured stellar properties: 
the primary mass, the mass ratio of the secondary to the primary, the
primary and secondary radii, the primary \teff, and the \teff\ difference 
between the two components. Throughout this review we refer to the ``primary" as the star
with the larger measured mass, and we
consider only models with solar composition.
We sought the best isochrone within a given model set that simultaneously
fit the two stars in a given EB; thus this test implicitly assumes coevality
for the two stars in the system.
Isochrones provided by the 
various groups were used without any additional modifications, when
possible. However, some models (i.e., Dartmouth 2008, Yale, and Brazil)
do not include a sufficiently young or a sufficiently well-spaced 
set of isochrones in the target age range (1--100 Myr). We therefore
computed isochrones from their evolutionary tracks with 
an age resolution of 0.2~Myr from 1 to 20~Myr, and 5 Myr from 20 to 
100~Myr. While the isochrones 
are not specifically provided at these ages by the model grids, 
interpolation is a standard procedure when applying the model grids
to observed properties of young stars.\footnote{We chose not
to request more extensive grids from authors of the models
so as to preserve the principle of using only publicly available 
models.}

Each EB was fit to a given isochrone grid using the following procedure. 
First, the mass spacing along each isochrone was standardized to a resolution
of 0.001~\msun\ to provide adequate sampling within one standard deviation 
of the quoted EB masses. Linear interpolation was sufficient
as the original mass resolution was around 0.05~\msun\ for most model sets. 
Next, for a given isochrone, the two stars of the EB were compared to each mass 
point along the isochrone, with residuals in the mass, radius, and \teff\ 
being calculated. To gauge the quality of fit, a goodness of fit statistic
for each star was computed, as well as a global fit statistic for every 
combination of primary and secondary mass. We chose to use a $\chi^2$ 
statistic,
\begin{equation}
    \chi^2 = \displaystyle\sum_{i=1}^N \left(\frac{\delta X_i}{\sigma_{X, i}}\right)^2.
\end{equation}
where $X_i$ $\in$ \{$M_A$, $q = M_B/M_A$, $R_A$, $R_B$, \teff$_{,\, A}$, 
$\Delta$\teff\ = \teff$_{,\, A}$ $-$ \teff$_{,\, B}$\} is one of the 
defined properties used in 
the fit, $\delta X_i$ is the difference between the observed and model 
isochrone value, $\sigma_{X, i}$ is that property's associated 
observational uncertainty, and $N$ is the total number of
parameters being fit. The $\chi^2$ statistic has the advantage of 
being an uncertainty weighted measure of the goodness of fit, thus 
forcing the isochrones to fit quantities that are the most precise. 

A best fit pair of points along the isochrone, representing the primary and
secondary star of the EB, was determined by locating the global $\chi^2$ 
minimum among all of the possible combinations. This procedure
was repeated for each isochrone of a given model set, yielding a list of
the best fit primary and secondary masses at a given age. To find the 
overall best fit isochrone for a given EB, the global $\chi^2$ minimum 
among this data set was determined. We enforced strict coevality;
the goodness of fit was determined by a single
isochrone at a time for both stars simultaneously.
Possible age spreads are considered in Sect.~\ref{sec:clusters}. 
The procedure was performed for each EB component using a restricted 
mass range along an isochrone between $5\sigma$ below
and $5\sigma$ above the observed mass. We chose $5\sigma$ as a cut-off to 
allow the models sufficient flexibility in finding a best fit, while minimizing 
computational time spent performing comparisons between mass points that would 
be guaranteed to have a $\chi^2 > 150$, and thus not likely to be the location
of a global $\chi^2$ minimum in our fitting routine. We tested that this 
restriction did not introduce any biases in our results by comparing results
of runs with and without the mass clipping for the Dartmouth models. No differences 
in the global minima occurred.

\section{Results: Comparison of Models versus PMS EB 
Measurements\label{sec:results}}

\subsection{Recovery of stellar masses in the H-R diagram\label{sec:hrd}}

\begin{figure*}[!htbp]
    \centering
    \includegraphics[width=0.95\linewidth]{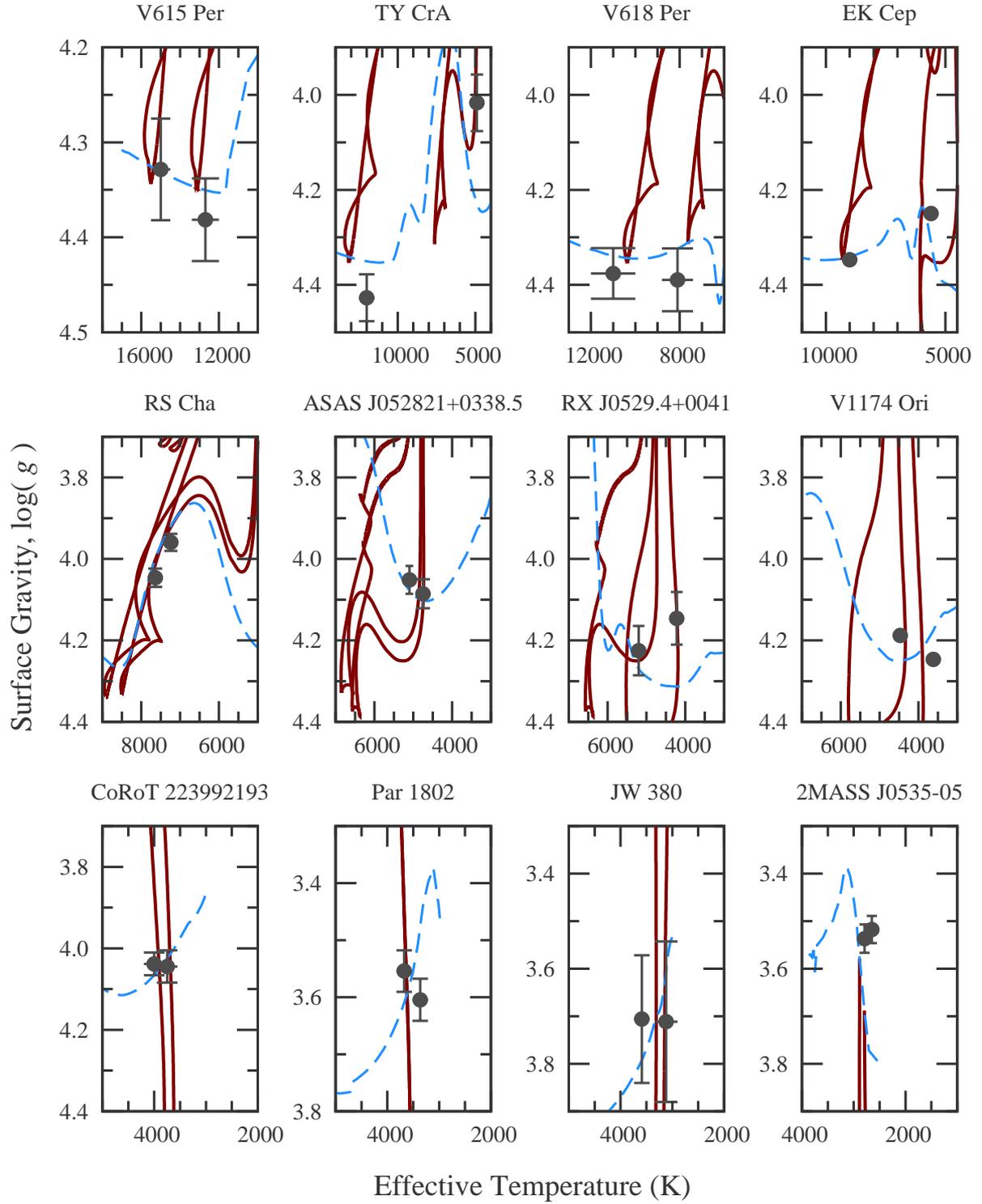}
    \caption{Fitting results for each of the benchmark EBs in a \teff--$\log g$ 
        HRD against Dartmouth 2014 mass 
        tracks (maroon, solid lines) and isochrones (light-blue, 
        dashed lines). Mass tracks were computed at the observed masses 
        listed in Table~\ref{tab:data1} while isochrones are shown at the 
        best fit age determined in Section~\ref{sec:fitting}.}
    \label{fig:hrd_grid}
\end{figure*}

Following previous reviews on the subject, we begin by considering the
question: how reliably may one expect to determine the mass of a PMS
star using the available stellar models? We perform this assessment
from the ``observer's perspective," that is, by considering what stellar
masses one would infer for each EB star if it were ``observed" in the 
H-R diagram and then compared to various stellar isochrones 
(Figure~\ref{fig:hrd_grid}). 

To perform this test, we searched for the best fit \teff\ and 
$L$ combination by minimizing the total $\chi^2$. Isochrones were 
searched within $1 \sigma$ of the quoted \teff. This is similar
to the approach in \citet{Hillenbrand2004} and \citet{Mathieu2007}, 
but here explicitly accounting for the observational uncertainties 
in \teff\ and $L$. 
Best fit $\chi^2$ values were 
in all cases $\approx 0.0$ as expected for a fit of
two free parameters (mass, age) to two constraints.
We then selected the mass at which the
minimum $\chi^2$ was found and compared it to the dynamically 
measured mass. Each star was treated individually; we did not 
enforce coevality. This in essence treats each individual star
as though it were observed on its own.

\begin{figure}[!ht]
    \centering
    \includegraphics[width=\linewidth]{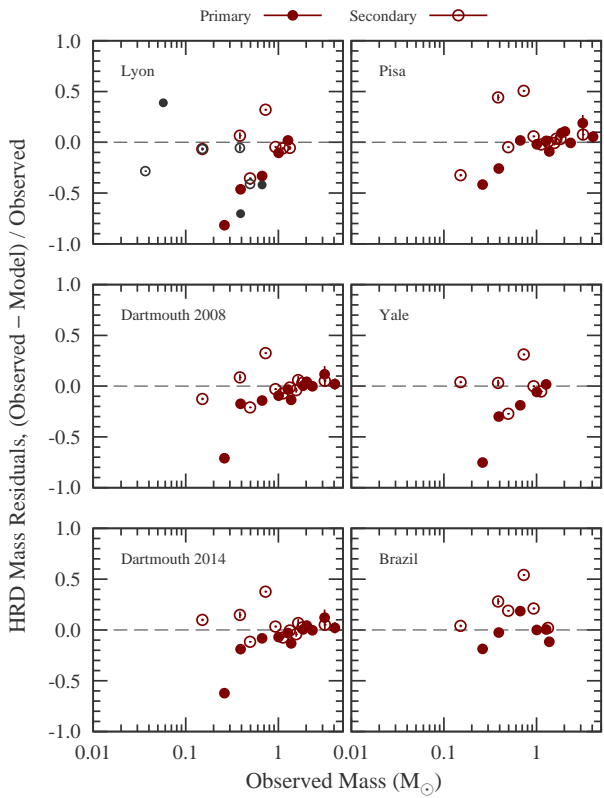}
    \caption{Mass residuals found from fitting observed PMS EB systems
             to models in the H-R diagram. Each panel shows the 
             fit results for a different model suite, with primary
             and secondary stars shown with different symbols as labeled.
             For the Lyon model comparison (upper left), dark symbols
             represent the models with non-solar $\amlt = 1.0$, which
             are the only models that extend below 0.1 \msun\ to fit
             the brown-dwarf EB, 2M\,0535$-$05.}
    \label{fig:hrd_mass}
\end{figure}

The results are shown in Figure \ref{fig:hrd_mass}, in which we
compare, for each of the models, the dynamically measured stellar
masses to the masses inferred by the model isochrones from the 
observed \teff\ and $L$ in the H-R diagram. We show the primary
stars and secondary stars separately. 
There is an apparent trend in most of the model sets in the
sense that the models over-predict the masses, and this tendency
increases toward the lowest masses. 
Interestingly, the tendency to
over-predict the masses is most prominent among the primary stars;
the secondaries appear fairly well distributed around the zero-point.
However, the secondaries do in all of the models exhibit a larger
dispersion.

The mean of the residuals and the r.m.s.\ deviations of
the primaries and secondaries (taken together) for each of the model sets are
summarized in Table \ref{tab:modelstats}. We divide each of the comparisons
at 1 \msun\ to convey the degree to which the lower mass stars
are more poorly reproduced. 
The Dartmouth models appear to exhibit the lowest overall scatter.
The Pisa models show a slightly larger dispersion than the Dartmouth
models at low masses, but have a comparable dispersion 
above 1 \msun. 
The Brazil models do not show the trend mentioned above of 
increasingly over-predicted masses toward 
lower masses, but they do show a larger dispersion than the 
Pisa and Dartmouth models. 


\begin{table*}[!t]
    \centering
    \caption{Statistics of HRD fit for high- and low-mass populations.\label{tab:modelstats}}
    \label{tab:hrd_stats}
    \renewcommand{\arraystretch}{1.25}
    \begin{tabular*}{0.95\linewidth}{@{\extracolsep{\fill}} l r c c c r c c}
    \noalign{\smallskip}\hline\hline\noalign{\smallskip}
                 & \multicolumn{3}{c}{Low-Mass Stars (M $<$ 1.0 \msun)}   & & \multicolumn{3}{c}{High-Mass Stars (M $\ge$ 1.0 \msun)}\\
    \cline{2-4}\cline{6-8}
    Model set            &    Mean    & Abs. Mean  & $\sigma$& &    Mean    & Abs. Mean &  $\sigma$ \\
    \noalign{\smallskip}\hline\noalign{\smallskip}
    Brazil               &    0.154  &   0.207   &  0.218 & &  $-0.023$ &   0.035 &  0.063 \\
    Dartmouth 2008       &  $-0.123$ &   0.225   &  0.295 & &  $-0.004$ &   0.051 &  0.067 \\
    Dartmouth 2014       &  $-0.044$ &   0.209   &  0.293 & &  $-0.002$ &   0.050 &  0.065 \\
    Lyon, $\amlt = 1.0$  &  $-0.342$ &   0.439   &  0.471 & &     ...   &    ...  &   ...  \\
    Lyon, $\amlt = 1.9$  &  $-0.212$ &   0.308   &  0.353 & &  $-0.050$ &   0.060 &  0.051  \\
    Pisa                 &  $-0.003$ &   0.259   &  0.339 & &    0.033  &   0.053 &  0.069 \\
    Yale                 &  $-0.141$ &   0.237   &  0.317 & &  $-0.031$ &   0.043 &  0.042  \\
    \noalign{\smallskip}\hline
    \end{tabular*}
    \vskip 5pt
    \parbox{0.93\linewidth}{
        {\bf Note: }{For the statistics compiled in the table, ``Mean'' refers
        to the direct mean fractional mass error, ``Abs. Mean'' is the mean absolute 
        fractional error, and $\sigma$ is the standard
        deviation about the mean computed using $N - 1$ in the denominator to 
        compensate for the small sample size.}}
\end{table*}

To be clear, this comparison is not quite fair to the stellar models, 
because it conflates any observational biases with true astrophysical 
effects, as we discuss below. However, to the extent that the problem
of inferring stellar masses from direct observables such as \teff\ and $L$
may be similarly affected by both observational and astrophysical effects
not represented in the stellar models, this comparison provides a fair
basis for quantifying the total errors that one may reasonably expect in
such mass estimations.

In summary (see Table~\ref{tab:modelstats} and Figure~\ref{fig:hrd_mass}), 
the accuracy with which one may expect to infer the true
stellar mass above 1 \msun\ is for most of the model sets quite
good, typically 1--10\% in the mean, $<$10\%
scatter, and without obvious systematics (though the sign of the mean
offsets does tend to indicate slightly over-estimated masses by the models). Below
1 \msun, the situation is markedly worse, with offsets and 
scatters of $\sim$40\%, and with a strong systematic tendency by most of
the models toward over-estimated masses, the over-estimation 
approaching $\sim$100\% at 0.1 \msun\ (Figure~\ref{fig:hrd_mass}). 
One exception to this trend is the Brazil model set, which yields no
large mean offset (an absolute mean deviation of 15\%) and a modest
scatter of 22\%. In any event it is clear that in general the H-R diagram
inferred masses with all of the model sets are highly reliable above 1 \msun\
but moderately to highly unreliable below 1 \msun.

We note that these findings differ qualitatively from those of
\citet{Hillenbrand2004} and \citet{Mathieu2007}, who similarly 
found generally good agreement above 1 \msun, but below 1 \msun\
found a tendency for the models to {\it under}-predict masses.
Note however that these previous studies used mainly non-EB PMS 
stars with masses determined via astrometric orbits or 
circumstellar disk rotation curves. Only three of the EBs in 
our sample are in common across these studies, and for these
three EBs we find very similar results as did those studies. In
addition, the previous studies considered mainly previous generation
PMS models---the Pisa, Brazil, and Dartmouth models were not yet
available---and in this study we have excluded most of the model
sets used in the previous studies for their use of grey
atmospheres and/or their lack of solar calibration. As a result, 
only the Lyon models are in common to this study and the 
\citet{Hillenbrand2004} and \citet{Mathieu2007} reviews. Therefore,
with a largely different set of benchmark EBs and a largely
different set of models considered, it may not be surprising that
we find qualitatively different results in the ability of the
models to recover the stellar masses in the H-R diagram. 

However, as discussed below, this is not the 
last word, as there are important physical effects to consider
that substantially alter the assessment of model performance
in the H-R diagram, a question to which we return in Sect.~\ref{sec:hrdredux}.

\subsection{Detailed fitting of individual systems\label{sec:fitting}}

As a more stringent and accurate test of the stellar models, we
have gone beyond the H-R diagram plane and have fit each of the model 
isochrones to six of the directly measured
properties for each EB, as described in Section \ref{sec:fitmethod}.

\begin{figure*}[!htbp]
    \begin{center}
        \includegraphics[width=0.23\linewidth]{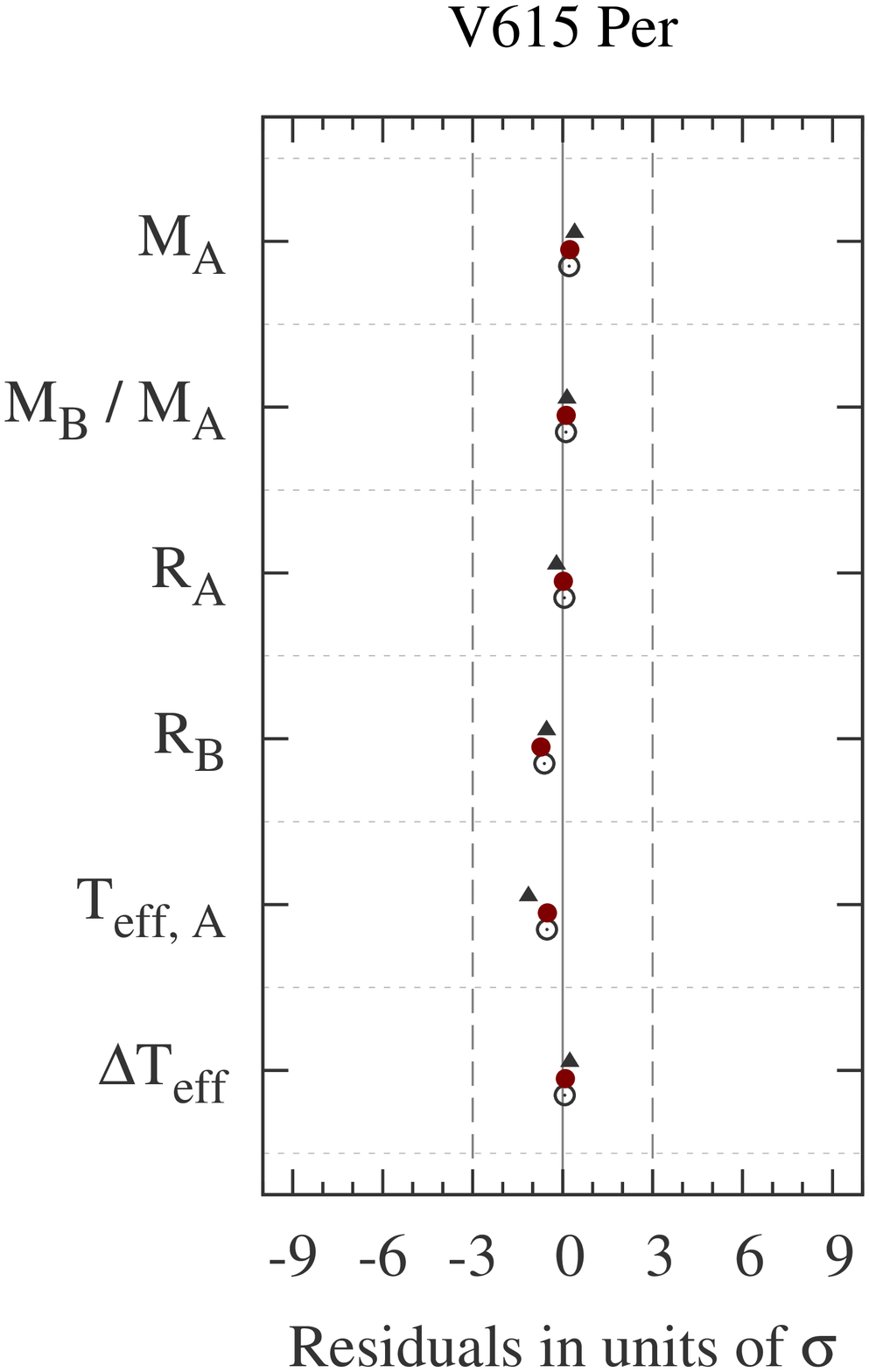}            \quad
        \includegraphics[width=0.23\linewidth]{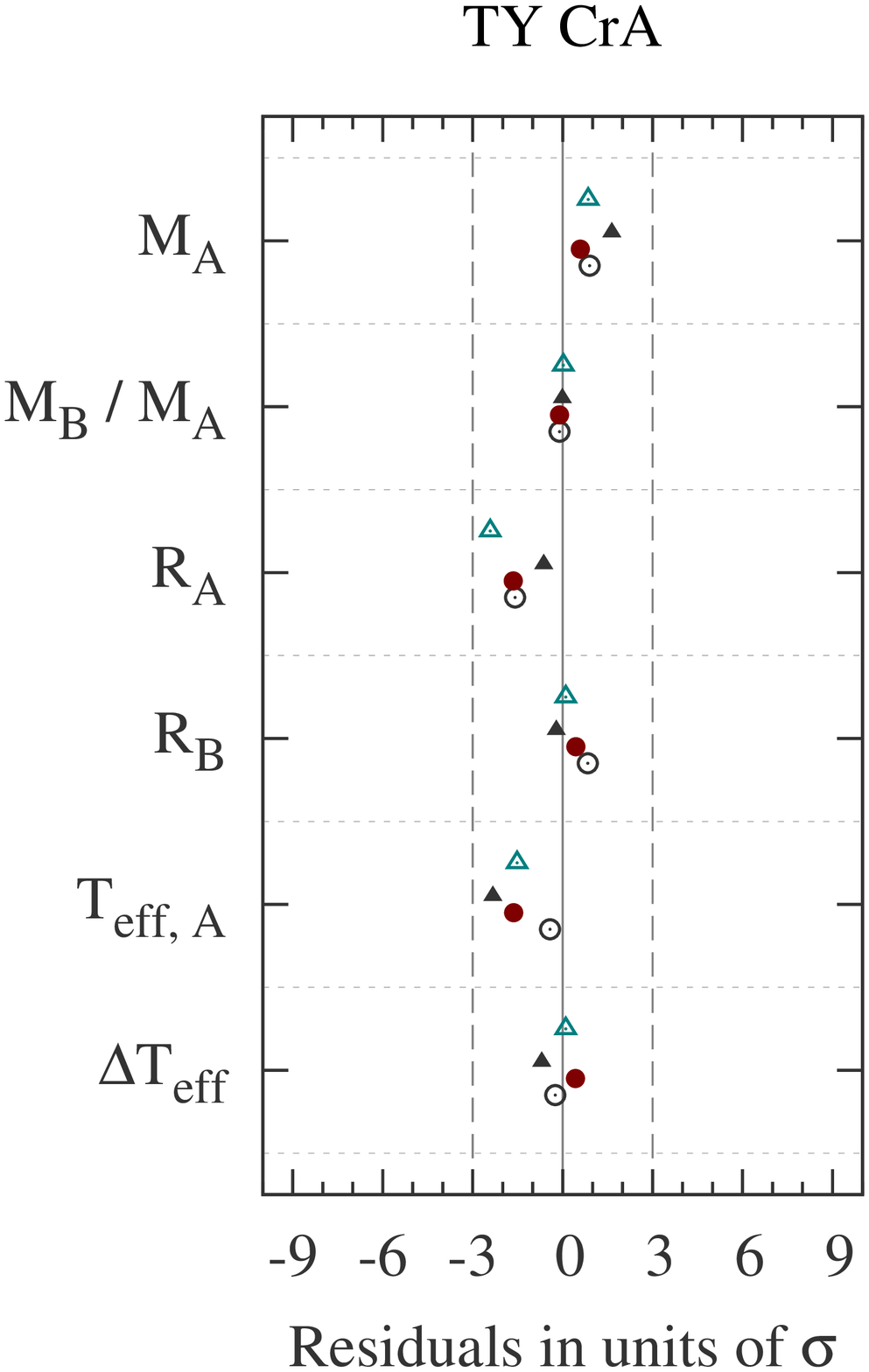}              \quad
        \includegraphics[width=0.23\linewidth]{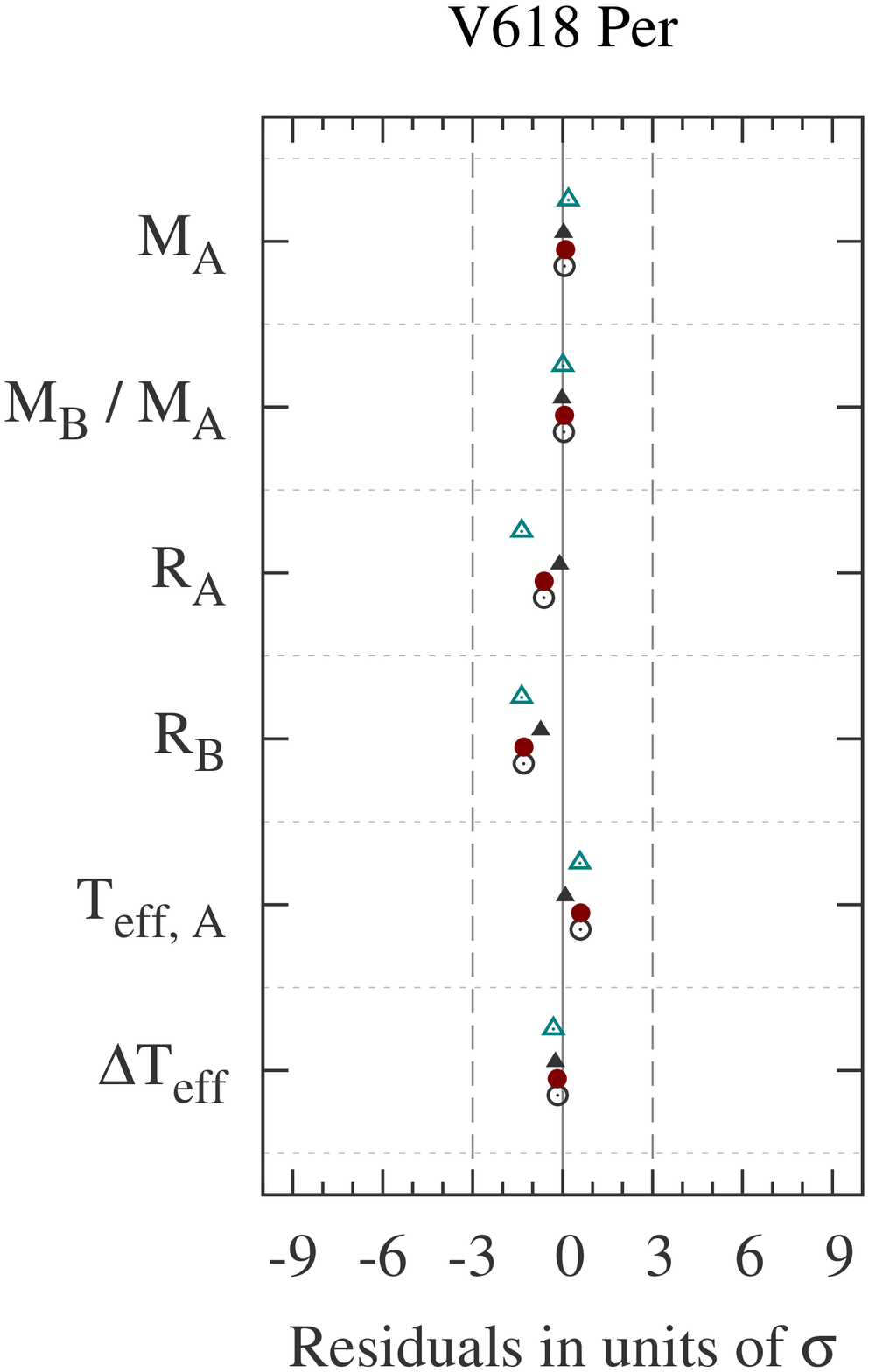}            \quad
        \includegraphics[width=0.23\linewidth]{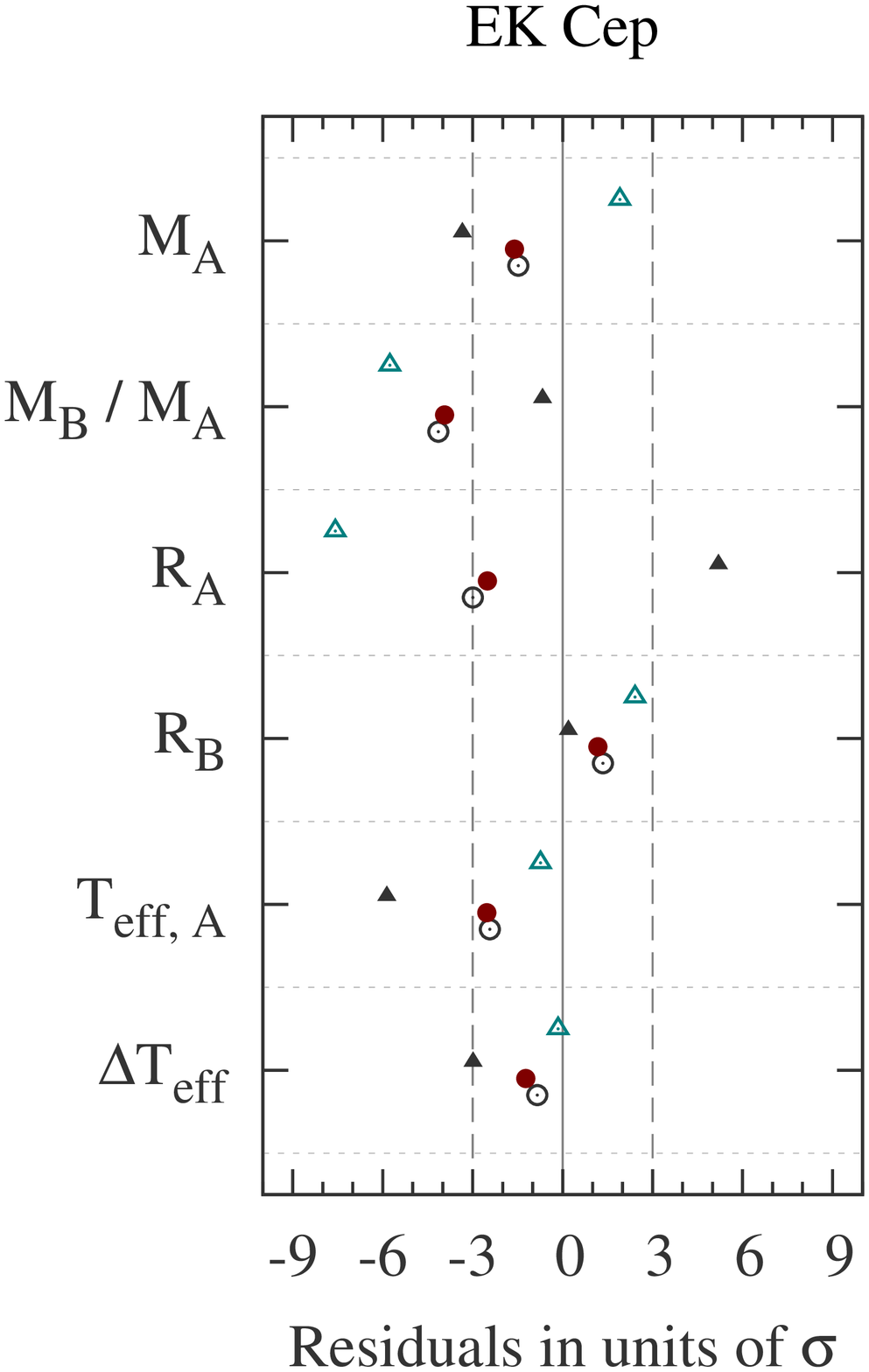}              \\
        \vspace{\baselineskip}
        \includegraphics[width=0.23\linewidth]{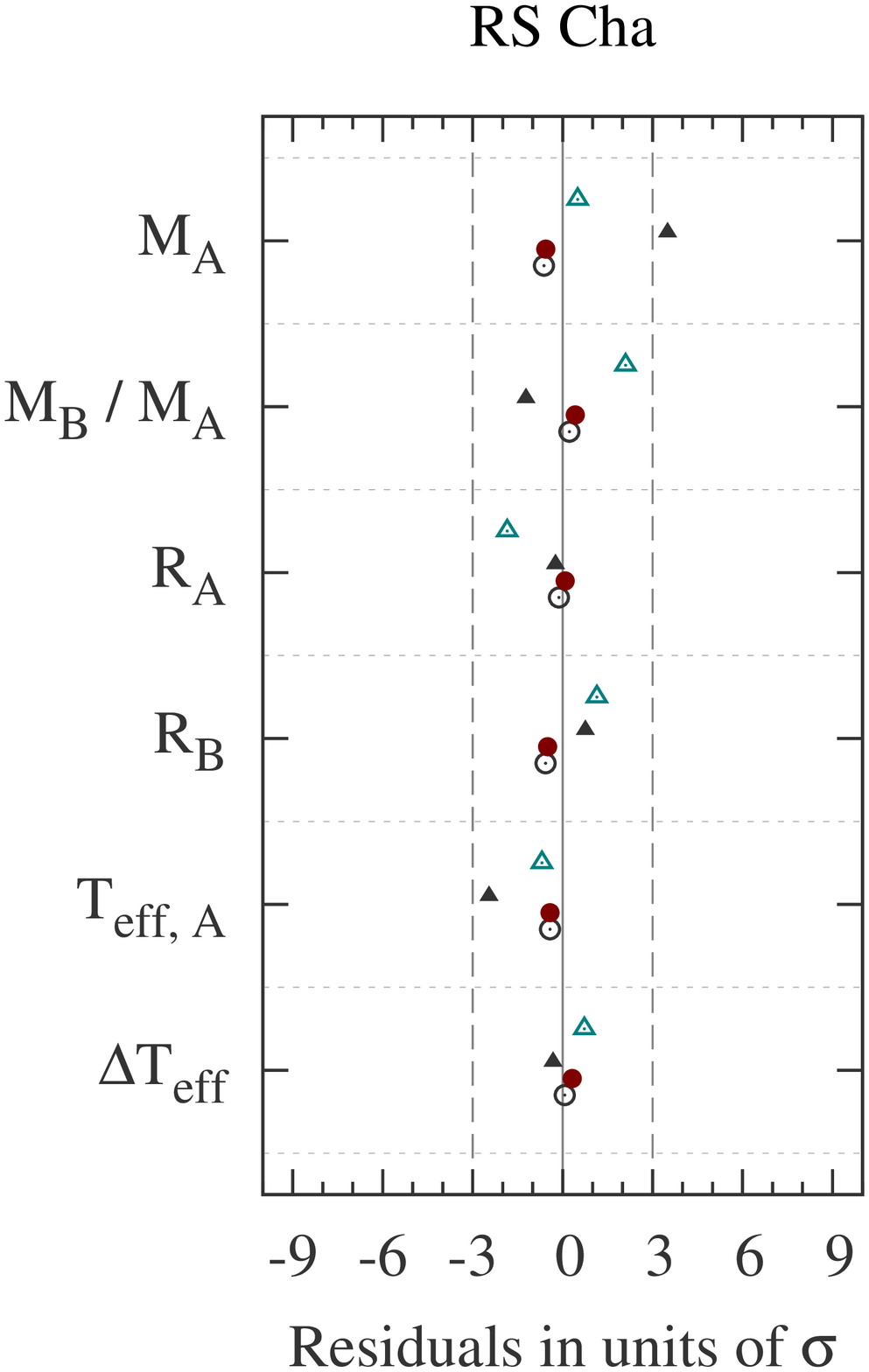}              \quad
        \includegraphics[width=0.23\linewidth]{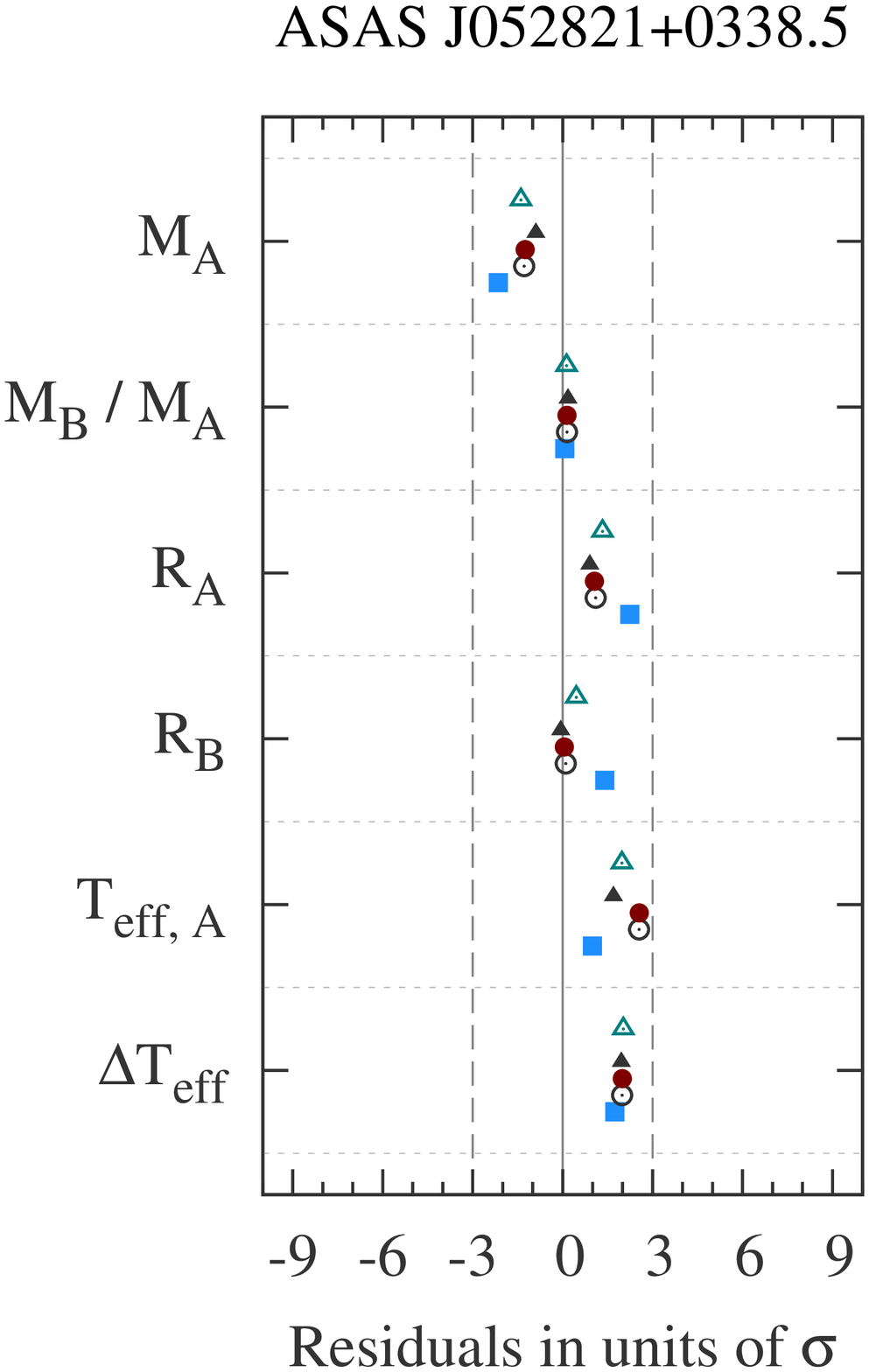} \quad
        \includegraphics[width=0.23\linewidth]{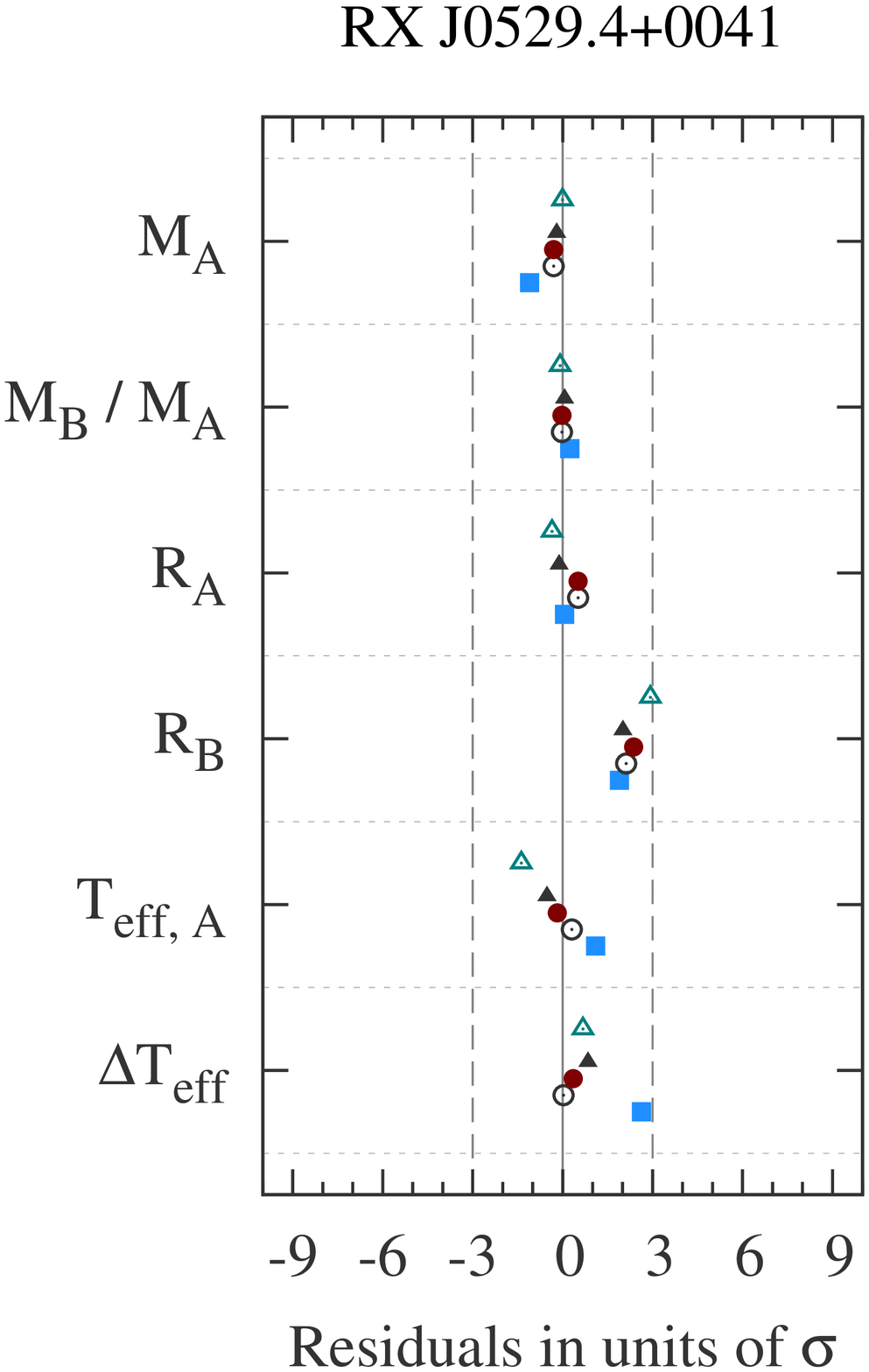}     \quad
        \includegraphics[width=0.23\linewidth]{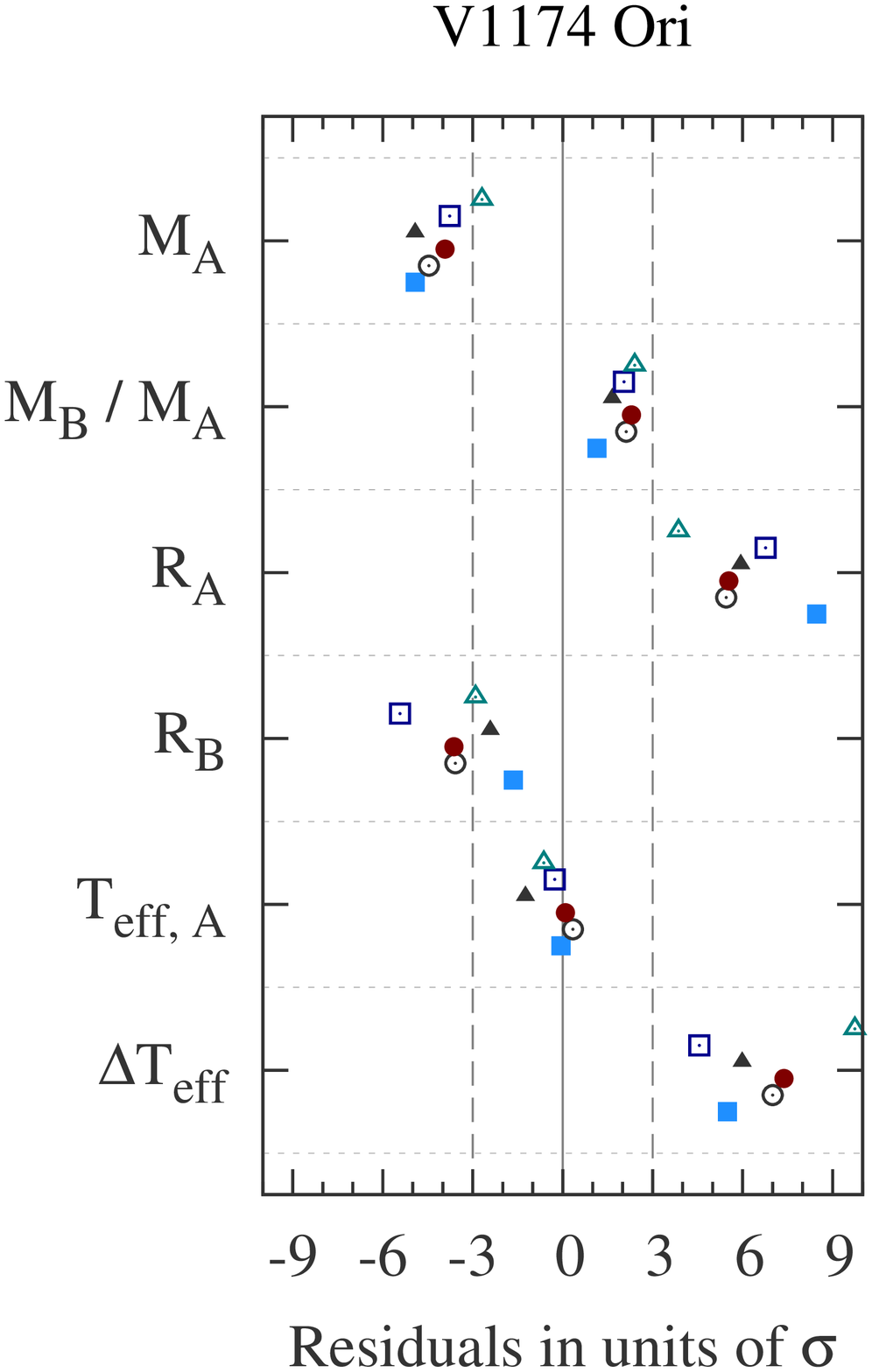}           \\
        \vspace{\baselineskip}
        \includegraphics[width=0.23\linewidth]{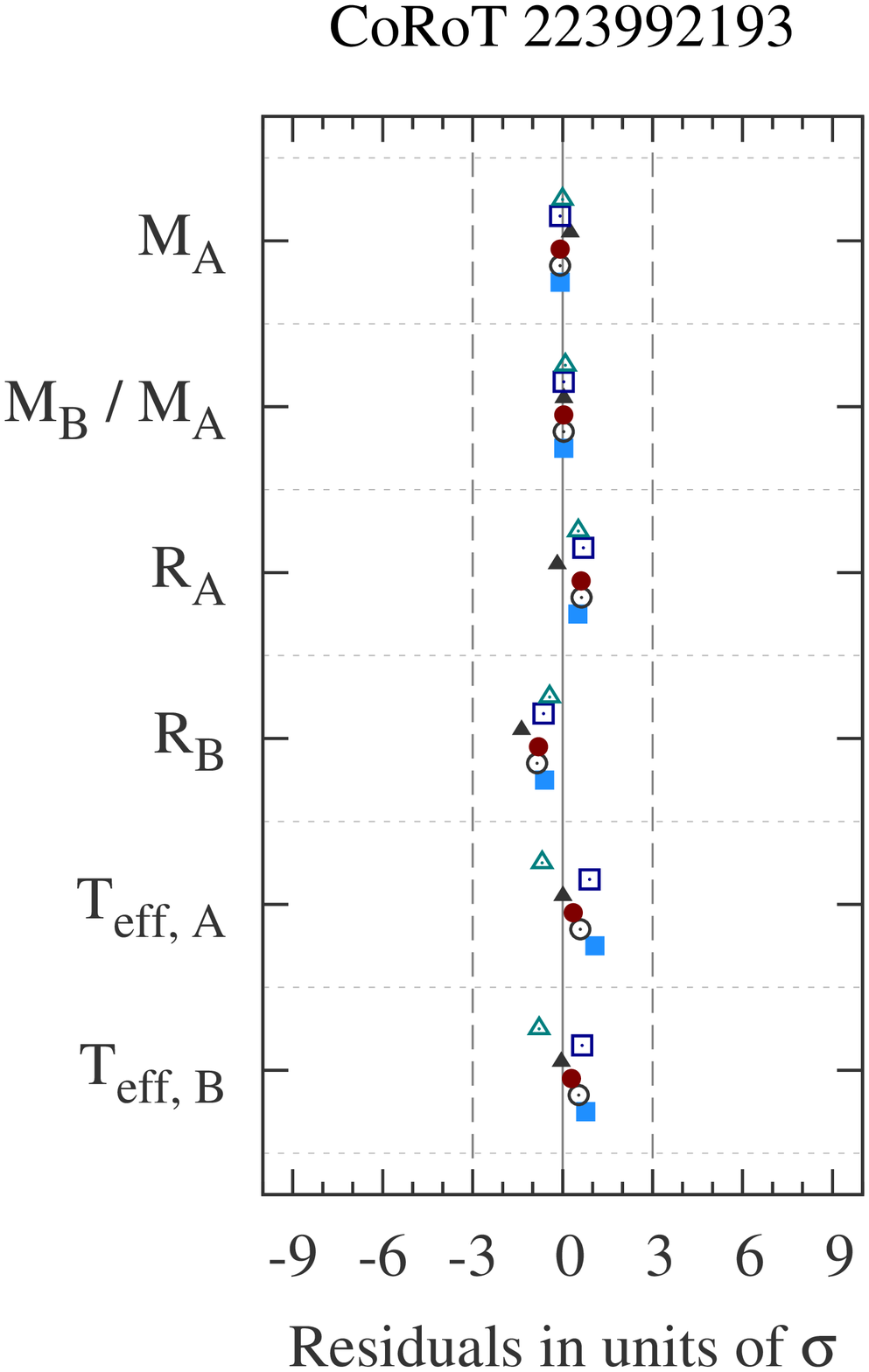}     \quad
        \includegraphics[width=0.23\linewidth]{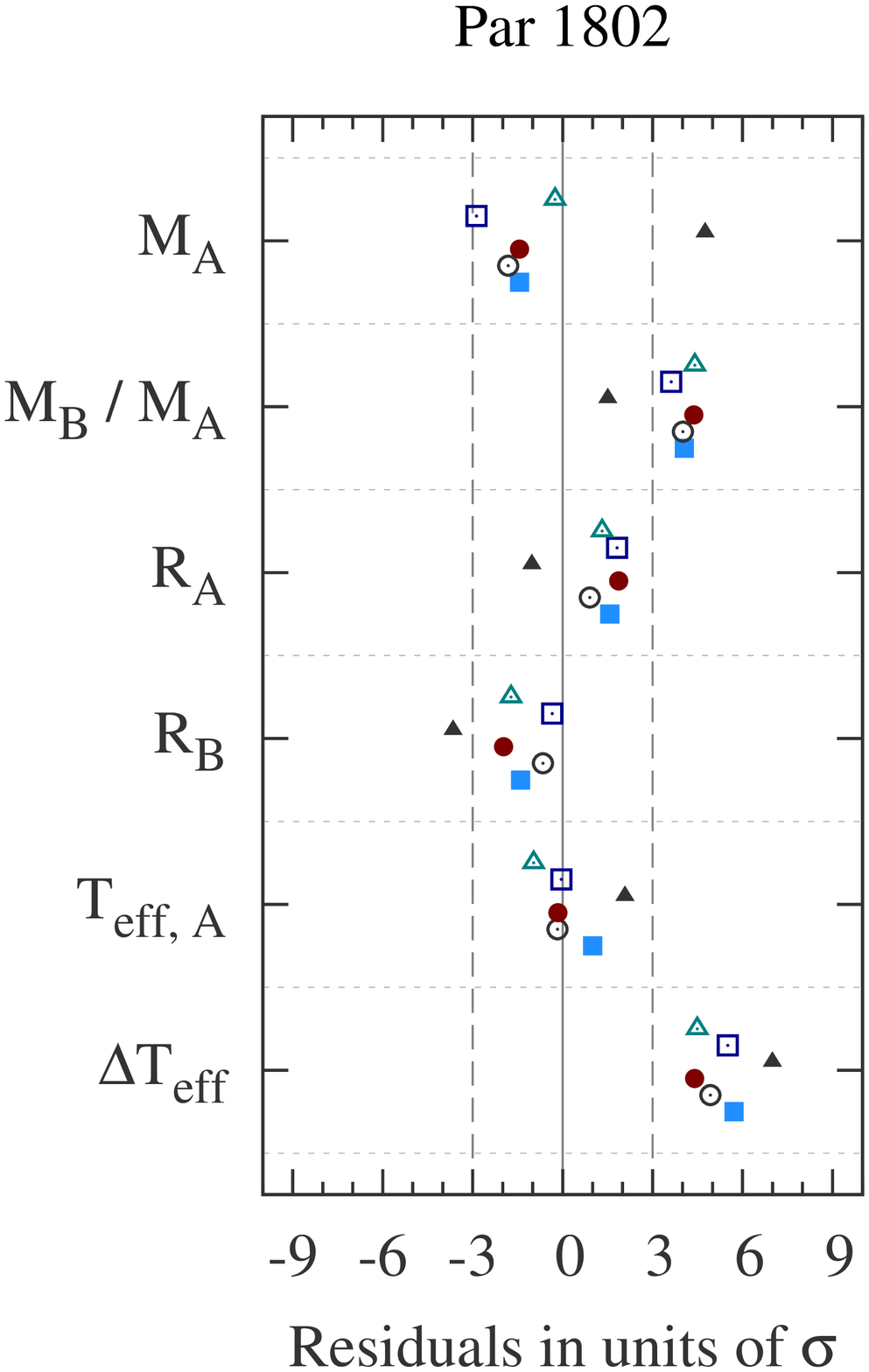}            \quad
        \includegraphics[width=0.23\linewidth]{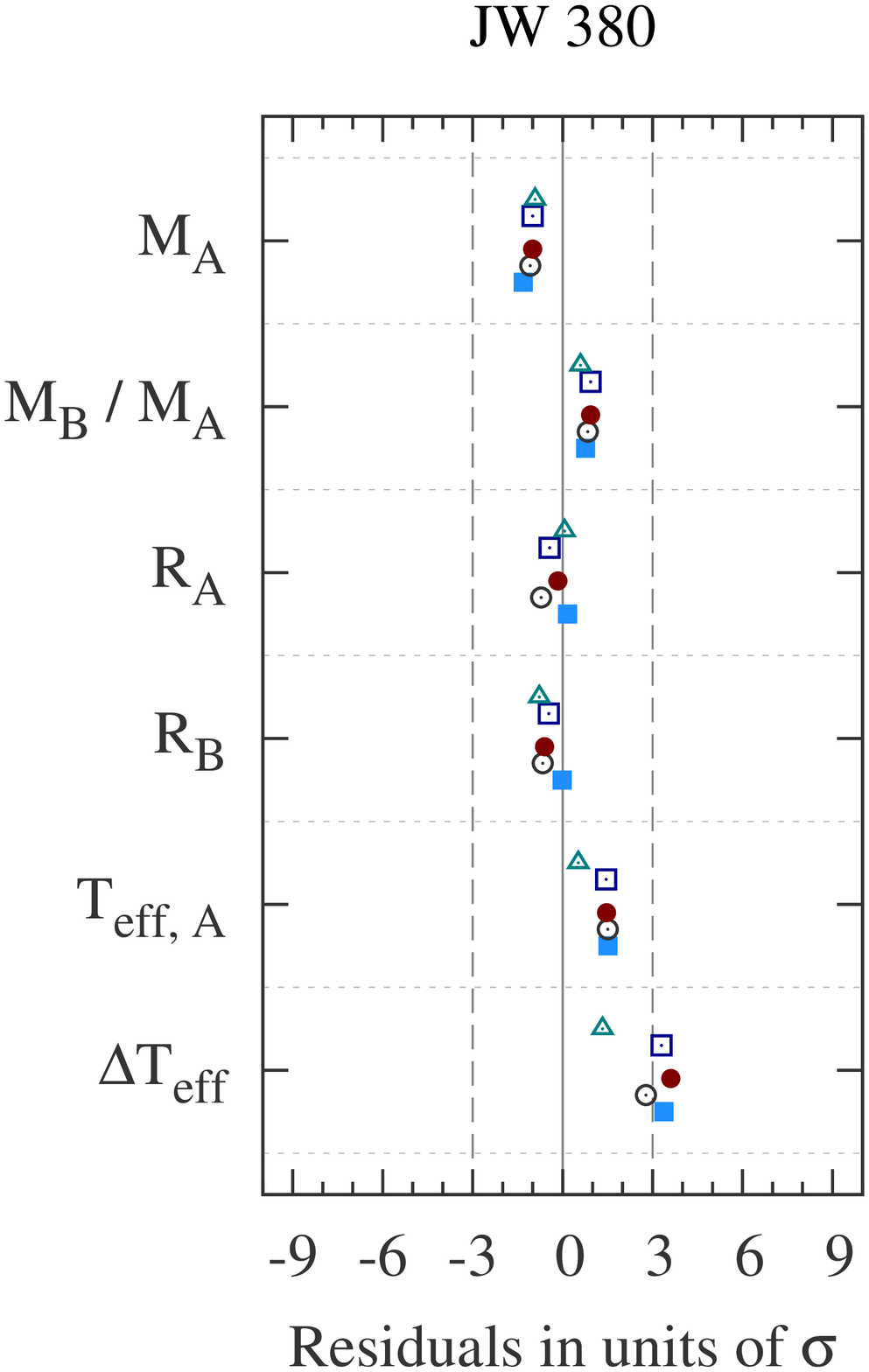}              \quad
        \includegraphics[width=0.23\linewidth]{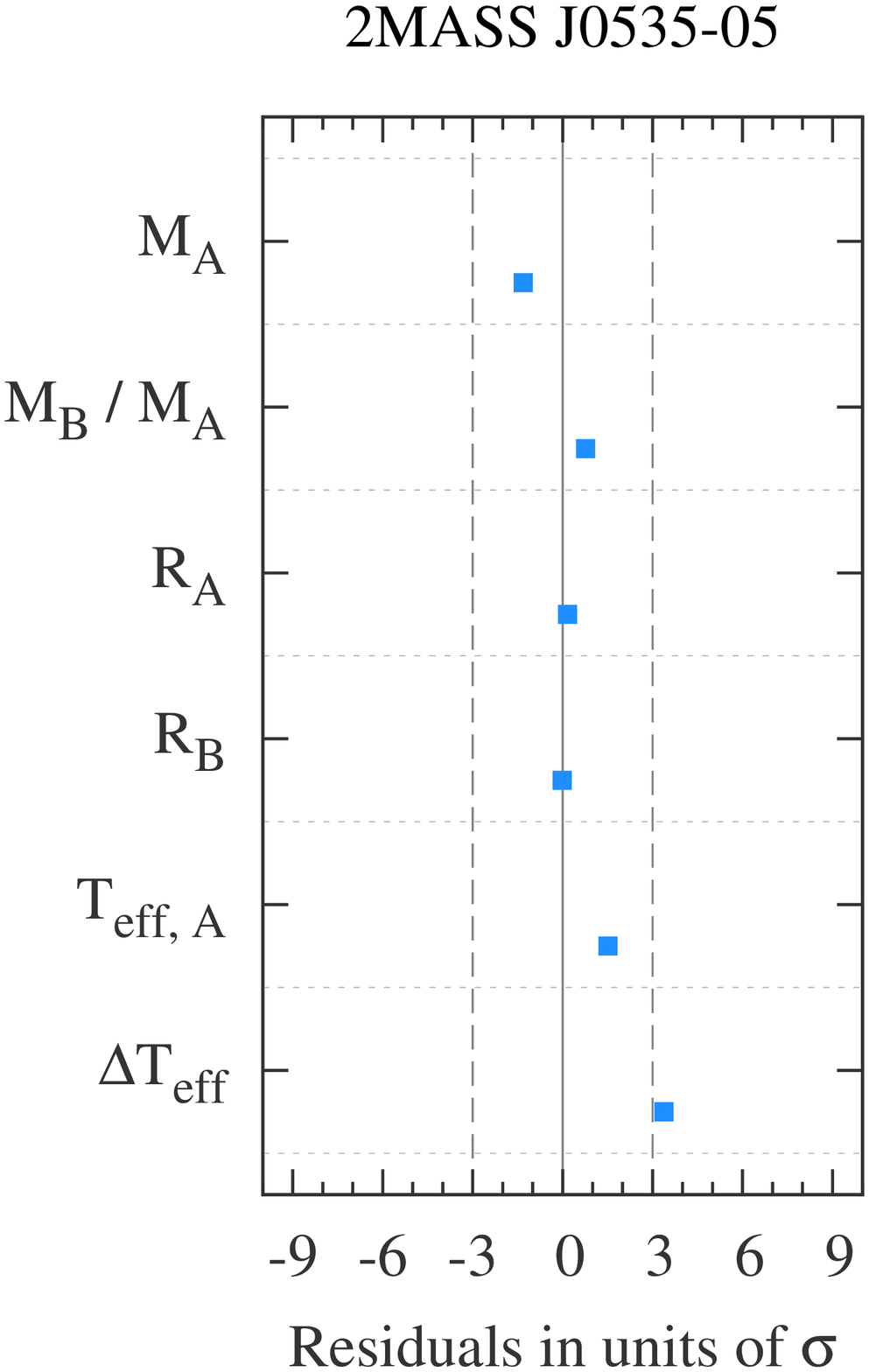}          \\
    \end{center}
    \caption{\label{fig:individual} 
        Fitting results for each of the benchmark EBs. Each model isochrone set
        (represented by different symbols) was fit to the six measured EB properties
        shown along the vertical axis. Note that for CoRoT\,223992193 we used 
        $T_{\rm eff,B}$ instead of $\Delta T_{\rm eff}$ as the temperature difference
        has not been reported. The horizontal axis shows the residuals of
        the best-fit isochrone in the sense of (data $-$ model), in units of the
        observational uncertainty, $\sigma$.
        Symbols are as shown in the legend of Figure \ref{fig:chi2mass}.
        Note that for the brown-dwarf EB, 2M0535--05, only one set of models was 
        applicable: the Lyon $\amlt$ = 1.0 models.
        }
\end{figure*}

An illustration of the quality of the fit of stellar evolution models
to the observations for each of the 13 EBs may be seen in
Figure~\ref{fig:individual}, in which we show the residuals (observed
minus predicted) for each of the fitted quantities normalized to their
corresponding observational errors. The vertical dashed lines
represent $\pm$3$\sigma$ limits. In each case we have compared the six
observables against all models from Table~\ref{tab:models_in} that allow
the comparison (e.g., that have a suitable mass range), and we have
represented each model with a different symbol. For MML\,53 the individual
component radii have not been measured, nor is an accurate temperature
difference available. Consequently the constraint on models is considerably
weaker than for other systems, which results in an artificially good match
to predictions with a low $\chi^2$ value. We therefore do not consider this
system in the discussion below.

%

The two incarnations of the Dartmouth models yield very similar
residuals in most cases, which is not surprising given that the
physical ingredients are largely the same.  In general the normalized
residuals from all other models are also quite comparable, though
there are exceptions such as the case of EK\,Cep, in which the Pisa
and Brazil calculations deviate in opposite directions compared to those from
Dartmouth.  Somewhat more than half of the systems may be considered
to be reasonably well fit by the models, with all normalized residuals
under 3$\sigma$.  V615\,Per, V618\,Per, and CoRoT\,223992193 are
particularly well matched. On the other hand, all models have great
difficulty fitting V1174\,Ori, as well as Par\,1802 to a lesser
degree, though they all fail in similar ways hinting at either a
common shortcoming in the physics of the models or perhaps
unrecognized systematic errors in one or more of the measured
quantities for those EBs. 
These two systems, along with EK\,Cep,
happen to
be the ones with the smallest formal relative errors in the individual
radii, all at or under 1\%. If those errors have been underestimated,
they could explain the larger residuals from the fit in these cases.

\begin{figure*}[!htbp]
    \centering
    \includegraphics[width=0.8\linewidth]{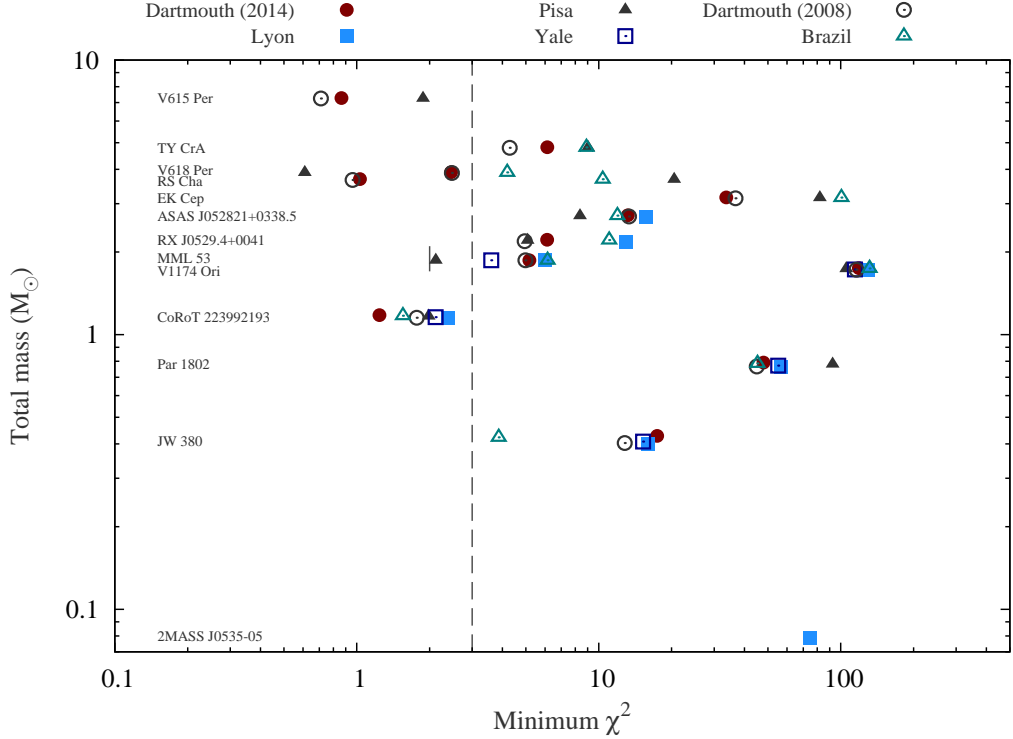}
    \caption{\label{fig:chi2mass} 
    Total $\chi^2$ of model isochrone fits (from Figure \ref{fig:individual}) 
    versus total EB system mass. The vertical line at $\chi^2 = 3$ demarcates
    systems that are well fit (to the left of the line) from those that are
    poorly fit (to the right of the line) based on the 
    number of degrees of freedom in the fits.
    }
\end{figure*}

The overall results of these fits are
visually summarized in Figure \ref{fig:chi2mass}, 
in which we plot the total system mass as a function of the $\chi^2$
of the best fit isochrones. There is a hint
of a tendency for the highest-mass EBs to be relatively well fit and for
the lowest-mass EBs to be more poorly fit, 
although a larger sample is highly desirable to confirm this.
Certainly there are some relatively high-mass systems that
are as poorly fit as the lowest-mass systems. Thus, while the source of the
system-to-system discrepancies with the model predictions may be mildly
dependent on system mass, evidently the discrepancies are caused principally
by other effects that we have not yet considered.

%

\begin{table*}[!htbp]
    \centering
    \caption{Best fit ages (Myr)\label{tab:ages}}
    \renewcommand{\arraystretch}{1.25}
    \begin{tabular*}{0.95\linewidth}{@{\extracolsep{\fill}} l c c c c c c | c}
        \noalign{\smallskip}\hline
        \hline\noalign{\smallskip}
        EB     &  Lyon  &  Dart~2008 & Dart~2014 & Pisa & Yale & Brazil & Gennaro$^{\dagger}$ \\
        \noalign{\smallskip}\hline\noalign{\smallskip}
        V615 Per                  &   ...  &   4.7  &   4.9  &  15.0  &   ...  &   ...  &   ...  \\
        TY CrA                    &   ...  &   3.6  &   7.7  &   7.0  &   ...  &   7.6  &   3.75 \\
        V618 Per                  &   ...  &  25.0  &  25.0  &  30.0  &   ...  &  19.8  &   ...  \\
        EK Cep                    &   ...  &  20.0  &  19.4  &  19.0  &   ...  &  18.8  &  18.95 \\
        RS Cha                    &   ...  &   7.7  &   7.7  &   7.0  &   ...  &   7.8  &   8.00 \\
        ASAS J052821+0338.5       &   6.3  &   4.1  &   4.0  &   4.0  &   ...  &   4.6  &   3.45 \\
        RX J0529.4+0041           &  10.0  &  10.4  &  10.8  &   9.0  &   ...  &  11.8  &   6.90 \\
        V1174 Ori                 &  10.0  &   8.2  &   7.8  &   8.0  &   7.8  &   7.0  &   7.40 \\
        MML 53                    &  20.0  &  17.4  &  16.3  &  13.0  &  15.2  &  13.6  &   ...  \\
        CoRoT 223992193           &   5.0  &   4.3  &   4.2  &   4.0  &   4.4  &   3.8  &   ...  \\
        Par 1802                  &   1.3  &   1.0  &   1.1  &   1.0  &   1.0  &   1.0  &   ...  \\
        JW 380                    &   2.5  &   1.4  &   2.1  &   ...  &   1.4  &   2.0  &   ...  \\
        2MASS J05352184$-$0546085 &   1.0  &   ...  &   ...  &   ...  &   ...  &   ...  &   ...  \\
        \noalign{\smallskip}\hline\noalign{\smallskip}
    \end{tabular*}
    \hspace{0.2cm}
    \parbox{0.93\linewidth}{$^{\dagger}$ Quoted ages are from \citet{Gennaro2012}, who used Bayesian
        statistics to derive a best fit age for each system. Uncertainties are not listed here. Ages
        were computed assuming Gaussian mass priors.}
\end{table*}

In Table~\ref{tab:ages} we list the best-fit ages we obtain for each
EB and each set of models. In general the scatter among the different
models for any given EB is typical of what is seen in the literature, and
the mean ages are also consistent with previous estimates. In
particular, the three EBs in the ONC display ages of 1--2\,Myr (their
canonical values). V615\,Per and V618\,Per, on the other hand, are not as
consistent with their nominal age of $\sim$13\,Myr. The first system seems
considerably younger according to the Dartmouth models, but not according
to the Pisa models. All model sets give ages for V618\,Per significantly
older than 13\,Myr. The reasons for these discrepancies are not obvious, although
we note that \cite{Southworth:04} found these two EBs to be better fit
with a lower metallicity than solar.
Finally, we list the ages determined by \citet{Gennaro2012} for six of the
EBs using a Bayesian analysis, finding that in general these Bayesian determined
ages are consistent with the ages determined from our 
$\chi^2$ minimization.

\subsection{Fitting including empirical corrections for activity\label{sec:activity}}

Being young, many of the stars in our PMS EB sample exhibit clear signs
of magnetic activity in the form of strong \ha\ and X-ray emission
(Table~\ref{tab:data2}). 
PMS stars may also exhibit strong \ha\ emission if they are actively
accreting. However, each of the EBs in our sample appears to be devoid
of circumstellar material (with the exception of CoRoT\,223992193), probably
because in order to be observed as an EB at all the system cannot have
a massive disk which would be viewed edge-on and consequently obscure the
central EB.

To examine the degree to which chromospheric activity effects might be 
responsible for the poor agreement between the observed stellar properties
and the theoretical stellar models for some of the systems, we have 
attempted to correct the directly observed \teff\ and $R$ values to 
what they would be if the stars were totally inactive. 
We converted the observed \ha\ 
equivalent widths and X-ray fluxes $F_{\rm X}$ to luminosities
(\lha\ and $L_{\rm X}$) using the distances in Table~\ref{tab:data2},
and in turn converted these into
corrections to the stellar radii ($\delta R$) and temperatures ($\delta$\teff) 
using the empirical relationships 
proposed by \citet{Stassun2012}.
Note that the \citet{Stassun2012} empirical relations are
based on active main-sequence field dwarfs and active main-sequence
EBs, and have not been broadly tested in the context of PMS stars
(see also Sec.~\ref{sec:disc_activity}).

To convert \ha\ equivalent widths to \lha\ we used standard 
NextGen stellar atmospheres \citep{Hauschildt1999a} to compute the surface continuum fluxes near 
\ha\ and multiplied these by the stellar surface areas (using the measured $R$)
and by the observed \ha\ equivalent width. In cases where only a single,
combined \ha\ equivalent width was reported for an EB, we assumed that each
component has the same intrinsic EW (equivalent to scaling
the observed EW by each star's \lbol\ and then correcting each
star's equivalent width for the continuum dilution by the other star). 
To convert $F_{\rm X}$ to $L_{\rm X}$, we assumed that each star in the EB contributes
half of the observed $F_{\rm X}$ and then used the nominal distance to the EB 
in the relation $L_{\rm X} = F_{\rm X} \times 4 \pi d^2$.
The sources we used for $F_{\rm X}$ are the ROSAT All-Sky Survey \citep{Voges:99},
the XMM-Newton Serendipitous Source Catalog \citep{Watson:09}, or measurements
we made ourselves based on publicly available observations from the Chandra
X-ray Observatory, and are indicated in each case in the references in
Table~\ref{tab:data2}.
For five of the components (RS\,Cha A and B, TY\,CrA A and B,
and Par\,1802 B), the measured \ha\ and/or X-ray emission
is very weak and as such these objects are below the range
of applicability defined by \citet{Stassun2012} for the
empirical relations; we do not attempt to correct these 
objects' properties.

\begin{figure*}[!t]
    \centering
    \includegraphics[width=0.80\linewidth]{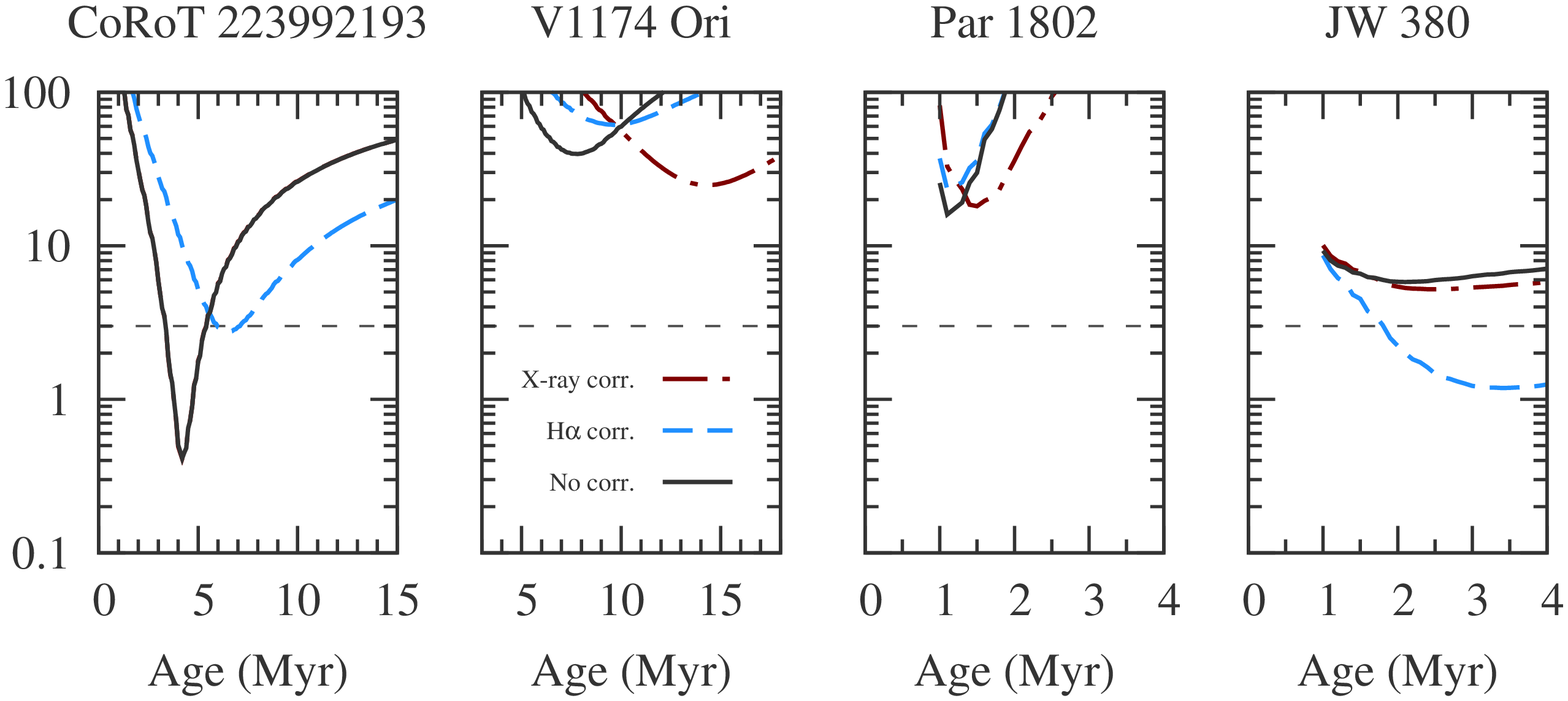} \\ \vspace{\baselineskip}
    \includegraphics[width=0.64\linewidth]{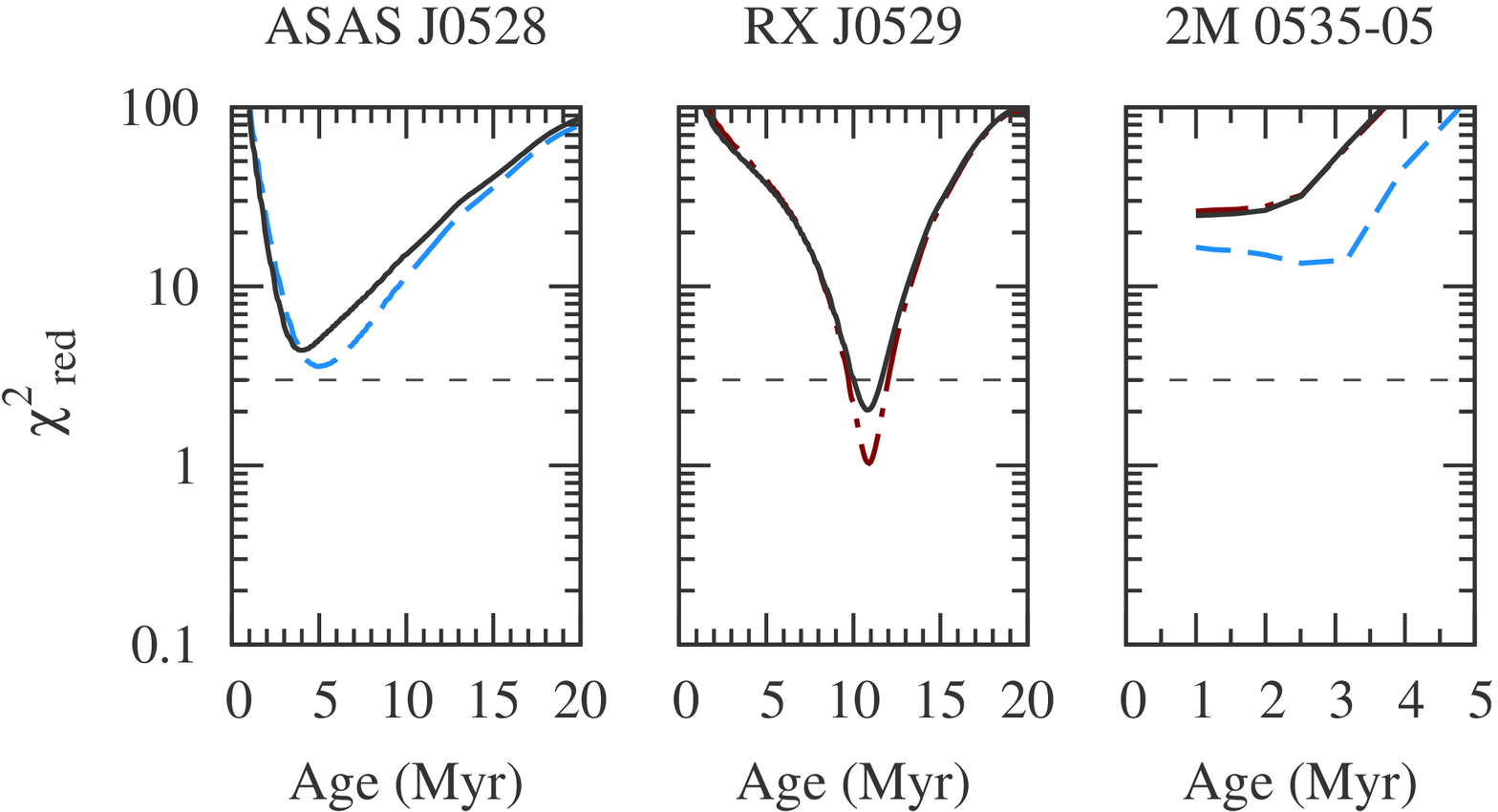}
    \caption{\label{fig:activitycorrections} 
    Change in the quality of the best-fit isochrones (black solid line) when the measured stellar
    radii and \teff\ are adjusted for the observed H$\alpha$ (maroon dash-dotted 
    line) and/or X-ray activity (light-blue dashed line)
    (Table \ref{tab:data2}) using the empirical relations of \citet{Stassun2012}.
    }
\end{figure*}

With the $\delta$\teff\ and $\delta R$ in hand for each star, we re-fit each
system using the same parameters and goodness-of-fit metrics as before. The
results are shown in Figure~\ref{fig:activitycorrections} for the EB systems that have
the requisite H$\alpha$ or X-ray measurements. In each case the
$\chi^2$ is shown as a function of the age of
the isochrone considered for the fit to the six measured quantities. For
this illustration we have used only the Dartmouth 2014 models. Solid lines
corresponding to fits with no corrections for activity are compared
separately with the goodness of fit after corrections based on either
H$\alpha$ or X-rays. It is not at all clear that the activity
corrections systematically improve the fits, as one might have
expected from the somewhat limited experience with similarly active
main-sequence EBs \citep[e.g.,][]{FC12b,FC13}. The corrections do improve the fits to the
brown dwarf system 2MASS J05352184$-$0546085 (hereafter 2M\,0535$-$05),
to ASAS\,052821+0338.5, and especially to JW\,380, but make
less of a difference for RX\,J0529.4+0041, which is already
well fit without the adjustments. The match to
CoRoT\,223992193 is actually made worse (although the quality of the
fit is still acceptable after the activity corrections), while that of
Par\,1802 is not improved and the results for V1174\,Ori are mixed,
depending on whether the corrections are based on H$\alpha$ or
X-rays. Perhaps the only pattern we see is the obvious one, which is that
activity corrections yield older ages because the activity-corrected
radii are smaller, making these young stars appear farther along in their
contraction phase.

\subsection{Lithium\label{sec:lithium}}

For seven of our EB systems there are measurements available of the
strength (equivalent width) of the lithium absorption line at
$\lambda$6708\,\AA, for one or both components. When reported in the
original sources the corresponding Li abundances have been taken as
published; otherwise we have computed them from the published equivalent
widths using the curve-of-growth tables of \cite{Pavlenko:96}, after
properly accounting for the light contribution of each component.
All values are listed in Table~\ref{tab:data2}. 

Note that we have not included
the secondary components of Par\,1802 and JW\,380 in the
compilation, even though they have published Li equivalent widths,
because these stars are cooler than the range of \teff\ included in
the \citet{Pavlenko:96} tables. The curve-of-grown tabulation provided
by \citet{Palla:07} does extend down to these very cool temperatures,
but does not quite reach the $\log g$ values required, and is computed
only under LTE.  Additionally, this more recent abundance scale seems
very different, and would yield values of $\log N{\rm (Li)}$ an order
of magnitude larger for some of our stars. For consistency with
previous work we have chosen to use the standard scale
of \cite{Pavlenko:96}, and we therefore exclude the secondaries of
Par\,1802 and JW\,380 from consideration.

\begin{figure*}[!ht]
    \includegraphics[width=0.9\linewidth]{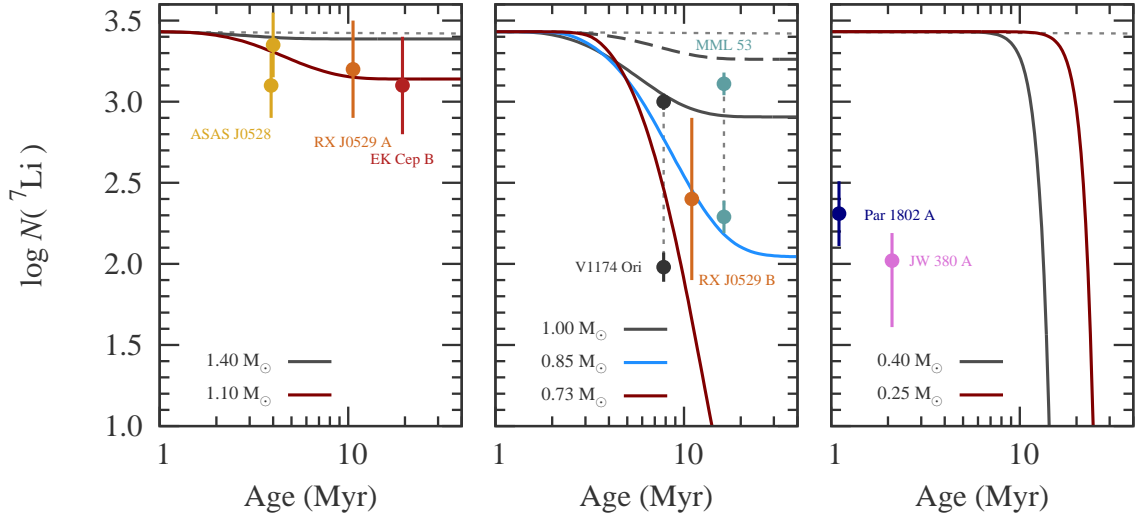}
    \caption{\label{fig:lithium} 
    Models of lithium depletion from the Dartmouth 2014 model set
    compared to the measured Li abundances for stars with $M > 1$ \msun\ (left),
    $0.5 < M < 1.0$ \msun\ (middle), and $M < 0.5$ \msun\ (right). The data
    are plotted at the best-fit age from the Dartmouth 2014 isochrone fits
    (Table \ref{tab:ages}).
    Model curves are shown for masses corresponding to specific stars in
    each panel for direct comparison. The primary and secondary components
    of V1174 Ori and MML 53 (middle panel) are connected with dotted lines.
    All models are non-magnetic, except the grey dashed track in the middle
    panel, which is a 1 \msun\ model with a 1~kG surface magnetic field.
    }
\end{figure*}

The measurements are compared with models of Li depletion in
Figure~\ref{fig:lithium}. We find theoretical predictions from standard
models to be broadly consistent with the observed abundances for the
higher-mass stars in our sample that are expected to be undepleted
(left panel, stars in the range 1.12--1.38\,\msun). 


For the two
components of V1174\,Ori (middle panel) there is good agreement with
non-magnetic models, whereas magnetic models seem to be excluded.
Reasonably good agreement is seen as well between the measured and
predicted Li abundance for
RX\,J0529.4+0041\,B and MML\,53\,A from standard models.
For V1174\,Ori, one would predict an age of 9.5~Myr based on Li depletion curves
compared to the 7.8~Myr age predicted from fitting the fundamental 
properties. Shifting the age to 9.5~Myr produces comparatively worse
agreement with the fundamental properties. At 9.5 Myr, models over-predict
the primary mass by 0.07~\msun\ ($5\sigma$) and the primary radius
by 0.11~\rsun\ ($10\sigma$). Marginal improvements are found with
the other properties in our fit. Looking instead at how the models 
perform at the exact quoted masses at 9.5 Myr, one finds the models under-predict the
primary and secondary radius by about 10\% and 3\% ($13\sigma$, $3\sigma$), 
respectively, while the primary \teff\ is under-predicted by 2\% ($<1\sigma$)
and the temperature difference is under-predicted by nearly 50\% ($7\sigma$).

In the right panel of Figure~\ref{fig:lithium} the measured
Li abundances for the primaries of Par\,1802 and JW\,380 are seen to
be significantly lower than predicted by either standard or magnetic
models for their mass and age, by more than an order of magnitude.
These are also the youngest systems in the sample with measured Li
abundances. 
In Sect.~\ref{sec:disc} we revisit these Li abundance patterns, and in
particular the apparent over-depletion of Par\,1802 and JW\,380,
including possible systematic errors in the absolute scale for the Li
abundances (see above).

The problematically low Li abundances for the lowest-mass
systems (Par\,1802, JW\,380) notwithstanding, we emphasize 
that broadly there is very good agreement between the observed and
predicted Li abundances in the EB sample (left and middle panels). In
particular, it is striking that both components of V1174\,Ori agree 
reasonably well with the expected Li abundance pattern for coeval stars 
at an age of $\sim$10 Myr, despite the other properties of the system being
very poorly fit by the same models (see Figure~\ref{fig:individual}). The same is
true of MML\,53, although the primary's Li is slightly elevated 
relative to the 1.0 \msun\ model prediction by $\sim 3\sigma$. The
primaries of MML\,53 and V1174\,Ori have nearly identical masses of 
$\approx$1.0 \msun, and comparable ages of $\sim$10 Myr, so might be
expected to have nearly identical Li abundances. Interestingly, the primary
of MML\,53 is more rapidly rotating than is the primary of V1174\,Ori
(by virtue of the shorter orbital period, and assuming tidal synchronization), thus it
is possible that the slightly higher Li abundance of the MML\,53 
primary is the result of 
rotationally induced activity retarding its Li destruction, as suggested
by \citet{Somers2014} to explain the spread of Li abundances in young clusters.


\subsection{Coevality of EBs in clusters\label{sec:clusters}}

\begin{figure}[!htbp]
    \centering
    \includegraphics[width=0.85\linewidth]{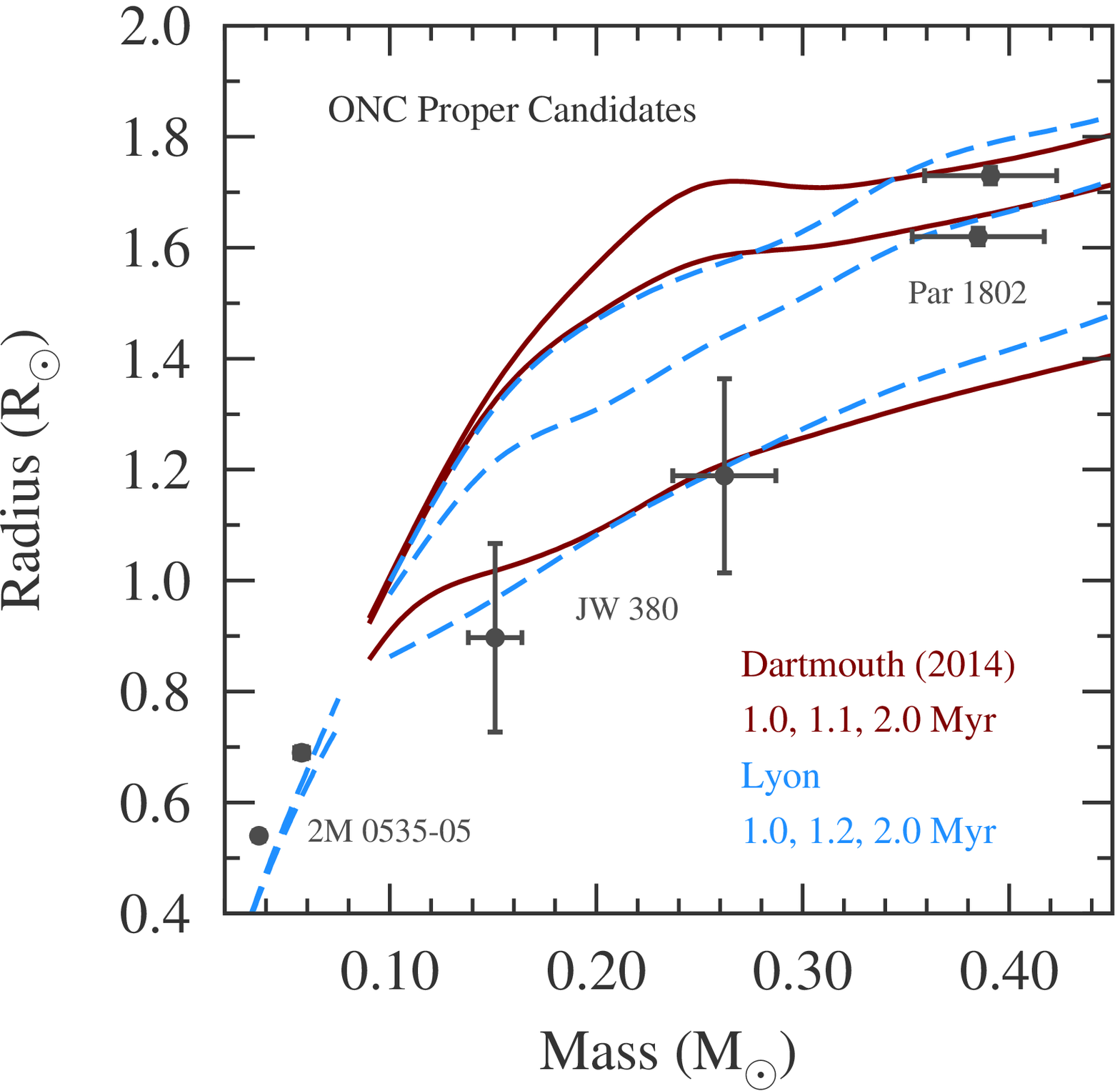} \qquad
    \includegraphics[width=0.85\linewidth]{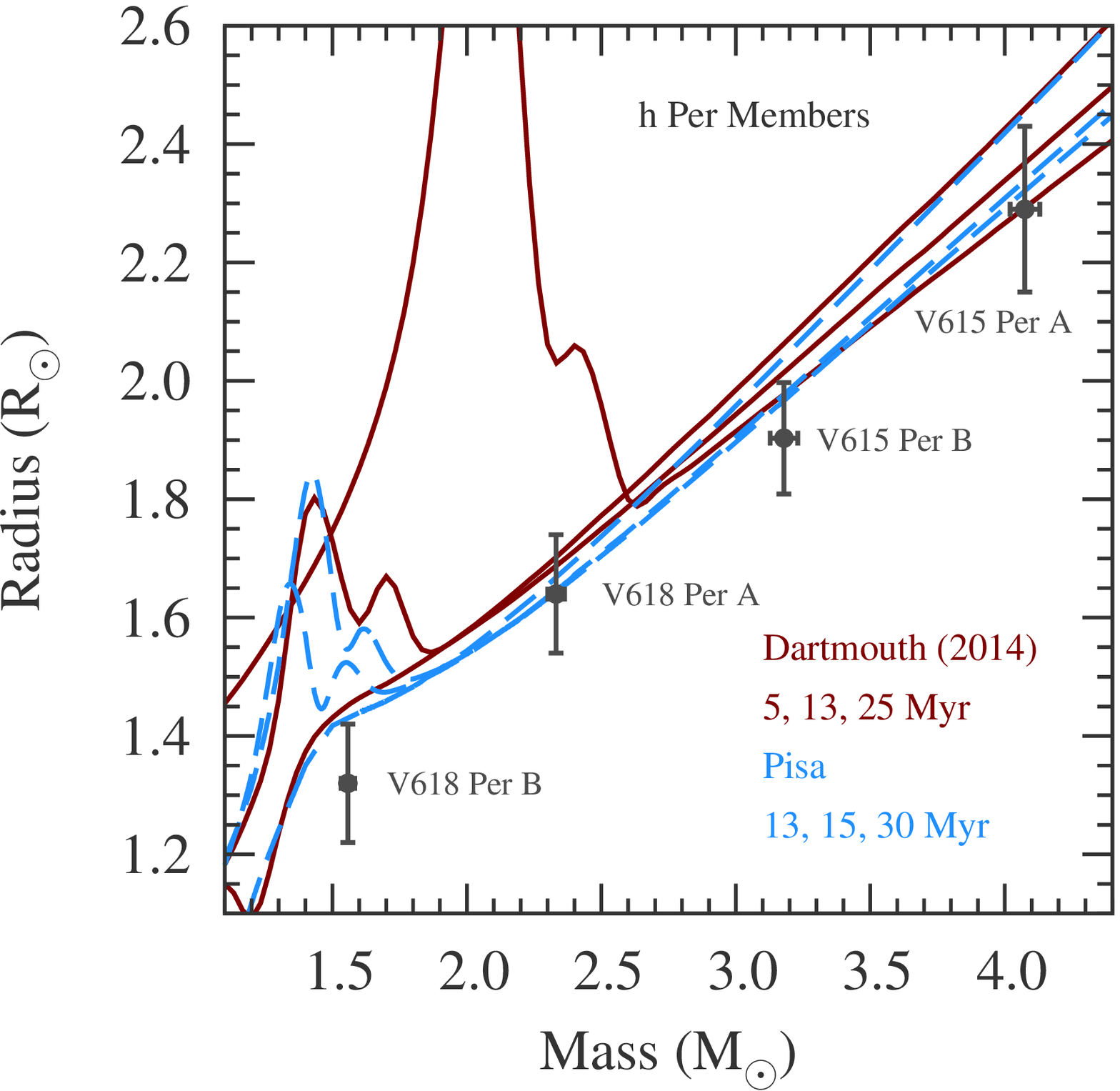} 
    \caption{\label{fig:cluster}
    Mass-radius diagram for the EBs in our sample that are members
    of cluster populations expected to be coeval. \emph{Left:}
    Orion Nebula Cluster systems compared against Dartmouth 2014 
    and Lyon ($\amlt = 1.9$) isochrones for a range of ages. The 2M\,0535$-$05
    system is compared against $\amlt = 1.0$ Lyon models, which are
    the only ones considered in this review that reach such low masses. 
    \emph{Right:} Similar
    diagram for the two h~Per systems, with a comparison against Dartmouth
    2014 and Pisa isochrones.}
\end{figure}

Two of the EBs in our sample (V615\,Per and V618\,Per) are members of the
young h~Persei cluster
(distance $\approx$2200 pc), 
and three others (Par\,1802, JW\,380, and
2M\,0535$-$05) are members of the ONC
(distance $\approx$420 pc). These EBs therefore permit a test of
the degree to which the stars in these two regions are truly coeval.
Figure \ref{fig:cluster} 
shows the result of comparing these EBs to single isochrones
near the nominal cluster ages of 1--2 Myr (ONC) and $\sim$13 Myr (Perseus).
For the ONC systems, there is no single isochrone that satisfactorily
fits all three systems at once. Collectively the stars are consistent with ages 
of 1--2 Myr,
although the JW\,380 system appears to be $\sim$0.5--1 Myr older than Par\,1802.
For the two Perseus systems, three of the four stars are consistent with a 
single cluster age of $\sim$13 Myr; however, the lowest mass star (V618\,Per\,B)
does not lie on the same isochrone, although the discrepancy is at less than the
2$\sigma$ level.

Thus, based on these comparisons, and to the extent that the measurements
are accurate, it appears that the members of these two
clusters are not strictly coeval, but exhibit an apparent age spread of 
$\sim$1 Myr. For the ONC, this represents a spread of $\sim$50\% in
age across the EB systems. Interestingly, however, the individual systems seem to be much more
coeval, with the inferred ages of the individual stars agreeing to $\sim$20\%.
This is similar to the findings of \citet{Kraus2009}, who showed using binaries
in Taurus that the stars within binaries are much more coeval ($\sim$40\%)
than are the binaries considered in aggregate ($\sim$150\%).


\section{Discussion\label{sec:disc}}

In the previous section, we examined the degree to which the 
measured properties of the benchmark EBs agree with the predictions 
of PMS stellar evolution models. We find very mixed results. 
The fits of the various stellar models to each system are
visually summarized in Figure \ref{fig:chi2mass}, where
we show the goodness-of-fit metric for each system as a
function of total system mass. Systems to the left of the
vertical line are those with a total $\chi^2$ that is 
equal to or better than that expected for a good fit.
%
A number of the systems in the sample can clearly be 
considered to be very well matched to most of the theoretical 
models for two coeval stars. These systems 
(to the left of the vertical line) include:
V618\,Per, V615\,Per, RS\,Cha, and CoRoT\,223992\-193.
At the same time, other systems are not well
matched by most (or any) of the theoretical models. These
systems (to the right of the vertical line) include:
TY\,CrA, EK\,Cep, ASAS\,052821+0338.5, RX\,J0529.4+0041,
V1174\,Ori, Par\,1802, JW\,380, and 2M\,0535$-$05.

As is evident from Figure~\ref{fig:chi2mass}, the tendency 
for some systems to be better fit by the stellar models is
not simply a function of the system mass. In this section,
we consider how the goodness of fit (or lack thereof) for
the different systems might be understood through the 
action of various possible physical effects, and we
close with a summary of the implications of this 
discussion for the efficacy of current theoretical models
in the context of observational studies of young stars
and star-forming regions.

\begin{figure*}[t]
    \centering
    \includegraphics[width=0.45\linewidth]{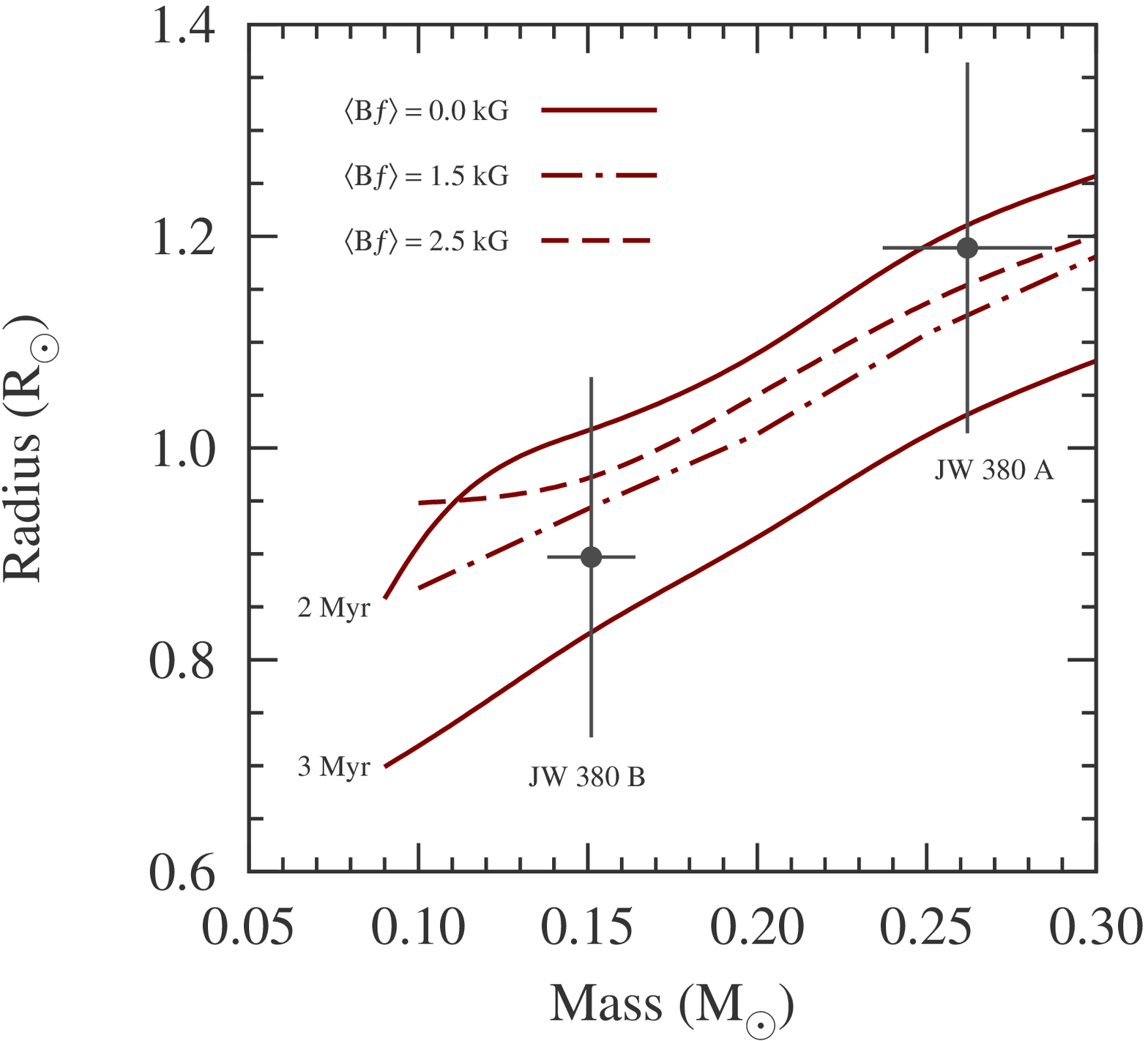} \quad
    \includegraphics[width=0.45\linewidth]{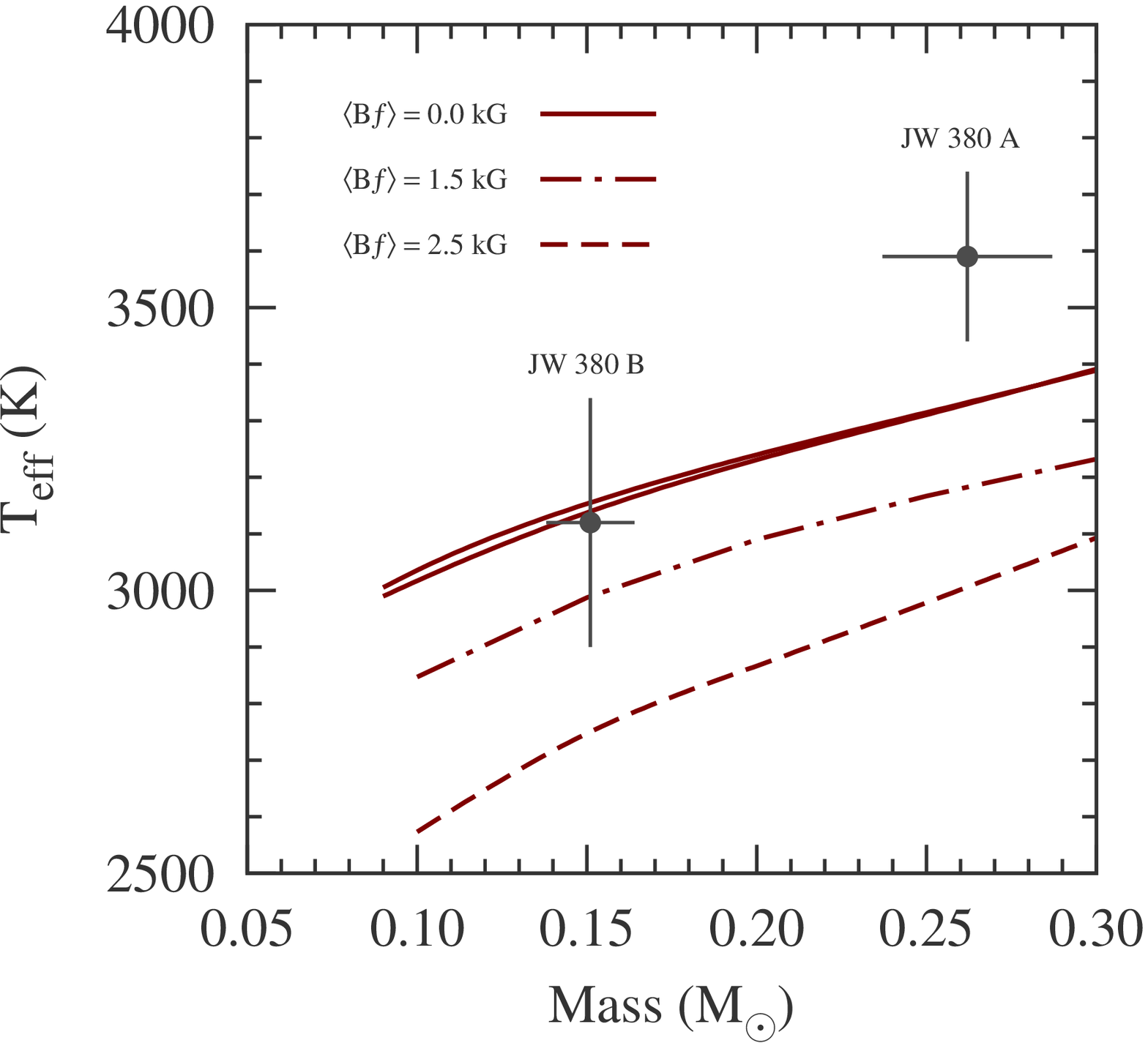} 
    \caption{Effect of magnetic fields on the predicted radii (left) and
    temperatures (right) for the components of JW\,380 according to the
    Dartmouth 2014 models,
    for three different field strengths as labeled, at an age of 3\,Myr.
    A non-magnetic isochrone for 2\,Myr is also shown.}
    \label{fig:mag_trk}
\end{figure*}

\subsection{Activity: Effects of Magnetic 
Fields\label{sec:disc_activity}}

Magnetic activity is an obvious culprit to consider for
explaining the discrepancies between the observed and
predicted properties for many of the PMS EBs. Most of the 
stars are observed to be active (in
\ha\ and/or X-rays; see Table~\ref{tab:data2}) and none 
of the stellar models currently incorporate magnetic 
field effects in their ``standard" versions.
In Sect.~\ref{sec:activity}
(see Fig.~\ref{fig:activitycorrections}),
we found that while a few of the systems 
(ASAS\,052821+0338.5, JW\,380, and 2M\,0535$-$05) 
do appear to be largely
``fixed" through the use of empirical activity adjustments
to the stellar radii and \teff\ \citep{Stassun2012},
the other active systems are either not significantly helped 
by these empirical corrections (V1174\,Ori, Par\,1802)
or else are actually made somewhat worse (CoRoT\,\-223992193). 

\citet{Stassun2012} previously showed that application of
their empirical activity corrections to 2M\,0535$-$05,
comprising two brown dwarfs at a nominal age of 1--2 Myr,
substantially improved that system's agreement with 
standard model predictions, so the improvement found
here for that system is not surprising. However,
although the observed reversal of \teff\ with mass present in this binary
\citep{Stassun2006,Stassun:07,Gomez2009} could be mitigated, the models 
still predicted a primary mass that was too low given the observed 
luminosity. Conversely, models of the secondary star 
predicted too large of a mass after activity corrections.
The reason
that the models still predict incorrect masses is that 
the activity corrections altered both the primary's \teff\  
and its radius, leaving the luminosity effectively constant,
whereas the luminosity must increase to find agreement with
the models. Issues with the primary could potentially be 
rectified by assuming that spots only influence the observed 
luminosity and \teff, but do not impart changes in the radii. 
Spots that evolve on a short timescale compared to the object's 
thermal timescale might be consistent with this picture. 
However, \citet{Mohanty2012} performed detailed spectral 
modeling of the 2M\,0535$-$05 system during eclipse to argue that
spots are probably not strong in the system and spot models
appear unable to explain the observed \teff\ reversal.
In any case, this would not provide better consistency with the 
secondary's mass. The disagreement may therefore point to yet 
unresolved systematic problems with sub-stellar structure 
models at PMS ages that even activity corrections cannot fully rectify.

The effects of strong surface and/or internal magnetic
fields have recently been incorporated into the Dartmouth
models using a physically consistent treatment within the 
framework of mixing length theory (MLT) \citep{LS95,FC12b}.
An example of what these models predict for the impact of magnetic 
fields in JW~380 is illustrated in Figure~\ref{fig:mag_trk}. 
Two 3~Myr magnetic model isochrones are shown, one where both
stars have a surface magnetic field $\langle$B$f\rangle = 1.5$~kG 
and the other with $\langle$B$f\rangle = 2.5$~kG. Peak magnetic
field strengths within the stars are 10--20~kG. The magnetic 
isochrones show the expected increase in stellar radii 
and reduction in \teff\ as convective flux is reduced. 
Although the magnetic
isochrones produce better consistency between the two stars 
at an age of 3~Myr in the $M$-$R$ plane, suppressing
the \teff s produces worse agreement in the $M$-\teff\ plane.
We find that this result is typical when trying to fit magnetic 
models to the PMS EB population. 


In summary, the effects of magnetic activity are 
expected to be ubiquitous among low-mass PMS stars such
as those that comprise our benchmark sample. 
However, the expected magnitude of the effect 
on the radii and \teff\ for most of 
the stars is not expected to be large enough to fully
explain the significant discrepancies seen in many of the EBs.
Activity effects alone do not appear to be the principal solution
to the problem of the poorly fit EBs. 

\begin{figure}[!htbp]
    \centering
     \includegraphics[width=0.95\linewidth]{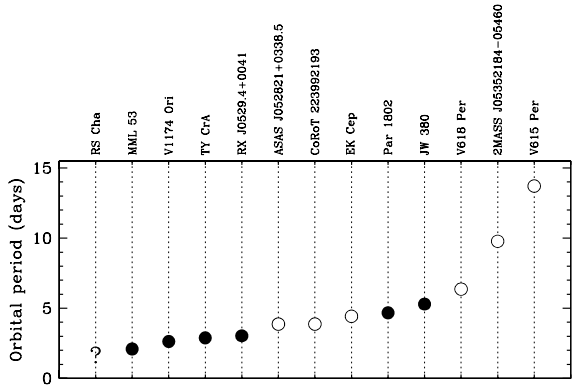}
    \caption{PMS EB sample sorted by orbital period. Filled circles represent
    systems with known tertiary components, and open circles those with
    no known third stars. Evidence for a tertiary in the case of RS\,Cha still needs to
    be confirmed.}
    \label{fig:tertiaries}
\end{figure}

\subsection{Lithium: Probing effects on internal stellar structure}

We have found that there is fairly good agreement between the
observed and predicted Li abundances in the EB sample
(see Section \ref{sec:lithium}).
This is interesting and perhaps counterintuitive considering the 
large discrepances between the
observed and predicted stellar properties for many of the EBs.
It is not clear how the stellar global properties (i.e., \teff\ and $R$)
can be altered without also 
causing larger discrepancies in Li abundance than observed. Here we
simply note this interesting observation and leave its explanation to
future investigation.

An additional factor that can influence the Li depletion, as well
as the global properties of PMS stars, is the efficiency of convection
\citep[see, e.g.,][]{Hillenbrand2004,Young:01},
typically parametrized in stellar evolution models in terms of the mixing
length parameter. While the models evaluated here (and indeed most of the
published models) adopt a value of $\amlt$ appropriate for the Sun,
\cite{Tognelli2012} have considered non-solar (lower) values of $\amlt$.
Their study included three of the EBs in the present sample as well as
several somewhat older clusters, but the results for the EBs were largely
inconclusive due to the small number of Li measurements and the fact that
the uncertainties are still relatively large in most cases.

Beyond the obvious choice of a solar-calibrated $\amlt$ for solar-mass
stars at solar age, the specific choice of $\amlt$ in other contexts 
has long been a source of debate as there
is a lack of empirical constraints to suggest what values are most 
appropriate in different evolutionary stages. 
Advancements in 3D radiation-hydrodynamic models and production of 
large grids of such models
may help guide the development of models that do not adopt a single value
for $\amlt$, but employ a value that varies with stellar properties as it
evolves \citep{Freytag2012,Magic2013,Magic2014}.


\subsection{Tertiary Stars: Effects of Dynamical and 
Tidal Heating}

A large number of the EBs in our benchmark sample are 
found to contain at least one additional stellar companion
(Table~\ref{tab:data1}). Indeed, six of the 13 systems
have known tertiary stars, for a high fraction of triples 
to binaries of almost 90\%\footnote{There is some evidence that
RS\,Cha may also be a triple from small changes in the 
systemic velocity and in the $O-C$ residuals of the eclipse
timings \citep{Bohm:09, Woollands:13}, but this has yet to be confirmed.}.
Such a high occurrence of triples is not unexpected for
this sample; most of the EBs have very short orbital 
periods, and in the field \citet{Tokovinin2006} finds that
the occurrence of triples among binaries with orbital
periods $<$3~d is as high as 96\%. Indeed, in the EB sample
studied here, there is a tendency for the tertiary 
companions to be present among the EBs with the shortest 
orbital periods and absent among those with the longest periods
(Figure \ref{fig:tertiaries}).

\begin{figure*}[!ht]
    \centering
    \includegraphics[width=0.8\linewidth]{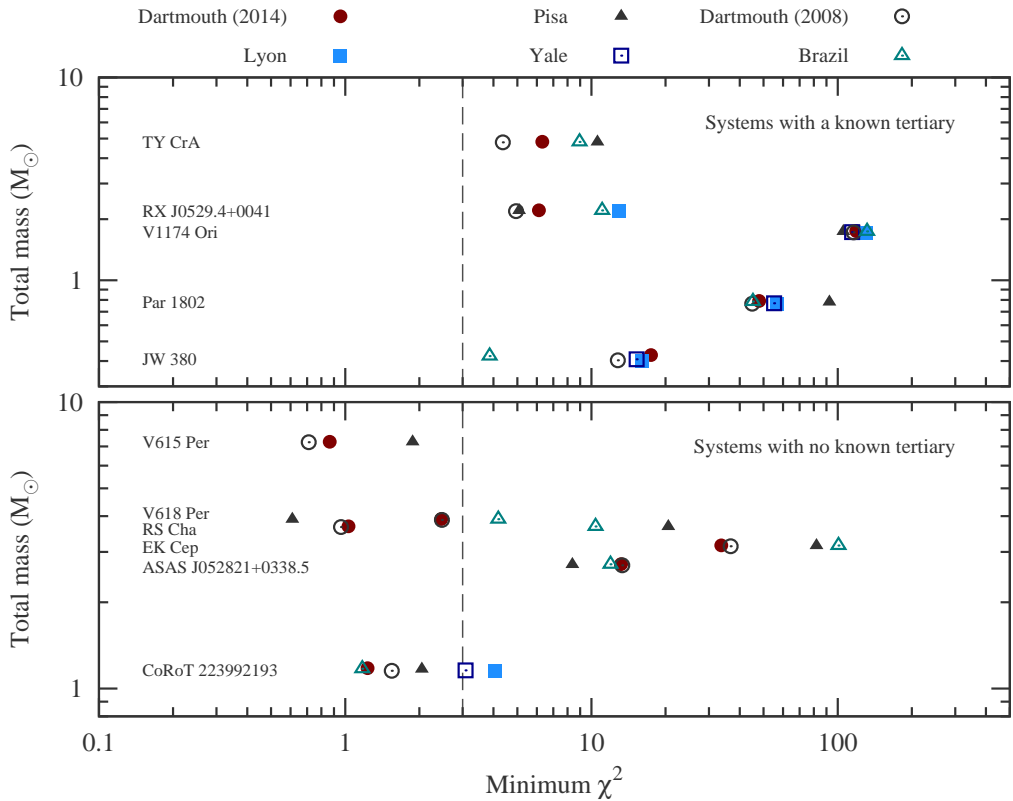}
    \caption{Best-fit $\chi^2$ values for stellar mass EBs in our sample 
     sorted according to the total mass of the system. The goodness of fit 
     is shown with different symbols for each model used in each case (key 
     on top). The vertical dashed line represents the expected $\chi^2$ 
     value based on the number of degrees of freedom, if the errors were 
     Gaussian. (Top) Systems with a known tertiary component. (Bottom) 
     Systems with no known tertiary component. The brown dwarf EB 2M0535--05,
     being substellar, is excluded from this diagram; see the text.}
    \label{fig:chi2tertiary}
\end{figure*}

Comparing the goodness of fit for the stellar mass EBs with 
known tertiaries to those without tertiaries
(Figure~\ref{fig:chi2tertiary}),
we find a striking difference, with the triple systems
being systematically poorly fit whereas the non-tertiary
systems are in general well fit by at least one of the models.
The exceptions to this trend are EK\,Cep and, to a lesser degree,
ASAS\,J052821+0338.5.\footnote{We exclude MML\,53 
here as it is not sufficiently well characterized to 
permit a stringent constraint; see Sect.~\ref{sec:fitting} 
and \ref{sec:ebnotes}.}
EK\,Cep is not known to have a tertiary component but
its properties are very poorly fit by all of the stellar models
(see Figure \ref{fig:individual}). 
In addition, the brown-dwarf EB 
2M\,0535$-$05 is not well fit by the models (only the Lyon
models go low enough in mass to attempt a fit) despite not having
a known tertiary. While improved through the application
of empirical activity corrections, it remains 
problematic, perhaps because of deficiencies in the models at
substellar masses (see Section \ref{sec:disc_activity}).
EK\,Cep and ASAS\,J052821+0338.5 remain problematic for
other as-yet unidentified reasons. Perhaps they are
triples after all but have not yet been identified as such.

Nonetheless, it appears that the tendency to triplicity
among the PMS EB sample cannot be ignored as a likely
important factor in their failure to be well fit by 
standard PMS stellar models. Broadly speaking, there are 
two possible explanations for this: observational systematics
associated with the presence of third light in the EB
analysis, and physical effects associated with the
additional dynamics of a three-body system. 

We believe that observational systematics are unlikely 
to be the dominant explanation. 
In each case in which contamination of the EB photometry by third
light has been suspected, the effect has been accounted for in the
analysis in one way or another, at least to first order (see \ref{sec:ebnotes}).
Any residual effects would mostly influence the inclination angle,
and have a much smaller impact on the relative radii and temperatures.
The absolute masses would be largely unaffected by residual errors in
the inclination angle. Additional blending of the spectral lines from
the unrecognized presence of a third star would also have a minimal
effect on the masses, which are determined mostly from observations at
the orbital quadratures, where line blending is the smallest.

In contrast, there are reasons on physical grounds to
expect that the dynamics of these three-body systems may
help to explain why these systems are not well fit by the
standard stellar models. For example, 
\citet{Reipurth2012} have demonstrated that the
rapid dynamical evolution of triple systems during the
PMS phase can explain the occurrence and hierarchical
nature of triples in the field. In those simulations, 
the triple systems initially begin non-hierarchically,
but then through a highly dynamical set of 
chaotic interactions, rapidly ``unfold" dynamically on
a timescale of 1--10 Myr, leaving behind a very tight
inner binary and a wide tertiary.

There are two dynamical effects in this picture that
could directly impact the binary and the properties of
the stars that comprise it. One effect is the input of
energy from the tertiary's orbit into one or both of the
inner binary stars. The second effect is the tidal 
interaction of the two inner binary stars if they
become tight enough at an early evolutionary stage when
their radii are large compared to their orbital separation.

We perform here a simple order-of-magnitude estimate to
evaluate potential effects of three-body dynamics on the 
properties of the stars in an inner binary, representing 
the EB that we observe, with an outer tertiary star. The 
conceptual argument is that the energy contained in the orbit
of the tertiary about the inner binary is of the same order
of magnitude as the total amount of energy that would need
to be injected into one or both of the stars of the inner
binary to produce the observed luminosity difference 
compared to the standard model predictions. This requires that 
the deposited energy is 
rapidly dissipated within the stars
from large-scales to small-scales 
so as to be able to quickly provide internal 
support against gravity 
\citep[e.g., via convective turbulence;][]{Zahn2005}.

Consider V1174\,Ori, perhaps one of the worst fit EBs
in the sample. We can compute the observed stellar
luminosities using the derived fundamental properties and
the Stefan-Boltzmann law. This yields $L_{\rm A} = 0.65 L_{\odot}$
and $L_{\rm B} = 0.17 L_{\odot}$ for the primary and secondary,
respectively. There is then a luminosity difference 
$\Delta L = 0.47 L_{\odot}$ between the two components. Stellar
models, on the other hand, predict a luminosity difference 
of only $\Delta L_{\rm model} \approx 0.24 L_{\odot}$ during the PMS
contraction of two stars with masses similar to V1174\,Ori. One 
can conclude that the primary is either too luminous or the 
secondary too faint for the system's age, or some combination 
of the two. 

For the purposes of this calculation, we assume that the primary
is too luminous by $\sim$0.2 $L_\odot$
and that the secondary is more-or-less ``normal.''
We also make the simplistic assumption that 
the tertiary star injects energy into
the primary star only, at a rate equal to the primary's observed 
over-luminosity.
If the tertiary has an orbital period of $P_{\rm orb} \sim 10^4$~yr,
as is not uncommon for tertiary stars in hierarchical triples, 
then at an earlier time in the dynamical evolution of the system its
orbit will have been shorter by, say, a factor of 100.
In any case, over the course of subsequent tertiary orbits of 
$10^2$--$10^4$ yr,
the primary will not have time to relax back to its original 
configuration on the Hyashi track between subsequent periastron passages
of the tertiary. This is because 
the Kelvin-Helmholtz timescale,
$t_{\rm KH}$, is $\sim$10$^7$~yr. The star will
contract somewhat after each periastron passage, but for simplicity we assume it 
does not. To maintain the primary at $L_{\rm A} = 0.65 L_{\odot}$, 
the tertiary must inject 
$Q_{\rm tidal} \sim 0.2 L_{\odot} \times 10^{2-4} {\rm yr} \sim 10^{42-44}$ erg
of heat in each orbital pass.
Here, $L_{\odot}$ is the solar luminosity and the last factor
is the tertiary orbital period.

By comparison, we can estimate the amount of energy contained in the tertiary's
orbit about the inner binary 
as the sum of the 
kinetic and potential energies of the inner binary and the tertiary: 
$E_{\rm orb} = -G M_{\rm binary} M_{\rm tertiary} / 2a$, where $a$ is the  
orbital semi-major axis. Using Kepler's third law we can rewrite this in terms of
the orbital period,
\begin{equation}
  E_{\rm orb} = -\frac{1}{2}\mu\left[\frac{2\pi G}{P} \left(M_{\rm binary} 
            + M_{\rm tertiary}\right) \right]^{2/3},
\end{equation}
where $\mu = M_{\rm binary}M_{\rm tertiary}/\left(M_{\rm binary} + M_{\rm tertiary}\right)$ 
is the reduced mass of the system, $M_{\rm binary}$ is the mass of the
inner binary, and $M_{\rm tertiary}$ is the mass of the tertiary.
Given an orbital period of $10^2$--$10^4$ yr and 
$M_{\rm binary} = 1.73 M_{\odot}$, we estimate
that $E_{\rm orb} \sim 10^{42}$--$10^{44}$ erg, depending on the tertiary mass.
Table~\ref{tab:orbital_energy} gives the range of energies that are required to be deposited
to generate the various luminosity differences we observe in V1174\,Ori, and in two
other systems. We list
values of the orbital energy 
assuming a tertiary mass $M_{\rm tertiary}$ of 0.05, 0.10, or $0.50 M_{\odot}$ 
orbiting the inner binary with a period of $10^2$, $10^3$, or $10^4$~yr.

\begin{table*}[!t]
    \centering
    \small
    \caption{Orbital energy (erg) for a tertiary companion of mass $M_{\rm tertiary}$ and
             orbiting with a period of $10^2$--$10^4$~yr.
             \label{tab:orbital_energy}}
    \begin{tabular*}{\linewidth}{@{\extracolsep{\fill}} l c c c c c c c c c c c c}
        \noalign{\smallskip}\hline
        \hline\noalign{\smallskip}
         EB      & $M_{\rm binary}$ & \multicolumn{3}{c}{$M_{\rm tertiary} = 0.05 M_{\odot}$} &
                                    & \multicolumn{3}{c}{$0.10 M_{\odot}$} &
                                    & \multicolumn{3}{c}{$0.50 M_{\odot}$} \\
                 \noalign{\smallskip}
                 \cline{3-5} \cline{7-9} \cline{11-13} 
                 \noalign{\smallskip}
                 & $(M_{\odot})$    & $10^2$~yr & $10^3$~yr & $10^4$~yr & & $10^2$~yr & $10^3$~yr & 
                                      $10^4$~yr & & $10^2$~yr & $10^3$~yr & $10^4$~yr \\
        \noalign{\smallskip}\hline\noalign{\smallskip}
        %
        V1174\,Ori & 1.73  &  2.9e+43 &  6.2e+42 &  1.3e+42 & & 5.7e+43 &  1.2e+43 &  2.7e+42 & & 2.7e+44 &  5.8e+43 &  1.2e+43  \\
        Par\,1802  & 0.77  &  1.7e+43 &  3.6e+42 &  7.7e+41 & & 3.3e+43 &  7.0e+42 &  1.5e+42 & & 1.4e+44 &  3.1e+43 &  6.7e+42  \\
        JW\,380    & 0.41  &  1.1e+43 &  2.3e+42 &  5.0e+41 & & 2.1e+43 &  4.5e+42 &  9.7e+41 & & 8.6e+43 &  1.8e+43 &  4.0e+42  \\
        \noalign{\smallskip}\hline\noalign{\smallskip}
    \end{tabular*}
\end{table*}

In general, the orbital energies available in the tertiary orbit are of the same order of magnitude
as the total amount of energy that typically needs to be injected into one of the 
stars of the inner binary. This is a highly idealized scenario and the actual tidal
interactions are far more complex, but it is tantalizing that the energies are so
similar. If the tertiary had an extremely strong and violent dynamical interaction
with a star in the inner binary, it is possible that the rate of energy transfer 
vastly exceeded the total luminosity of the star in the inner binary. This could 
then deposit the requisite quantity of energy in the star in only a few encounters.
However,
it could also be that weaker, more regular encounters occurring over several Myr 
have a more lasting impact and allow the star to maintain a higher luminosity for
longer due to continual injections of energy. 


Finally, once the tertiary has sufficiently tightened the inner
binary orbit, producing the short-period EB that we observe,
tidal interaction between the two EB stars can potentially 
continue to inject heat into one or both of the EB stars. For
example, \citet{Gomez2012} showed in the case of Par\,1802
that tidal interaction between the two EB stars over the
past $\sim$1 Myr can account for the over-luminosity of 
$\sim 10^{26}$~W by the primary over the nearly equal-mass
secondary. Par\,1802 is one of the EB systems in our sample
possessing a known tertiary component. While \citet{Gomez2012}
did not attempt to model the dynamical evolution of the system as a 
triple, they concluded that the direct tidal 
interaction of the EB stars was likely driven into its current
configuration by the action of the tertiary in the recent past.
Therefore it appears plausible, both qualitatively and in terms
of the quantitative energetics discussed above, that the
tertiaries in many of the PMS EBs have influenced the evolution
of the stars either directly, through injection of tertiary
orbital energy, or indirectly through tightening of the inner
EB to the point that binary tidal interaction injects sufficient
heat to alter the observed stellar properties.


\subsection{Accretion: Effects of Accretion History
\label{sec:accretion}}

Some theoretical studies
have argued that the detailed
accretion histories of low-mass stars can be 
important in setting the physical properties of the 
stars during the PMS contraction phase 
\citep[e.g.,][]{Siess1996,Hartmann1997,Tout1999,Baraffe:10,Hosokawa2011,Baraffe2012}.
If true, calculations suggest that the effect could be quite dramatic,
with a very ``bursty" accretion history 
producing a change in the stellar radius by a factor of $\sim$3 or
more at $\sim$1 Myr \citep[e.g.,][]{Baraffe:10}. 
In addition, the higher 
internal temperatures would lead to dramatically enhanced 
Li depletion,
and thus could explain the apparently anomalously low
Li abundances reported for some members of young
star-forming regions \citep[e.g.,][]{Palla:07}. 
%
Thus, we may consider a scenario in which the EBs in our
sample possessing tertiary components have undergone more
bursty accretion episodes in the past, leading to 
discrepant stellar properties
(undersized radii, increased \teff, and under-luminosity)
compared to standard models.
The apparently highly Li depleted JW\,380 might be taken as 
circumstantial evidence of accretion effects. 
If we assume the system is younger than estimated by standard 
stellar 
models (1 Myr instead of 2 Myr, for instance), then one finds 
that both 
stars exhibit smaller radii than predicted by models. Furthermore, 
the primary is hotter (by about 200\,K) and less luminous than 
one would then expect from a 1 Myr isochrone. 
However, coupling this interpretation with the apparently 
high Li depletion depends critically on
the adopted PMS curve-of-growth and Li abundance
scale, which has large model uncertainties at the very
cool \teff\ of this low-mass system.
%
The same concern regarding the Li abundance scale at cool \teff\ 
applies to Par\,1802, 
although in any case this system 
does not appear to exhibit properties 
consistent with episodic accretion predictions.
In particular, despite the primary and secondary stars having a
mass ratio very near unity, 
the primary is much hotter, larger, and more luminous than the secondary.
To cause the secondary to be so much smaller than the primary, 
it would need to have undergone stronger accretion bursts than the primary, 
in which case the secondary should be significantly hotter than the primary,
which is the opposite of what is observed.
None of the other tertiary-hosting EBs in our sample exhibit strong Li depletion. 
Overall, there are no clear ``smoking gun" signals in our EB sample of the 
predicted effects of accretion history. 

\subsection{Overall performance of PMS stellar models, and
the H-R Diagram Revisited\label{sec:hrdredux}}

In Section \ref{sec:hrd} we considered how well the various PMS stellar
models are able to reproduce the known masses of the EB stars in the
H-R diagram,  as this is a basic use of the stellar models for 
observational studies of young stars and star-forming regions. 
We found that in general the model-inferred masses are accurate to
$\sim$10\% for masses above 1 \msun, but considerably less reliable
(mass errors of 50--100\%) below 1 \msun\ and with a tendency for the
models to over-predict the masses for these low-mass stars.
While these trends were found to be true for most or all of the 
models, there we did find some differences. For example, the Pisa
and Brazil model sets exhibited the lowest scatter in the mass
discrepancies below 1 \msun, and especially the Brazil models showed
the smallest mean offset in the inferred stellar masses 
(Table \ref{tab:modelstats}).

The relatively good performance of the Pisa models in particular might 
not be surprising considering that these models were spec\-ifically 
developed for application to the PMS. The relatively good
performance of the Brazil models is interesting, as to our 
knowledge these have not been widely utilized in the literature. 
These models are the only representative of the ATON code among
the model sets we have examined in this paper, and they appear
to perform on par with the other more widely used model sets.
Perhaps this is due to the limited set of outputs provided in 
this model series for direct comparison to observables (notably
colors and magnitudes).

However, and  most importantly, there is a fundamental limitation 
to firmly assessing the efficacy of the models with the EB sample
that we have considered in this paper. It is clear that EBs,
and in particular those with tertiary companions (at least half
of the sample), may not represent a directly comparable test of
standard models of single stars.

\begin{figure}[!ht]
    \centering
    \includegraphics[width=0.9\linewidth]{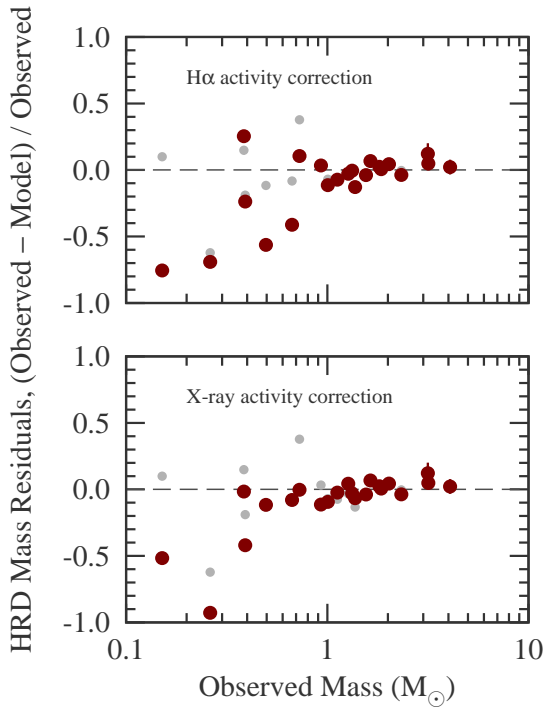}
    \caption{Normalized mass residuals between the measured EB masses
    and those inferred by the Dartmouth 2014 models in the H-R diagram,
    as a function of measured mass, after correction for activity effects
    on $R$ and \teff\ following \cite{Stassun2012}. Results are shown
    separately for activity corrections based on \ha\ (top) and X-rays
    (bottom). Residuals without corrections are
    represented with grey symbols.}
    \label{fig:hrd_activity}
\end{figure}

Therefore, we revisit the performance of the stellar models
in predicting the measured EB stellar masses in the H-R diagram,
but now accounting for activity effects and the presence of tertiaries.
We again fit each of the model isochrones to the individual EB stars 
as in Section \ref{sec:hrd}, but now adjust the observed 
stellar \teff\ and $R$ for activity effects as in Section
\ref{sec:activity}. 
The result is shown in Figure \ref{fig:hrd_activity} 
where we find that while there is some improvement in the model
inferred masses for some objects, in general activity effects
alone cannot fully reconcile the discrepancies in the inferred
masses. This is especially noticeable at low masses, where 
empirically correcting the data for magnetic activity further 
degrades agreeement between models and observations.

\begin{figure}[!ht]
    \centering
    \includegraphics[width=0.9\linewidth]{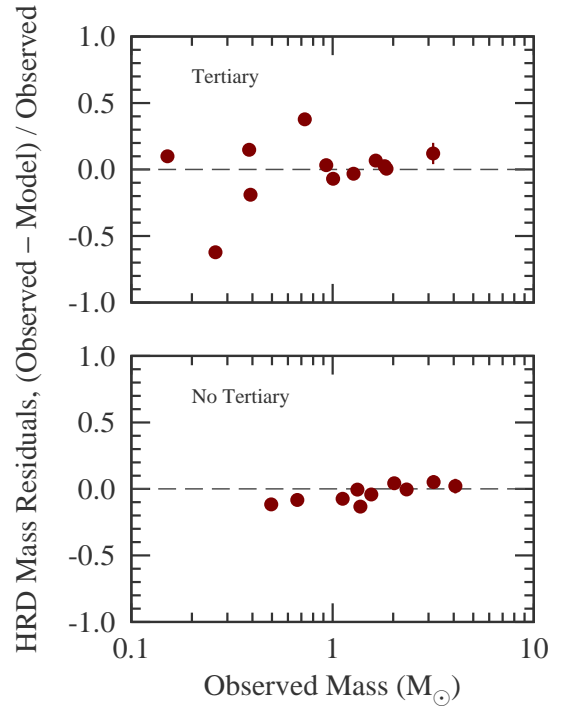}
    \caption{Normalized mass residuals from the Dartmouth 2014 models
    as in Figure~\ref{fig:hrd_activity}, shown here without corrections
    for activity effects, and displayed separately for systems with and
    without known tertiary components.}
    \label{fig:hrd_tertiary}
\end{figure}

Next, we separate the EB sample into those systems bearing tertiaries
versus those without, in Figure \ref{fig:hrd_tertiary}. 
Here we do not adjust the observed stellar
properties for activity effects. 
We observe a striking result, namely that the stars not in
triple systems have their masses very well reproduced by the
models, to better than $\sim$10\% over the full range of stellar
masses sampled (from 4.0 \msun\ down to 0.5 \msun), whereas those in 
triple systems constitute all of the highly discrepant cases, with mass 
errors of 50\% or more as noted earlier. 

There is a hint in the no-tertiary
sample of a tendency toward slightly over-predicted stellar 
masses at the low-mass end. Unfortunately, the lowest-mass stars
are all in triple systems, and therefore we cannot determine with
this sample whether the trend of over-predicted masses continues 
to the lowest masses, or whether the large mass discrepancies
are entirely the result of effects associated with the tertiaries
in those systems.

\section{Conclusions and Future Work\label{sec:summary}}

We have performed a detailed assessment of the ability of various
standard (solar-calibrated) PMS stellar evolution model sets to accurately reproduce the
observed properties of a benchmark sample of 13 PMS EBs with
masses in the range 0.04--4.0 \msun\ and with nominal ages ranging from
$\sim$1 to 20 Myr. The fundamental properties of the EBs, and the salient
physical ingredients of the models, have been carefully compiled for 
future reference and use by the community.
Through this exercise we have learned
much about the interesting and complex physics of the PMS 
stage---particularly in PMS stars that are influenced by their 
companions---complex\-ities that are definitively not yet incorporated
in any standard PMS models.



Crucially, we have found that the presence of tertiary companions
in many---perhaps most---of the PMS EBs appears in some way to 
corrupt the agreement of the observed stellar properties with the
standard model predictions for young single stars. We have considered
several mechanisms by which the tertiaries might do this, 
with direct injection of heat into one or both of the EB stars
appearing to be the most plausible explanation. Indeed, we find that
the energies of the tertiary orbits are comparable to that needed to
potentially explain the scatter in the EB properties, perhaps through
tidal interaction.

We find that the three EBs in the ONC are individually coeval
to better than $\sim$20\%, whereas in the aggregate they show an age
spread of $\sim$50\%. In other words, the two stars in each EB system
are more coeval than are the EBs relative to one another. 
This may be the result of scatter in the EB properties due to
the tertiaries. Alternatively,
this could
be taken as evidence for a real age spread across the ONC cluster.
The apparently high Li depletion of JW\,380 
could 
be taken as additional evidence of an age for this system 
that is higher than average for the cluster.
However, work is needed to firm up the Li abundance scale at such low 
\teff\ for PMS stars to verify that the Li abundance in this system
(and in Par\,1802) is in fact highly depleted. 

More generally, we find broadly very good agreement between the observed 
and predicted Li abundances in the EB sample, including for EBs that
are otherwise very poorly fit by the stellar models in terms of the
other physical properties. This suggests that whatever the
mechanism is by which the EB stars' surface properties are affected,
the internal structure of the stars is not sufficiently altered
to cause comparably large problems with the Li abundances. At the
same time, detailed comparisons between EB stars of comparable mass
and age (i.e., V1174\,Ori A and MML\,53 A, both with mass $\sim$1.0 \msun\
and age $\sim$10 Myr) show slightly different Li abundances, with
the more rapidly rotating MML\,53 being less depleted, suggesting 
that rotationally suppressed Li depletion may be inducing spreads in
Li abundances as early as $\sim$10 Myr \citep{Somers2014}.

The aforementioned results refer only to models with a solar-calibrated 
$\amlt$, since those are the most commonly available. Use of a different 
value for $\amlt$ will have an impact on the ability of standard models 
to fit the EB data \citep[e.g.,][]{Tognelli2012,Hillenbrand2004,Young:01}. 
In this review we have adopted the solar-calibrated $\amlt$ primarily for
the sake of uniformity so that we might better tease out the other
dominant effects that bear upon the performance of the models against
measured stellar properties.
Advancements in 3D radiation-hydrodynamic models 
are needed to guide the development of stellar evolution models that 
employ a value for $\amlt$ that varies 
in a physically appropriate way as the stars evolve. 

As surveys of young stars and star-forming regions continue, 
there will
be opportunities to enhance the sample of the benchmark PMS EBs. 
Some of the existing PMS EBs need to have complete analyses
performed, in particular MML\,53 (which still does not have 
definitive individual radii determined) and the recently 
discovered EBs in the ONC by \citet{Morales2012}.
There is also in general a need for more careful and consistent
measurement of activity tracers in these EBs. 
In this paper, we have attempted to recover the 
activity information from the primary literature on the EB
sample used here, but this was difficult in some cases. 
Still, we remain convinced that activity effects
alone are unlikely to fully reconcile the discrepancies.
The reason is that for a subset of the EBs we do already
have reliable activity measurements, and the model versus data differences are not 
uniformly improved using the empirical activity corrections
that have been developed based on main-sequence samples
\citep[e.g.][]{Stassun2012}. 
It appears that PMS
stars are instead dominated by other effects tied in some
way to the presence of tertiary companions, as we have discussed 
(e.g., tidal interaction).

Similarly, while our main analysis 
made use of
the \teff\ differences measured directly from the light curve
solutions, our assessment of the models versus the data in the
H-R diagram necessarily still depends on the absolute \teff\ 
of the primary stars, which
could be subject to systematic uncertainties due to 
different \teff\ scales adopted by various authors. Finally,
in general the original EB discovery papers have not performed 
detailed analyses of the metallicities of the EBs, so we have
resorted to assuming solar metallicity for most of the systems
in our analysis. This is probably a reasonable assumption, 
especially for the EBs that are members of young clusters whose
overall metallicities do have some observational basis for
being very nearly solar. But in general metallicity is 
expected to be a factor in the predicted stellar properties,
so this should ideally be carefully investigated on a 
system-by-system basis in the future. Along the same lines, efforts
should be made whenever possible to measure the Li abundance
for both components of young EBs, or at least to report their
equivalent widths along with the light ratios between the stars, to
enable the abundances to be computed.

While the focus of PMS EB discovery papers is usually to
establish the absolute masses, radii, and temperatures of the stars,
observers are strongly encouraged to report also other intermediate
quantities such
as mass ratios, temperature ratios (or differences), and radius ratios,
which can usually be determined to higher accuracy and precision than the absolute
properties and consequently provide more stringent constraints on models,
as we have shown here. Reliable measurements for PMS EB stars are still so
few in number that it is particularly important also to fully document any efforts
to control systematic errors, or at least to assess their importance,
in order to enable a critical judgment of the reliability of the analysis. 


Just as observers should strive to publish all of the available
information that can be pulled from the data, so too should modelers. 
Aside from the most basic quantities ($R$, \teff, $L$), modelers
should strive to publish a wide array of secondary quantities, 
such as lithium abundance and properties of the stellar convection
zone, including convective turnover times---both local and globally 
averaged. Such quantities as the radius of gyration and internal
structure constants may also be of interest to those testing stellar
models. Additionally, modelers are encouraged to develop extensive,
high resolution model grids, in both mass and metallicity, that 
minimize the need for interpolation into sub-grid regimes. This 
includes making available both detailed mass tracks and isochrones.
Observers would also benefit from development of grids that cover
``non-standard'' parameter spaces \citep[e.g.,][]{Tognelli2011,Spada2013}, 
such as varying $\amlt$, deuterium abundances, solar composition, 
magnetic fields, and accretion physics. Of course, 
computation of such extensive grids is daunting given 
the number of tunable parameters, but effort toward this ideal should
be made nonetheless.

Finally, it is clear that the triplicity of these systems, and
the resulting dynamics, cannot be ignored in future efforts to understand
and fully utilize these most fundamental stellar benchmarks.
Indeed, when we consider only EB stars that are not members of 
triple systems, we find that current PMS stellar models are able
to faithfully recover the measured stellar masses to better than
$\sim$10\% in the H-R diagram, down to 0.5 \msun. It is possible that there 
is a systematic tendency for the models to over-predict the stellar
masses in the H-R diagram at low masses, but unfortunately the
current benchmark EB sample is dominated by triple systems at
the lowest masses, precluding our ability to properly assess
this possibility. 
We do not believe that observational issues are to blame for
the poor fits of the stellar models to the EBs in triple
systems, but rather it is very likely that there are real
astrophysical effects driving this fundamental finding.
Failing to account for the presence of triples 
in the EB sample leads to inferred masses in
the H-R diagram that are incorrect by 50--100\%.

We hope that these insights will motivate additional 
theoretical work into the effects and observable consequences
of magnetic fields and activity, 
of accretion history, and especially into the dynamical evolution
of binary and triple systems including the effects of N-body
and tidal interaction. To the extent that most or all young 
stars likely experience accretion early in their evolution
and/or dynamical interactions especially in cluster environments,
these issues will likely continue to be salient in our 
understanding of PMS stellar evolution and in our ability to
test and calibrate the theoretical stellar models that are so
central to our paradigm of star formation.

\section*{Acknowledgments}
KGS acknowledges NSF grants AST-1009810 and AST-0849736.
GT acknowledges partial support from NSF grant AST-1007992.
GAF acknowledges NSF grant AST-0908345 and the William H. Neukom 
1964 Institute for Computational Science at Dartmouth College,
which both supported the development of the magnetic Dartmouth
stellar evolution code.
This research has made use of NASA’s Astrophysics Data System, 
the SIMBAD database and the VizieR catalog access tool, both
operated at CDS, Strasbourg, France, and the 
ROSAT data archive tools hosted by the High Energy Astrophysics 
Science Archive Research Center (HEASARC) at NASA’s Goddard Sp\-ace 
Flight Center. The research has also made use of data obtained from
the Chandra Data Archive and the Chandra Source Catalog, and software
provided by the Chandra X-ray Center (CXC) in the application
packages CIAO, ChIPS, and Sherpa. We thank Scott Wolk for his
assistance in the use of those tools.

\section*{References}

\bibliography{papers,ref1_appendix}

\appendix

\section{Notes on Individual PMS EB Systems\label{sec:ebnotes}}


\begin{table*}[!htbp]
\small
\caption{Activity and other properties of young EBs\label{tab:data2}}
\begin{tabular}{l @{}c c c c c c c}
\noalign{\smallskip}\hline
\hline\noalign{\smallskip}
 Star & Age (Myr) & Comp & $\log N{\rm (Li)}$$^a$ & H$\alpha$ EW (\AA)$^b$  & $\log L_{\rm H\alpha}/L_{\rm bol}$ & $\log L_{\rm X}/L_{\rm bol}$ & [Fe/H]$^c$ \\
      & Dist (pc) &      &       & $F_{\rm X}$ (erg\,s$^{-1}$\,cm$^{-2}$)$^d$  &                  &                              & References$^e$ \\
\noalign{\smallskip}\hline\noalign{\smallskip}

V615 Per                &  13   & A  &    ...    &    ...   &   ...  &   ...  &    (0.0)  \\
                        & 2200  & B  &    ...    &    ...   &   ...  &   ...  &     1,2      \\

\noalign{\medskip}

TY CrA                  &   3   & A  &    ...    &    ...   &   ...  &  $-5.37$$^f$ &     ...   \\
                        &  100  & B  &    ...    &  $1.6\times 10^{-12}$ &   ...  &  $-3.94$$^f$ &   3,4,5        \\

\noalign{\medskip}

V618 Per                &  13   & A  &    ...    &    ...   &   ...  &   ...  &    (0.0)  \\
                        & 2200  & B  &    ...    &    ...   &   ...  &   ...  &     1,2      \\

\noalign{\medskip}

EK Cep                  &  20   & A  &    ...    &    ...   &   ...  &   $-4.80$$^f$  & $+0.07 \pm 0.05$$^g$  \\
                        &  150  & B  &  $3.1\pm 0.3$  &    $6.6\times 10^{-14}$   &   ...  &   $-3.82$  &  2,6,7,8,9,10,11              \\

\noalign{\medskip}

RS Cha                  &   6   & A  &    ...    &    ...   &   ...  &  $-5.47$$^f$ & $+0.17 \pm 0.01$    \\
                        &  100  & B  &    ...    &  $3.0\times 10^{-13}$ &   ...  &  $-5.46$$^f$ &  6,12,13,14,15             \\

\noalign{\medskip}

ASAS                 &  10   & A  & $3.10 \pm 0.20$ &   ... / 1  &   ...  &   $-3.24$  & $-0.15 \pm 0.20$$^h$  \\
J052821+0338.5          &  280  & B  & $3.35 \pm 0.20$ &$5.0\times 10^{-13}$   &  $-3.98$ &  $-3.06$  &  16,17             \\

\noalign{\medskip}
 
RX J0529.4+0041         &  10   & A  &  $3.2 \pm 0.3$  &    ...   &   ...  &  $-3.39$ &     (0.0)     \\
                        &  325  & B  &  $2.4 \pm 0.5$  &  $3.4\times 10^{-13}$ &   ...  &  $-2.97$ &  18,19           \\

\noalign{\medskip}

V1174 Ori               &  10   & A  & $3.00 \pm 0.05$$^i$ & 0.4 / 4.2 &  $-4.42$ &  $-3.12$ &    (0.0)    \\
                        &  390  & B  & $1.98 \pm 0.09$$^i$ & $2.1\times 10^{-13}$ &  $-3.80$ &  $-2.55$ &    6,20         \\

\noalign{\medskip}

MML 53                  &  15   & A  & $3.11 \pm 0.07$$^i$ &   ...   &   ...  &   $-3.49$  &     ...     \\
                        &  136  & B  & $2.29 \pm 0.10$$^i$ &  $9.5\times 10^{-13}$   &   ...  &  $-3.01$  &  17,21,22           \\

\noalign{\medskip}

CoRoT                   & 3--6  & A  &    ...     &   5.6   &  $-3.45$ &   $-3.19$  &    ($-0.15$)     \\
223992193               &  800  & B  &    ...     &   $2.5\times 10^{-14}$   &  $-3.60$ &   $-2.94$  &   11,23,24          \\

\noalign{\medskip}

Par 1802                & 1--2  & A  & $2.31 \pm 0.13$$^i$ &    1    &  $-4.40$ &  $-3.31$ &    (0.0)    \\
                        &  420  & B  &  ...                               & $8.8\times 10^{-14}$ &  $-4.49$$^f$ &  $-3.10$ &  25,26           \\

\noalign{\medskip}

JW 380                  & 1--2  & A  &  $2.0 \pm 0.3$$^i$   &   13    &  $-3.32$ &  $-3.71$ &    (0.0)    \\
                        &  420  & B  &      ...         & $1.5\times 10^{-14}$ &  $-3.51$ &  $-3.22$ &   27,28          \\

\noalign{\medskip}

2MASS                   & 1--2  & A  &    ...    & 32.6 / 4.8 &  $-3.47$ &  $-3.89$ &    (0.0)    \\
J05352184$-$0546085     &  420  & B  &    ...    &  $1.0\times 10^{-15}$ &  $-4.30$ &  $-3.76$ &  29           \\
\noalign{\smallskip}\hline
\end{tabular}
\vskip 5pt
{$^a$ Abundance on the usual scale in which $\log N{\rm (H)} = 12$.} \\
{$^b$ Two values are listed when measured separately for the primary and secondary.} \\
{$^c$ Values in parentheses are assumed for the parent cluster or association of the binary.} \\
{$^d$ Values from the sources indicated in the last column, or measured here directly from publicly available Chandra observations. In the case of ASAS\,J052821+0338.5
the location of the ROSAT source is nominally 11.6$^{\prime\prime}$ from the binary position, still within the 14$^{\prime\prime}$ ROSAT error circle; for MML\,53 the ROSAT source is 5.7$^{\prime\prime}$ from the binary
position, well within the ROSAT error circle.} \\
{$^e$ Sources for the data presented here as well as in Table~\ref{tab:data1}.} \\
{$^f$ Value outside of the range in which the activity correction relations of \citet{Stassun2012} are valid.} \\
{$^g$ Metallicity for the secondary. The primary star has [Fe/H] $= -0.24$ from a single line.} \\
{$^h$ Average metallicity for the two components ([Fe/H] $= -0.2 \pm 0.2$ and [Fe/H] $= -0.1 \pm 0.2$, respectively).} \\
{$^i$ Computed here or in the original sources under NLTE using the tables of Pavkenko \& Magazz\`u (1996).} \\
{\bf References:} 1.~\cite{Southworth:04}; 2.~\cite{Popper:80}; 3.~\cite{Casey:98};
4.~\cite{Casey:95}; 5.~\cite{Vaz:98}; 6.~\cite{Torres:10}; 7.~\cite{Popper:87};
8.~\cite{Hill:84}; 9.~\cite{Tomkin:83}; 10.~\cite{Martin:93}; 11.~\cite{Watson:09};
12.~\cite{Andersen:75}; 13.~\cite{Clausen:80}; 14.~\cite{Ribas:00}; 15.~\cite{Alecian:05};
16.~\cite{Stempels:08}; 17.~\cite{Voges:99}; 18.~\cite{Covino:04}; 19.~\cite{Covino:01};
20.~\cite{Stassun:04}; 21.~\cite{Hebb:11}; 22.~\cite{Hebb:10}; 23.~\cite{Gillen:14};
24.~\cite{King:00}; 25.~\cite{Gomez:12}; 26.~\cite{Cargile:08}; 27.~\cite{Irwin:07};
28.~\cite{DaRio:09}; 29.~\cite{Gomez:09}.

\end{table*}


We describe here the particulars of the 13 pre-main sequence eclipsing
binaries discussed in the main paper, explaining the sources of the
mass, radius, and effective temperature measurements presented in
Table~\ref{tab:data1}. Because our goal is to provide meaningful
comparisons with stellar evolution models, our primary concern is the
accuracy of the results for each system, rather than the stated
precision in the original publications. Consequently, we have examined
each binary critically, and in some cases we have re-derived the
stellar properties and/or adjusted their uncertainties to better
reflect both the quality and quantity of the observations, as well as
any complicating factors in the analysis. One of the most obvious for
this sample is the presence of distortions in the light curve due to
spots, which can evolve with time and can seriously compromise the
accuracy of the radii \citep[see, e.g.,][]{Windmiller:02}. In about half
of the systems the photometry is also contaminated to various degrees
by ``third light''. A more subtle problem that is often overlooked is
degeneracies between several of the fitted parameters, most notably
between the relative radii. The radius \emph{sum} is usually well
determined, but the \emph{ratio} is much harder to establish
accurately. This is commonly seen in partially eclipsing systems with
similar components, and the best way to overcome the problem is to
make use of an external constraint such as a spectroscopic light ratio
\citep[see, e.g.,][]{Andersen:91}, by either imposing it on the light
curve solution, or at least checking for consistency between the
photometric and spectroscopic values. This has not always been
possible for the systems below. As a result of these difficulties, 
and despite our best efforts to take them into account, the
uncertainties that are reported may still be somewhat optimistic in some cases
as they do not fully account for all systematic errors. This should be kept
in mind when evaluating the fits to stellar evolution models discussed in
the text.

In three cases (EK\,Cep, RS\,Cha, V1174\,Ori) a similarly critical
examination of the original sources and revision of the stellar
parameters was carried out by \citet{Torres:10}, and we have simply
adopted their values here.

In addition to compiling the usual fundamental properties of the
binary components ($M$, $R$, \teff), we have made an effort to extract
from the original bibliographic sources measurements that are
differential in nature, because they are usually more accurate as well
as more precise, and therefore provide better and stronger constraints
on stellar evolution models. One of these is the temperature
difference between the primary and secondary, $\Delta T_{\rm eff}$.
This quantity can typically be determined much more accurately from
the light curve solutions than spectroscopically because it is tied
directly to difference in the depths of the eclipses, which is
relatively easy to measure in most cases. Individual temperatures, on
the other hand, are more prone to systematics stemming from
uncertainties in the absolute temperature scale for PMS stars. While
errors for $\Delta T_{\rm eff}$ are only rarely reported, we have
reconstructed them here from the published information or by adopting
a conservative estimate of the uncertainty in the temperature ratio,
usually 0.01 unless otherwise noted.  The ratio of the masses ($q
\equiv M_{\rm B}/M_{\rm A}$) is typically also better determined than
either of the individual masses, and was taken directly from the
original sources, when published, or else derived from the velocity
semi-amplitudes and their uncertainties. We have not collected or made
use of the radius ratios from the literature because of the potential
for systematics noted above, because they are seldom reported, and
because it is not possible to recover their true errors from those of
the individual radii.

\paragraph{\bf V615\,Per and V618\,Per}
These two EBs are members of the young open cluster h\,Per, for which
various studies in the literature indicate a metallicity near solar,
although with some scatter. The light curves analyzed
by \citet{Southworth:04} are uncomplicated, and the solutions were
constrained using the spectroscopic light ratio, at least for
V615\,Per. We have adopted the absolute masses, mass ratios, and radii
from that source, along with their uncertainties. Similarly for the
primary temperature, which is based on spectral disentangling. The
temperature difference (and resulting secondary \teff) was derived by
us on the basis of the measured ratio of the central surface
brightness ($J_s$) from the light curve fits along with the
limb-darkening coefficients and the calibration by \cite{Popper:80},
and is considered more accurate than can be determined from the
spectra.

\paragraph{\bf TY\,CrA}
The primary is a Herbig Be star. The light curve analysis of this
system is complicated not only by the presence of significant light
from a third, spatially unresolved spectroscopic component, but also
by contamination from the reflection nebula NGC\,6726/7 in which the
object is embedded. The velocity of the third star and of the
center-of-mass of the binary are variable, and there is also evidence
of a light-travel time effect in the eclipse timing residuals.  The
secondary eclipse is very shallow ($\sim$0.03 mag).  Additionally, the
light curves \citep{Vaz:98} show variability on several timescales
from days to years. We have adopted the absolute masses, radii, and
primary \teff\ from the work of \citet{Casey:98}, in whose analysis an
attempt was made to constrain the radius ratio with external
information from their spectra. We have taken the radius errors from
the same source. However, our independent analysis of the original
radial velocities from \citet{Casey:95} indicates significantly larger
uncertainties for the masses, so we adopted those more conservative
errors here. The primary \teff\ of \citet{Casey:98} is based on color
indices in the Str\"omgren system, and that of the secondary comes
directly from the light-curve analysis. The error in the temperature
difference was estimated as indicated above.  Finally, adaptive optics
imaging by \citet{Chauvin:03} has revealed a close companion at a
separation of 0.29$^{\prime\prime}$. If physically associated, it would be an M4
star, making the system at least quadruple.

\paragraph{\bf EK\,Cep}
The light curves of this system show no complications. The masses and
radii have been taken directly from the compilation
of \citet{Torres:10}, which updates the work of \citet{Popper:87}, who
in turn based his results on the light curves of \citet{Hill:84} and
the spectroscopy of \citet{Tomkin:83}. The secondary eclipse appears to
be total, or nearly so, and the primary eclipse possibly annular,
which tends to alleviate degeneracies between the individual component
radii. The primary \teff\ and its error have also been adopted
from \citet{Torres:10}.  The secondary temperature was revised slightly
here as done above for V615\,Per and V618\,Per. \citet{Martin:93}
derived an estimate of the chemical abundance of the cooler secondary
star of ${\rm [m/H]} = +0.07 \pm 0.05$, based on five metal lines. The
system shows measurable apsidal motion.

\paragraph{\bf RS\,Cha}
This is classified as a Herbig Ae system and a member of the young
$\eta$\,Cha cluster. We have adopted the masses and radii from the
compilation of \citet{Torres:10}; they are based on the spectroscopic
work of \citet{Andersen:75} and photometric analysis
of \citet{Clausen:80}.  Non-radial ($\delta$\,Sct-type) pulsations have
been detected in the primary and secondary of RS\,Cha, both
photometrically \citep{Clausen:80} and
spectroscopically \citep{Alecian:05, Bohm:09}. The effects on the
light curves are small and were accounted for in the analysis
of \citet{Clausen:80}, which also makes use of the spectroscopic light
ratio to lift the degeneracy in the radius ratio.  We have adopted the
primary \teff\ from \citet{Ribas:00}, which we have then
combined with the temperature ratio reported by \citet{Clausen:80} to
infer the secondary \teff, as well as $\Delta T_{\rm eff}$ along
with its corresponding error. Claims of changes in the center-of-mass
velocity of the binary \citep{Woollands:13} and non-linear variations
in the $O-C$ eclipse timing residuals \citep{Bohm:09} that might be
indicative of a third body in the system require confirmation (see
\citealt{Andersen:75} and \citealt{Alecian:05}). 
The latter study provided an estimate of the metallicity as 
${\rm [Fe/H]} = +0.17 \pm 0.01$, with a rather small uncertainty.

\paragraph{\bf ASAS\,J052821+0338.5}
This system is likely a member of the Orion\,OB1a region onto which it
is projected.  The light curves show obvious distortions presumably
due to spots.  The analysis of \citet{Stempels:08} explored two
different treatments of these distortions (rectification, and direct
modeling), and found rather significant differences in the relative
radii and \teff\ from the two approaches. The authors noted that
the radius ratio is rather poorly determined in this case, likely due
to the partial nature of the eclipses and the lack of a spectroscopic
constraint on the light ratio.  Their final results, which we have
adopted here, are based on the rectified light curves. However, in
view of the sensitivity of the results to the methodology, our
concerns about the accuracy of the radius ratio, and the fact that our
independent reanalysis of the spectroscopy yields slightly different
velocity semi-amplitudes (particularly for the primary), we have
conservatively increased the uncertainties in the masses and
especially those in the radii over those reported in the original
analysis. The primary \teff\ we have adopted comes from spectral
disentangling performed by \citet{Stempels:08}. The secondary \teff\
and corresponding $\Delta T_{\rm eff}$ derive from the light curve
analysis. We have assigned an error to the latter quantity based on an
assumed uncertainty in the \teff\ ratio of 0.03, which is larger
than in other cases to account for the issues described above. Rough
metallicities were reported for the primary and secondary as ${\rm
[m/H]} = -0.2 \pm 0.2$ and ${\rm [m/H]} = -0.1 \pm 0.2$. We have
adopted the average here for the system.

\paragraph{\bf RX\,J0529.4+0041}
This EB in the Orion\,OB1a region has a visual companion at
approximately 1.3$^{\prime\prime}$, which is also seen spectroscopically
\citep{Covino:01} and is likely physically associated as it shares the
same radial velocity. The light curves are significantly affected by
spots, in addition to the contamination from third light. The analysis
of \citet{Covino:04} attempted to correct for both effects through a
combination of direct spot modeling and rectification. The eclipses
are partial and the stars rather similar, which combined with the
other difficulties just described makes it challenging to reach high
accuracy in the relative radii. Although a spectroscopic light ratio
is available that might help remove the degeneracy, it does not appear
that this piece of external information was used in this case. The
possibility of systematic errors in the radii therefore remains, as is
also mentioned by the authors. We have adopted the masses for the two
components and their uncertainties as published, as well as the
nominal radii, but we have increased the radius errors to be
conservative. \citet{Covino:04} inferred the primary \teff\ from its
spectral type and several calibrations, while the secondary
\teff\ comes directly from the light curve solution. We take those
values here as published, with their corresponding uncertainties,
along with an error for $\Delta T_{\rm eff}$ of 50\,K as reported. The
metallicity is assumed here to be that of the parent population, i.e.,
near solar.

\paragraph{\bf V1174\,Ori}
This Orion nebula cluster system is considered a likely member of the
Orion\,OB1c subgroup. We have adopted the masses, radii, and
temperatures from the compilation of \citet{Torres:10}, who provided
slight revisions of the values in the original work
of \citet{Stassun:04}. The primary temperature is based on an assumed
spectral type of ${\rm K4.5} \pm 0.1$ and a standard calibration. The
uncertainty in $\Delta T_{\rm eff}$ listed in Table~\ref{tab:data2}
was derived by us from an assumed error of 0.01 in the temperature
ratio.  As in other cool systems the effects of spots are obvious, but
were accounted for in the modeling of \citet{Stassun:04}.  Evidence for
a third star in the system comes from extra light required to properly
fit the $I$-band light curve, as well as a significant color excess
that increases toward the red.

\paragraph{\bf MML\,53}
This EB is a probable member of the Upper Cent\-aurus-Lupus
sub-association, in the region of the Sco-Cen OB complex. The light
curves display significant distortions due to spots. Rectified
versions were used in the preliminary analysis of \citet{Hebb:10},
although full details were not reported. A third, spatially unresolved
star was discovered spectroscopically by these authors, which is
likely physically associated with the EB and also affects the light
curves at the level of about 15\%. A subsequent analysis by
\citet{Hebb:11} presented a spectroscopic orbit and minimum masses for
the components, which include contributions to the errors from the
potential effects of spots. A definitive study of this binary is still
needed, preferably with more complete phase coverage in the
spectroscopy. For the present paper we have derived the absolute
masses from the $M \sin^3 i$ values of \citet{Hebb:11} and the
inclination angle of 83.1$^\circ$ reported in their earlier study, to
which we have assigned an uncertainty of 1$^\circ$. Individual radii
have not been reported, likely because of the difficulty in
determining the radius ratio with the photometric material at hand. A
spectroscopic constraint on the light ratio would also be helpful, but
is lacking. The only quantity pertaining to the star sizes that we are
able to derive for MML\,53 based on the published information is the
radius sum, $R_{\rm A} + R_{\rm B}$, which we calculated from the sum
of the relative radii (with an assumed 3\% error), the projected
semimajor axis, and the inclination angle. The \teff s reported
by \citet{Hebb:10} rely on a joint fit of the light curve and a single
spectrum, and were considered by those authors to be preliminary. We have
assigned them an error of 100\,K.

\paragraph{\bf CoRoT\,223992193}
This system is deemed a member of the young open cluster NGC\,2264,
for which an estimate of the metallicity has been given as ${\rm
  [Fe/H]} \approx -0.15$ \citep{King:00}. The high-quality and
continuous 23.4-day light curve from {\it CoRoT\/} shows obvious
rotational modulation from spots that are seen to change with
time. These distortions were removed in the analysis of
\citet{Gillen:14} prior to fitting for the photometric elements. To
help eliminate correlations between the relative radii the authors
constrained the fit using their measured spectroscopic light ratio. We
have adopted their masses and radii here as published, together with
their formal uncertainties. The effective temperatures of both stars
were determined by those authors by comparing their spectra with
synthetic templates via cross-correlation. We have assigned them
conservative uncertainties of 200\,K. The original study does not
provide information on the temperature difference from the light-curve
solution.

\paragraph{\bf Par\,1802}
Another member of the Orion Nebula Cluster. The light curves show
intrinsic variability due to spots, as well as contamination from the
light of a third unresolved star (also possibly spotted), implying the
system is triple. Both effects were accounted for in the light-curve
solutions of \citet{Gomez:12}.  The parameters and uncertainties we
list in Table~\ref{tab:data2} were adopted from that work, which uses
the spectroscopic material from \citet{Cargile:08}. The primary
\teff\ is based on an adopted luminosity ratio and a combined
\teff\ for the system based on the combined M2 spectral type. We
have computed the uncertainty in $\Delta T_{\rm eff}$ from that of the
\teff\ ratio given by \citet{Gomez:12} and a more conservative
error of 0.01 in that quantity. The system has the peculiarity that
even though the components are of similar mass their temperatures are
rather different, leading to a large difference in luminosity in
excess of 60\%, perhaps due to tidal heating 
\citep[see also][]{Stassun:08}.

\paragraph{\bf JW\,380}
A likely member of the Orion nebula cluster. In their analysis of this
EB \citet{Irwin:07} detected a spatially unresolved spectroscopic
companion that is most likely physically associated, making this a
triple system. Its contribution to the total light ($\sim$13\%) was
accounted for in the fit.  The light curve was also rectified prior to
fitting in order to rid it of significant distortions due to spots.
The authors noted the strong degeneracy in the radius ratio, from the
fact that the luminosity ratio was not constrained in the fit, and
they cautioned about the possibility of systematic errors in their
light-curve results beyond the formal uncertainties. Here we have
adopted their absolute masses (and uncertainties) as published. We did
the same for the radii, but in view of their warning we adopted the
largest of the (strongly) asymmetric error bars. Although
\citet{Irwin:07} did not attempt to infer the temperatures, they did
quote a value for the primary based on its assumed spectral type
(M1.5). We have adopted this value, and assigned it a conservative
uncertainty of 200\,K. The secondary temperature (along with $\Delta
T_{\rm eff}$ and its uncertainty) were inferred from the reported
temperature ratio, to which we have attached a more conservative error
of 0.02 than originally indicated.

\paragraph{\bf 2MASS\,J05352184$-$0546085}
This is a rare pair of eclipsing brown dwarfs in the Orion nebula
cluster. In the most recent analysis of \citet{Gomez:09} the effects of
spots apparent in the light curves were accounted for in the modeling;
they imply a rather large spot coverage on the primary ($\sim$65\%). This
system shows a surprising temperature reversal, in the sense that the
more massive and larger primary star is cooler than the secondary (see
also \citealt{Stassun:06} and \citealt{Stassun:07}).  We have adopted here the
masses and radii as reported in the \citet{Gomez:09} study, along with
their uncertainties. We have also taken the primary temperature (based
on the assumed spectral type of ${\rm M6.5} \pm 0.5$) and its error
from these authors, and computed the secondary temperature and $\Delta
T_{\rm eff}$ using the reported temperature ratio with a more
conservative error of 0.01.

\section{Physical Ingredients of Stellar Models, 
and Notes on Discarded Models\label{sec:tracknotes}}

The physical ingredients of the accepted and rejected model
sets are summarized in Tables \ref{tab:models_in} and
\ref{tab:models_ex}, respectively. Notes on the rejected
model sets are provided below.

\paragraph{\bf ATON} 
Models by  \citet{DAntona1994,DAntona1997} are still
adopted for studies of low-mass PMS systems. Although these models
lack non-grey surface boundary conditions, they are still the only
PMS model set to implement a non-local treatment of convection. This
treatment, known as Full Spectrum of Turbulence \citep[FST;][]{Canuto1991}, makes these
models relevant for discussions of low-mass stellar physics. There
have been updates to the ATON code \citep[version 3.x][]{Ventura2008}, 
which include, among other improvements, the addition of
non-grey surface boundary conditions that use both standard mixing
length theory and FST \citep{Montalban2004,DVD09}. Unfortunately, the only
set of publicly available models for the updated ATON 3.x code are 
for metal-poor stars. Solar metallicity tracks, though they have 
been computed, have not been made publicly available.

\paragraph{\bf Grenoble}
Of the model sets excluded from our analysis, the
Grenoble models by \citet{Siess2000} are arguably the most widely
adopted. A primary reason for the exclusion of these models is their
unique adoption of ``semi-grey'' surface boundary conditions. These
semi-grey surface boundary conditions are an analytical fit to the 
thermal structure of non-grey atmospheres. 
The analytical fit to non-grey atmosphere data reproduces
the original non-grey atmosphere structures to within about 20\%, suggesting
non-adiabatic conditions in the outer layers of cool stars are being
captured, at least in part. Comparing the \citet{Siess2000} mass 
tracks to Dartmouth model mass tracks, which adopt non-grey surface
boundary conditions (see Figure~\ref{fig:trackoverview}), we find that
the \citet{Siess2000} tracks predict systematically hotter effective
temperatures. This is the same effect one expects from the use of grey
$T(\tau)$ atmosphere relations at low-masses \citep{CB97}, leading us 
to believe that the semi-grey approach may be more akin to grey surface 
boundary conditions than non-grey. The morphology of the \citet{Siess2000} 
0.2 \msun\ track rather closely matches that of the \citet{BCAH98} 
models, although significant differences can be seen at 0.5~\msun, where
the \citet{Siess2000} models deviate quite noticeably. Our evaluation was 
further complicated by the fact that the 1.0~\msun\ mass track does
not reproduce the properties of the Sun, as evidenced in 
Figure~\ref{fig:trackoverview}(b), despite a solar calibration having been
performed \citep{Siess2000}.

\paragraph{\bf Padova} 
The latest version of the Padova stellar evolution code include all 
physics relevant for PMS work at higher masses \citep{Girardi2000,
Bressan2012}. They adopt the latest solar abundance values \citep{Caffau2011} 
and are available for a wide range of stellar masses and metallicities.
However, their adoption of an Eddington $T(\tau)$ grey surface boundary
condition gives them limited applicability for rigorous tests of the
validity of physical inputs at lower masses.

\paragraph{\bf BaSTI}
The Bag of Stellar Tracks and Isochrones (BaSTI) code is another 
stellar evolution code that has been updated in the past decade 
\citep{Pietrinferni2004}. As was the case with the Padova code, BaSTI is
applicable to a wide range of astrophysical problems, particularly 
population synthesis studies. It is excluded from the present study
due to its use of grey surface boundary conditions \citep{KS66}. This,
as has been mentioned previously, gives the code limited applicability 
for detailed studies near the bottom of the H--R diagram.

\paragraph{\bf Palla \& Stahler}
Models by \citet{Palla1993,Palla1999} were unique for their use of an empirically
calibrated stellar birthline used to define initial conditions for young stars. 
This is in contrast to most stellar evolution codes that assume an arbitrarily 
large radius at a very young age (e.g., 1000 years) as an initial condition. As 
with many older codes, these models use grey boundary conditions and are not 
publicly available.

\paragraph{\bf Swenson et al.}
Based on the Cambridge STARS code \citep{Eggleton1971,Eggleton1972}, the
models by \citet{Swenson1994} were designed for studying lithium abundances
in young stars. The original models were never updated beyond the original
release and are increasingly difficult to acquire. Their legacy lives on, in
part, through the Victoria stellar evolution models \citep[e.g.,][]{VandenBerg2000}.
At the time of their creation, non-grey atmosphere codes that were suitable
for low-mass and very-low-mass stars were just coming of age \citep{Allard1995},
so these models were released with grey atmospheres only.

\clearpage

\begin{landscape}
\begin{table}[!htbp]
    \centering
    \small
    \caption{Properties of pre-main-sequence evolutionary track sets adopted for
             use in this review.\label{tab:models_in}}
    \begin{tabular}{l c c c c c c c c}
    \noalign{\smallskip}\hline
    \hline\noalign{\smallskip}
    Track & Code & Masses & Metallicities & Surface Boundary Conditions &
        Radiative Opacities & EOS & Convection & D, Li Burning$^a$ \\
    Set   & Lineage & (\msun) & (solar) & (fit point; mixture) &
        (mixture) &  & (solar $\amlt$) & ($X_D \times 10^{5}$) \\
    \noalign{\smallskip}\hline\noalign{\smallskip}
    
    Lyon  & Lyon & 0.02 -- 1.5 & $-0.5$, $0.0$ & non-gray, \citet{Allard1995} &
       OPAL; AF94 & SCvH95 & MLT & Separate, Yes \\
     & & ($\amlt$ dependent) & (GN93; 0.019) & ($\tauf = 100$; GN93) & (GN93) & & 
     (1.9) & (2.0) \\
     
    & & \\
    
    Dartmouth 2008 &  YREC  &  0.08 -- 5.0  &  $-2.5 \to 0.5$ &  non-gray, \citet{Hauschildt1999a,Hauschildt1999b} &
             OPAL; F05 & FreeEOS & MLT & No, Separate \\
      & & & (GS98; 0.0169) & non-gray, \citet{Castelli2003} & (GS98) & 
             CK95 & (1.938) & (0.0) \\ 
      & & & & ($T(\tauf) =$ \teff; GS98) \\
    
    & & \\ 
    
    Dartmouth 2014 & YREC &  0.08 -- 5.0  &  $-1.0 \to 0.5$ &  non-gray, \citet{Hauschildt1999a,Hauschildt1999b} &
             OPAL; F05  & FreeEOS & MLT & Yes, Separate \\
      & & & (GS98; 0.0169) & non-gray, \citet{Castelli2003} & (GS98) &
         CK95 & (1.884) & (2.0) \\ 
      & & & & ($\tauf = 10$; GS98) & \\
      
    & & \\
    
    Brazil & ATON & 0.085 -- 3.8 & 0.0 & non-gray, \citet{Allard1995} & OPAL; AF94 & 
            OPAL96 & MLT & Yes, Separate \\
    & & & (GS99; 0.0175) & ($\tauf = 10$; GN93) & (AG89; GN93) & MHD & (2.0) & (2.0) \\
    
    & & \\
    
    Yale  &  YREC & 0.1 -- 1.25  & $-1.5 \to 0.3$ & non-gray, \citet{Allard2011} & OPAL; 
        F05 & SCvH95 & MLT & No, Separate \\ 
      & & & (GS98; 0.0163) & ($\tau = \ldots$; GN93) & (GS98) & OPAL05 & (1.875) & (0.0) \\
    
    & & \\
    
    Pisa  &  FRANEC & 0.20 -- 7.0  & $Z = 0.0002 \to 0.03$ & non-gray, \citet{Brott2005} 
          & OPAL; F05 & OPAL06 & MLT & Yes$^*$, Yes$^*$ \\
     & & & (AGS05; 0.01377) & ($\tauf = 10$; GN93) & (AGS05) & & (1.68) & (2.0, 4.0) \\
     & & & & non-gray, \citet{Castelli2003} & \\
     & & & & ($\tauf = 10$; GS98) \\
    
    \noalign{\smallskip}\hline
    \end{tabular}
    \vskip 5pt
    \parbox{0.95\linewidth}{
    
    {$^a$}\ {Deuterium (D) and lithium (Li) burning was assessed on two 
            levels. First, do the models include light element burning (Yes or No)?
            Second, if they do include D and Li burning, is it performed within 
            a larger nuclear reaciton network (Yes), outside of the regular nuclear 
            reaction network (Separate), or if it is unclear from reading the 
            literature (Yes$^*$).} \\
    
    {\bf References}\ 
            {{\it Track set}: Lyon, \citet{BCAH98}; Dartmouth 2008, \citet{Dotter2008};
               Dartmouth 2014, Feiden et al.\ (in prep); Brazil, \citet{Landin2010};
               Yale, \citet{Spada2013}; Pisa, \citet{Tognelli2011}; 
            {\it Heavy element mixture}: AG89, \citet{AG89}; GN93, \citet{GN93}; 
            GS98/99, \citet{GS98}; AGS05, \citet{Asplund2005}; AGSS09, 
            \citet{Asplund2009}; C11, \citet{Caffau2011};
            {\it Opacities}: AF94, \citet{AF94}; F05,
               \citet{Ferguson2005}; OPAL, \citet{Iglesias1996};
            {\it EOS}: EFF: \citet{Eggleton1973}; MHD, \citet{MDH88}; 
               PTEH, \citet{Pols1995};
               CK95, \citet{Chaboyer1995}; SCvH95, \citet{scvh95};
               OPAL, \citet{Iglesias1996}; OPAL06, \citet{Rogers2002};
               FreeEOS, \citet{Irwin2007};
            {\it Convection}: MLT, \citet{vitense53}, \citet{Mihalas1978},
               \citet{Henyey1965};} 
    }
\end{table}
\end{landscape}

\clearpage

\begin{landscape}    
\begin{table}[!htbp]
    \centering
    \small
    \caption{Properties of pre-main-sequence evolutionary track sets that were
             not adopted for this review.\label{tab:models_ex}}
    \begin{tabular}{l c c c c c c c c}
    \noalign{\smallskip}\hline
    \hline\noalign{\smallskip}
    Track & Code & Masses & Metallicities & Surface Boundary Conditions &
        Radiative Opacities & EOS & Convection & D, Li Burning$^a$ \\
    Set   & Lineage & (\msun) & (solar) & (fit point; mixture) &
        (mixture) &  & (solar $\amlt$) & ($X_D \times 10^{5}$) \\
    \noalign{\smallskip}\hline\noalign{\smallskip}
    
    B12 & Padova & 0.10 -- 12.0 & $0.0005 \to 0.07$ & gray, Eddington $T(\tau)$ &
        OPAL; MA09 & FreeEOS & MLT & Yes, Yes \\
    & & & (C11; 0.015) & ($\tauf = 2/3$; C11) & (C11) & & (1.74) & (?) \\
    
    & & \\
    
    DM97    & ATON &  0.015 -- 3.0 & $Z = 0.01,\, 0.02$ & gray, \citet{Henyey1965} &
             OPAL; AF94 & MHD & MLT, FST & Separate, Separate \\
      & & & (AG89; 0.019) & ($\tauf = 2/3$; AG89) & (AG89; GN93) & OPAL96 & (1.5) & 
      (1.5, 2.5, 4.5) \\
      
    & & \\
    
    DVD09 &  ATON & 0.10 -- 1.5 & $-2.0 \to 0.0$ & non-gray, \citet{Allard1995} & 
        OPAL; F05 & SCvH95 & MLT, FST & Yes$^{*}$, Yes$^{*}$\\
    MDKH04 & & & (GS99; 0.0175) & ($\tauf = 3$; GN93) & (GS99) & OPAL & (1.6) & 
       (2.0, 4.5) \\
      & & & & non-gray, \citet{Heiter2002} \\
      & & & & ($\tauf = 10$; GN93) \\
    
    & & \\
    
    PCSC04 & BaSTI & 0.50 -- 10.0 & $Z = 0.0001 \to 0.04$ & gray, \citet{KS66} & 
        OPAL; AF94 & FreeEOS & MLT & No, Yes \\
      & & ($Z$ dependent) & (GN93; 0.0198) & ($\tauf = 2/3$, GN93) & (GN93) & 
        & (1.931) & (0.0) \\
    
    & & \\
    
    PS99  & PS93 & 0.10 -- 6.0 & 0.0 & gray, Eddington $T(\tau)$ & OPAL; AF94 & 
        EFF & MLT & Separate, No \\
      & & & (0.02) & ($\tauf = 2/3$) & (GN93) & PTEH & (1.5) & (2.5) \\
    
    & & \\
    
    SDF00 & Grenoble & 0.10 -- 7.0 & $Z = 0.01 \to 0.04$ & semi-gray$^b$, \citet{Plez1992} &
        OPAL; AF94 & Augmented & MLT & Yes$^*$, Yes$^*$ \\
      & & & (GN93; 0.02) & semi-gray$^b$, \citet{Bell1976} & (GN93) & PTEH & (1.605) & 
        (2.0) \\
      & & & & semi-gray$^b$, \citet{Kurucz1991} \\
      & & & & ($\tauf = 10$, AG89) \\ 
      
    & & \\
    
    SFRI94 & STARS & 0.15 -- 5.0 & $-0.5$, 0.0 & gray, Eddington $T(\tau)$ & OPAL; A92 & Augmented & MLT & No, Yes \\
      & & & (AG89; 0.0188) & ($\tau_{\rm fit} = \ldots$; AG89) & (AG89) & EFF & (1.957) & \\

    \noalign{\smallskip}\hline 
    \end{tabular}
    \vskip 5pt
    \parbox{0.95\linewidth}{
    
    {$^a$}\ {Deuterium (D) and lithium (Li) burning was assessed on two 
            levels. First, do the models include light element burning (Yes or No)?
            Second, if they do include D and Li burning, is it performed within 
            a larger nuclear reaciton network (Yes), outside of the regular nuclear 
            reaction network (Separate), or if it is unclear from reading the 
            literature (Yes$^*$).} \\
            
    {$^b$}\ {\citet{Siess2000} use neither a fully gray or fully non-gray approach. 
            They calculate an analytical fit to the non-gray atmospheres to provide 
            their surface boundary conditions. However, this fit does not completely 
            reproduce the original non-gray atmosphere data (within 20\%), thus we 
            deem it a ``semi-gray'' approach.} \\

    {\bf References}\
        {{\it Track set}: B12, \citet{Bressan2012}; DM97, \citet{DAntona1994,
            DAntona1997}; DVD09, \citet{DVD09}; MDKH04, \citet{Montalban2004};
            PCSC04, \citet{Pietrinferni2004}; PS93/99, \citet{Palla1993,Palla1999}; 
            SDF00, \citet{Siess2000}; SFRI94, \citet{Swenson1994};
            {\it Heavy element mixture}: AG89, \citet{AG89}; GN93, \citet{GN93}; 
            GS98/99, \citet{GS98}; AGS05, \citet{Asplund2005}; AGSS09, 
            \citet{Asplund2009}; C11, \citet{Caffau2011};
            {\it Opacities}: AF94, \citet{AF94}; F05, \citet{Ferguson2005}; 
            OPAL, \citet{Iglesias1996}; MA09, \citet{Marigo2009};
            {\it EOSs}: EFF: \citet{Eggleton1973}; MHD, \citet{MDH88}; 
            PTEH, \citet{Pols1995};
            OPAL, \citet{Iglesias1996}; OPAL06, \citet{Rogers2002}; FreeEOS, 
            \citet{IrwinFEOSV};
            {\it Convection}: MLT, \citet{vitense53}, \citet{Mihalas1978},
            \citet{Henyey1965}; FST, \citet{Canuto1991};}
    }
\end{table}
\end{landscape}

\end{document}